\documentclass[article]{IEEEtran}
\usepackage{array}
\usepackage{amsfonts}
\usepackage{amsmath}
\usepackage{amssymb}
\usepackage{bm}
\usepackage{float}
\usepackage{enumerate}
\usepackage{color}
\usepackage{breqn} %for dmath
\usepackage{algorithm}
\usepackage{algorithmic}
\usepackage{stmaryrd}
\usepackage{mathtools}
\mathtoolsset{showonlyrefs}

\usepackage{tikz}
\usepackage{graphicx}
\usepackage{epsf}
\usepackage{epstopdf}
\usepackage{cite}
\usepackage{xcolor,pgf,pstricks,pgfplots}
\usepackage{subfigure}

\usepackage{tkz-berge}
\usepackage{tikz}
\usetikzlibrary{decorations.markings, arrows, automata}
\usepackage{calc}

\newrgbcolor{ColorLink}{0 .4 .6}
\newrgbcolor{ColorCite}{0 .5 .3}

\usepackage[plainpages=false,pdfpagelabels,colorlinks=true,linkcolor=ColorLink,citecolor=ColorCite]{hyperref}

\newtheorem{theorem}{Theorem}
\newtheorem{lemma}{Lemma}
\newtheorem{definition}{Definition}

\newcommand\scalemath[2]{\scalebox{#1}{\mbox{\ensuremath{\displaystyle #2}}}}

\usetikzlibrary{positioning,chains,fit,shapes,calc, arrows, shadows,trees}

\tikzstyle{vertex}=[circle, draw, inner sep=0pt, minimum size=5pt]
\tikzset{symbol/.style={rectangle, draw, very thick,
minimum size=10mm, rounded corners=1mm}}
\tikzset{symbol2/.style={rectangle , draw,  thick,
minimum size=35mm, rounded corners=1mm}}

\newtheorem{remark}{Remark}
\newtheorem{example}{Example}

\begin{document}

%\sloppy

\newcommand{\llb}{\llbracket}
\newcommand{\rrb}{\rrbracket}

\def\argmax{\operatornamewithlimits{arg\,max}}
\def\argmin{\operatornamewithlimits{arg\,min}}
\newcommand{\diag}{\mathop{\rm diag}\nolimits}
\newcommand{\tr}{\mathop{\rm Tr} \nolimits}
\newcommand{\SINRd}{\mathop{\rm \mathsf{SINR}}\nolimits}
\newcommand{\eig}{\mathop{\rm eig} \nolimits}
\newcommand {\cp}{\mathbf{p}}

\newcommand {\usinr}{\boldsymbol{\beta}_{\ut}} % uplink SINR vector 	
\newcommand {\dsinr}{\boldsymbol{\beta}_{\dt}} % downlink SINR vector 	

\newcommand {\F}{\mathrm{F}}
\newcommand {\C}{\mathrm{C}}
\newcommand {\kcl}{k_{\C,\ell}}
\newcommand {\kfl}{k_{\F,\ell}}
\newcommand {\kfmax}{k_{\F}}
\newcommand {\kcmin}{k_{\C}}
\newcommand {\kf}{k_{\F}}
\newcommand {\kc}{k_{\C}}

\def\L{\Lambda}

\newcommand {\ct}{\mathrm{c}}
\newcommand {\dt}{\mathrm{d}}
\newcommand {\ut}{\mathrm{u}}
\newcommand {\eff}{{\text{\textnormal{eff}}}}
\newcommand {\sic}{{\text{\textnormal{SIC}}}}
\newcommand {\dpc}{{\text{\textnormal{DPC}}}}

\newcommand {\upc}{\bm{\rho}_{\ct,\ut}} 	% uplink coding power vector
\newcommand {\pw}{\rho} 	         % single power entry

\newcommand {\upt}{\bm{\rho}_{\text{\textnormal{t}},\ut}} 	% uplink transmitting  power vector
\newcommand {\dpt}{\bm{\rho}_{\text{\textnormal{t}},\dt}} 	% downlink transmitting power vector

\newcommand {\upm}{\mathbf{P}_{\ut}} 	 % uplink coding power matrix
\newcommand {\dpm}{\mathbf{P}_{\dt}} 	 % downlink coding power matrix	

\newcommand {\ptu}{P_{\text{\textnormal{total}}}} % total power
\newcommand {\ptd}{P_{\text{\textnormal{total}}}}

\newcommand{\pul}{P_{\ut,\ell}} % individual power
\newcommand{\pdl}{P_{\dt,\ell}}

\newcommand {\T}{^{\mathsf{T}}}

\newcommand {\wvt}{\mathbf{\tilde{w}}}
\newcommand {\tvt}{\mathbf{\tilde{t}}}

\newcommand {\ewise}{\odot} % element wise multi

\newcommand {\umut}{} % uplink permutation
\newcommand {\dmut}{} % downlink permutation

\newcommand {\perm}{\mathcal{P}_{\pi}}
\newcommand {\perminv}{\mathcal{P}_{\pi^{-1}}}

\newcommand{\lth}{\ell^{\text{th}}}
\newcommand{\mth}{m^{\text{th}}}
\newcommand{\jth}{j^{\text{th}}}
\newcommand{\ith}{i^{\text{th}}}
\newcommand{\ex}{\mathbb{E}}
\newcommand{\pr}{\mathbb{P}}

%Fields, etc.
\newcommand{\CC}{\mathbb{C}}
\newcommand{\RR}{\mathbb{R}}
\newcommand{\ZZ}{\mathbb{Z}}
\newcommand{\NN}{\mathbb{N}}

% Vectors
\newcommand{\av}{{\mathbf a}}
\newcommand{\bv}{{\mathbf b}}
\newcommand{\cv}{{\mathbf c}}
\newcommand{\dv}{{\mathbf d}}
\newcommand{\ev}{{\mathbf e}}
\newcommand{\fv}{{\mathbf f}}
\newcommand{\gv}{{\mathbf g}}
\newcommand{\hv}{{\mathbf h}}
\newcommand{\iv}{{\mathbf i}}
\newcommand{\jv}{{\mathbf j}}
\newcommand{\kv}{{\mathbf k}}
\newcommand{\lv}{{\mathbf l}}
\newcommand{\mv}{{\mathbf m}}
\newcommand{\nv}{{\mathbf n}}
\newcommand{\ov}{{\mathbf o}}
\newcommand{\pv}{{\mathbf p}}
\newcommand{\qv}{{\mathbf q}}
\newcommand{\rv}{{\mathbf r}}
\newcommand{\sv}{{\mathbf s}}
\newcommand{\tv}{{\mathbf t}}
\newcommand{\uv}{{\mathbf u}}
\newcommand{\wv}{{\mathbf w}}
\newcommand{\vv}{{\mathbf v}}
\newcommand{\xv}{{\mathbf x}}
\newcommand{\yv}{{\mathbf y}}
\newcommand{\zv}{{\mathbf z}}
\newcommand{\zerov}{{\mathbf 0}}
\newcommand{\onev}{{\mathbf 1}}

% Bold greek letters
\newcommand{\alphav}{\boldsymbol{\alpha}}
\newcommand{\betav}{\boldsymbol{\beta}}
\newcommand{\gammav}{\boldsymbol{\gamma}}
\newcommand{\deltav}{\boldsymbol{\delta}}
\newcommand{\etav}{\boldsymbol{\eta}}
\newcommand{\lambdav}{\boldsymbol{\lambda}}
\newcommand{\epsilonv}{\boldsymbol{\epsilon}}
\newcommand{\nuv}{\boldsymbol{\nu}}
\newcommand{\muv}{\boldsymbol{\mu}}
\newcommand{\zetav}{\boldsymbol{\zeta}}
\newcommand{\phiv}{\boldsymbol{\phi}}
\newcommand{\psiv}{\boldsymbol{\psi}}
\newcommand{\thetav}{\boldsymbol{\theta}}
\newcommand{\tauv}{\boldsymbol{\tau}}
\newcommand{\chiv}{\boldsymbol{\chi}}
\newcommand{\upsilonv}{\boldsymbol{\upsilon}}
\newcommand{\omegav}{\boldsymbol{\omega}}
\newcommand{\xiv}{\boldsymbol{\xi}}
\newcommand{\sigmav}{\boldsymbol{\sigma}}
\newcommand{\piv}{\boldsymbol{\pi}}
\newcommand{\rhov}{\boldsymbol{\rho}}
\newcommand{\Gammam}{\boldsymbol{\Gamma}}
\newcommand{\Lambdam}{\boldsymbol{\Lambda}}
\newcommand{\Deltam}{\boldsymbol{\Delta}}
\newcommand{\Sigmam}{\boldsymbol{\Sigma}}
\newcommand{\Phim}{\boldsymbol{\Phi}}
\newcommand{\Pim}{\boldsymbol{\Pi}}
\newcommand{\Psim}{\boldsymbol{\Psi}}
\newcommand{\Thetam}{\boldsymbol{\Theta}}
\newcommand{\Omegam}{\boldsymbol{\Omega}}
\newcommand{\Xim}{\boldsymbol{\Xi}}

%Tilde vectors
\newcommand{\atv}{\mathbf{\tilde{a}}}
\newcommand{\btv}{\mathbf{\tilde{b}}}
\newcommand{\ctv}{\mathbf{\tilde{c}}}
\newcommand{\etv}{\mathbf{\tilde{e}}}
\newcommand{\ftv}{\mathbf{\tilde{f}}}
\newcommand{\gtv}{\mathbf{\tilde{g}}}
\newcommand{\htv}{\mathbf{\tilde{h}}}
\newcommand{\itv}{\mathbf{\tilde{i}}}
\newcommand{\jtv}{\mathbf{\tilde{j}}}
\newcommand{\ktv}{\mathbf{\tilde{k}}}
\newcommand{\ltv}{\mathbf{\tilde{l}}}
\newcommand{\mtv}{\mathbf{\tilde{m}}}
\newcommand{\ntv}{\mathbf{\tilde{n}}}
\newcommand{\otv}{\mathbf{\tilde{o}}}
\newcommand{\ptv}{\mathbf{\tilde{p}}}
\newcommand{\qtv}{\mathbf{\tilde{q}}}
\newcommand{\rtv}{\mathbf{\tilde{r}}}
\newcommand{\stv}{\mathbf{\tilde{s}}}
\newcommand{\ttv}{\mathbf{\tilde{t}}}
\newcommand{\utv}{\mathbf{\tilde{u}}}
\newcommand{\wtv}{\mathbf{\tilde{w}}}
\newcommand{\vtv}{\mathbf{\tilde{v}}}
\newcommand{\xtv}{\mathbf{\tilde{x}}}
\newcommand{\ytv}{\mathbf{\tilde{y}}}
\newcommand{\ztv}{\mathbf{\tilde{z}}}
\newcommand{\lambdatv}{\boldsymbol{\tilde{\lambda}}}

%Vector estimates
\newcommand{\shv}{\mathbf{\hat{s}}}
\newcommand{\thv}{\mathbf{\hat{t}}}
\newcommand{\uhv}{\mathbf{\hat{u}}}
\newcommand{\vhv}{\mathbf{\hat{v}}}
\newcommand{\whv}{\mathbf{\hat{w}}}
\newcommand{\muhv}{\boldsymbol{\hat{\mu}}}
\newcommand{\nuhv}{\boldsymbol{\hat{\nu}}}
\newcommand{\chihv}{\boldsymbol{\hat{\chi}}}

%Bar vectors
\newcommand{\wbv}{\mathbf{\bar{w}}}

% Matrices
\newcommand{\Am}{{\mathbf A}}
\newcommand{\Atm}{{\mathbf {\tilde{A}}}}
\newcommand{\Bm}{{\mathbf B}}
\newcommand{\Cm}{{\mathbf C}}
\newcommand{\Dm}{{\mathbf D}}
\newcommand{\Em}{{\mathbf E}}
\newcommand{\Fm}{{\mathbf F}}
\newcommand{\Gm}{{\mathbf G}}
\newcommand{\Gtm}{{\mathbf {\tilde{G}}}}
\newcommand{\Hm}{{\mathbf H}}
\newcommand{\Id}{{\mathbf I}}
\newcommand{\Jm}{{\mathbf J}}
\newcommand{\Km}{{\mathbf K}}
\newcommand{\Lm}{{\mathbf L}}
\newcommand{\Mm}{{\mathbf M}}
\newcommand{\Nm}{{\mathbf N}}
\newcommand{\Om}{{\mathbf O}}
\newcommand{\Pm}{{\mathbf P}}
\newcommand{\Qm}{{\mathbf Q}}
\newcommand{\Rm}{{\mathbf R}}
\newcommand{\Sm}{{\mathbf S}}
\newcommand{\Tm}{{\mathbf T}}
\newcommand{\Um}{{\mathbf U}}
\newcommand{\Wm}{{\mathbf W}}
\newcommand{\Vm}{{\mathbf V}}
\newcommand{\Xm}{{\mathbf X}}
\newcommand{\Ym}{{\mathbf Y}}
\newcommand{\Zm}{{\mathbf Z}}

% Matrix estimates
\newcommand{\Shm}{{\mathbf{\hat{S}}}}
\newcommand{\Uhm}{{\mathbf{\hat{U}}}}
\newcommand{\Vhm}{{\mathbf{\hat{V}}}}

% Bar matrices
\newcommand{\Abm}{{\mathbf {\bar{A}}}}
\newcommand{\Lbm}{{\mathbf {\bar{L}}}}

% Calligraphic
\newcommand{\Ac}{{\mathcal A}}
\newcommand{\Bc}{{\mathcal B}}
\newcommand{\Cc}{{\mathcal C}}
\newcommand{\Dc}{{\mathcal D}}
\newcommand{\Ec}{{\mathcal E}}
\newcommand{\Fc}{{\mathcal F}}
\newcommand{\Gc}{{\mathcal G}}
\newcommand{\Hc}{{\mathcal H}}
\newcommand{\Ic}{{\mathcal I}}
\newcommand{\Jc}{{\mathcal J}}
\newcommand{\Kc}{{\mathcal K}}
\newcommand{\Lc}{{\mathcal L}}
\newcommand{\Mc}{{\mathcal M}}
\newcommand{\Nc}{{\mathcal N}}
\newcommand{\Oc}{{\mathcal O}}
\newcommand{\Pc}{{\mathcal P}}
\newcommand{\Qc}{{\mathcal Q}}
\newcommand{\Rc}{{\mathcal R}}
\newcommand{\Sc}{{\mathcal S}}
\newcommand{\Tc}{{\mathcal T}}
\newcommand{\Uc}{{\mathcal U}}
\newcommand{\Wc}{{\mathcal W}}
\newcommand{\Vc}{{\mathcal V}}
\newcommand{\Xc}{{\mathcal X}}
\newcommand{\Yc}{{\mathcal Y}}
\newcommand{\Zc}{{\mathcal Z}}

% math
\def\argmax{\operatornamewithlimits{arg\,max}}
\def\argmin{\operatornamewithlimits{arg\,min}}

\colorlet{color1}{blue}
\colorlet{color2}{teal}
\colorlet{color3}{orange}
\colorlet{color4}{green!50!black}
\colorlet{color5}{orange!5!brown}
\colorlet{color6}{blue!40!white}
\colorlet{color7}{red}

% colors
\newrgbcolor{BoxRed}{1 .6 .6}
\newrgbcolor{BoxBlue}{.8 .8 1}
\newrgbcolor{BoxGreen}{.8 1 .8}
\newrgbcolor{BoxOrange}{1 .65 0}
\newrgbcolor{BoxPurple}{.87 .79 .85}
\newrgbcolor{LineGray}{.6 .6 .6}
\newrgbcolor{AxisGray}{.3 .3 .3}
\newrgbcolor{LineBlue}{.2 .6 .9}
\newrgbcolor{LineRed}{.9 .3 .3}
\newrgbcolor{LineGreen}{.3 .8 .3}
\newrgbcolor{LinePurple}{.8 .3 .8}

% revision change color
\newcommand{\bl}[1]{\textcolor{blue!95!black}{#1}}

\title{Uplink-Downlink Duality for Integer-Forcing}
\author{\thanks{W. He and B. Nazer were supported by NSF grants CCF-1253918 and CCF-1302600. The work of S. Shamai has been supported by the by the S. and N. Grand Research Fund, and the European Union's Horizon 2020 Research and
Innovation Programme, grant agreement no. 694630. This work was presented at the 2014 Communication Theory Workshop, 2014 IEEE International Symposium on Information Theory and the 15th IEEE International Symposium on Signal Processing Advances in Wireless Communications in 2014.}Wenbo He, Bobak Nazer,  \IEEEmembership{Member, IEEE}, and Shlomo Shamai (Shitz), \IEEEmembership{Fellow, IEEE} \thanks{W. He was with the Department of Electrical and Computer Engineering, Boston University, Boston, MA and is now with The Mathworks, Inc., Natick, MA, email: \texttt{whe02@bu.edu}, B. Nazer is with the Department of Electrical and Computer Engineering, Boston University, Boston, MA, email: \texttt{bobak@bu.edu}, and S. Shamai (Shitz) is with the EE Department, Technion, Haifa, Israel, email: \texttt{sshlomo@ee.technion.ac.il}.}}

\markboth{IEEE Trans Info Theory, to appear}{~}

\maketitle

\begin{abstract}  
Consider a Gaussian multiple-input multiple-output (MIMO) multiple-access channel (MAC) with channel matrix $\mathbf{H}$ and a Gaussian MIMO broadcast channel (BC) with channel matrix $\mathbf{H}\T$. For the MIMO MAC, the integer-forcing architecture consists of first decoding integer-linear combinations of the transmitted codewords, which are then solved for the original messages. For the MIMO BC, the integer-forcing architecture consists of pre-inverting the integer-linear combinations at the transmitter so that each receiver can obtain its desired codeword by decoding an integer-linear combination. In both cases, integer-forcing offers higher achievable rates than zero-forcing while maintaining a similar implementation complexity. This paper establishes an uplink-downlink duality relationship for integer-forcing, i.e., any sum rate that is achievable via integer-forcing on the MIMO MAC can be achieved via integer-forcing on the MIMO BC with the same sum power and vice versa. Using this duality relationship, it is shown that integer-forcing can operate within a constant gap of the MIMO BC sum capacity. Finally, the paper proposes a duality-based iterative algorithm for the non-convex problem of selecting optimal beamforming and equalization vectors, and establishes that it converges to a local optimum.
\end{abstract}

%\begin{keywords} 
%MIMO, duality, broadcast, multiple-access, integer-forcing, dirty-paper coding, power allocation, optimization
%\end{keywords} 

\section{Introduction} \label{s:intro}

The capacity region of the Gaussian MIMO MAC is well-known~\cite[Sec. 10.1]{tseviswanath} and can be attained via joint maximum likelihood (ML) decoding. While joint ML decoding is optimal, its implementation complexity  grows exponentially with the number of users. This has lead to considerable interest in linear receiver architectures~\cite{verdu,lv89,mh94}, which rely only on single-user decoding algorithms. A conventional linear receiver consists of a linear equalizer that generates an estimate of each user's codeword followed by parallel single-user decoders. However, even with optimal minimum mean-squared error (MMSE) estimation, linear receivers fall short of the MIMO MAC sum capacity. This gap can be closed via successive interference cancellation (SIC), provided that the transmitters operate at one of the corner points of the capacity region. The full capacity region can be attained via SIC combined with either time sharing~\cite{vg97,wfgv98} or rate splitting~\cite{ru96}. 

Although the Gaussian MIMO BC is non-degraded, its capacity region can be established via its uplink-downlink duality with the Gaussian MIMO MAC~\cite{cs03,vjg03,vt03,yc04,wss06}. Specifically, uplink-downlink duality refers to the fact that any rate tuple that is attainable on the Gaussian MIMO MAC with channel matrix $\mathbf{H}$ is also attainable on the ``dual'' Gaussian MIMO BC with channel matrix $\mathbf{H}\T$ using the same sum power and vice versa. The capacity region is attained via dirty-paper coding~\cite{costa83}, which requires high implementation complexity at the transmitter. As with the MIMO MAC, significant effort has gone towards characterizing the performance of linear transmitter architectures for the MIMO BC (see, for instance,~\cite{yg06,wes08}), which are suboptimal in general. As demonstrated by~\cite{vt03}, linear transceiver architectures also satisfy uplink-downlink duality. That is, given equalization and beamforming matrices for a MIMO MAC, we can achieve the same rate tuple on the dual MIMO BC with the same sum power by exchanging the roles of the equalization and beamforming matrices. 

Integer-forcing is a variation on conventional linear transceiver architectures that can attain significantly higher sum rates. Rather than using the equalization and beamforming matrices to separate users' codewords, an integer-forcing transceiver employs them to create an integer-valued effective channel matrix. The single-user decoders are then used to recover integer-linear combinations of the codewords. By selecting an integer-valued effective matrix that closely approximates the channel matrix, this transceiver can reduce the effective noise variances seen by the decoders, leading to higher rates. In the MIMO MAC, these integer-linear combinations are solved for the original codewords~\cite{zneg14}. In the MIMO BC, the transmitter applies the inverse linear transform to its messages prior to encoding, so that each integer-linear combination corresponds to the desired message of that user~\cite{hc12,hc13}. Here, we demonstrate that integer-forcing transceivers satisfy uplink-downlink duality for the sum rate in the sense of~\cite{vt03}. At a high level, this means that the sum rate achievable for decoding the integer-linear combinations with integer coefficient matrix $\mathbf{A}$ over a MIMO MAC with channel matrix $\mathbf{H}$ is also achievable for decoding the integer-linear combinations with integer coefficient matrix $\mathbf{A}\T$ over a MIMO BC with channel matrix $\mathbf{H}\T$ by exchanging the roles of the equalization and beamforming matrices. One technical obstacle is that the effective noise variances may be associated with different users in the dual channel, which in turn means we can only guarantee duality in terms of the sum rate.

We also present two applications of uplink-downlink duality: a constant-gap optimality result for downlink integer-forcing and an iterative algorithm for optimizing beamforming and equalization matrices. To motivate the first application, prior work~\cite{oen14,ncnc16} established that integer-forcing can operate within a constant gap of the MIMO MAC sum capacity using only ``digital'' successive cancellation, assuming that channel state information (CSI) is available at the transmitters. Using duality, we demonstrate that integer-forcing can also operate within a constant gap of the MIMO BC capacity using only ``digital'' dirty-paper coding, again assuming CSI is available at the transmitter. For the second application, it is well-known that simultaneously optimizing beamforming and equalization matrices corresponds to a non-convex problem. However, for both the MIMO MAC and BC, finding the optimal equalization matrix for a fixed beamforming matrix has a closed-form solution. Therefore, a natural algorithm is to iterate between a problem and its dual, updating the equalization matrix at every iteration. For example, the maxSINR algorithm~\cite{gcj11} relies on a variation of this idea  to identify good interference alignment solutions. Here, we propose an iterative algorithm for optimizing the beamforming and equalization matrices used in integer-forcing and show that it converges to a local optimum. Recent follow-up work has used a variation of our algorithm to identify good integer-forcing interference alignment solutions~\cite{ehn15}.

%%%%%%%%%

\subsection{Related Work}

Prior work on integer-forcing~\cite{zneg14,oen13,hc12,hc13} has focused on the important special case where all codewords have the same effective power. This constraint is implicitly imposed by the original compute-and-forward framework~\cite{ng11IT}. In order to establish uplink-downlink duality, we need the flexibility to allocate power unequally across codewords. We will thus employ the expanded compute-and-forward framework~\cite{ncnc16}, which can handle unequal powers. Our achievability results draw upon capacity-achieving nested lattice codes, whose existence has been shown in a series of recent works~\cite{zf96,loeliger97,ftc00,zse02,elz05,oe12eilat}. We refer interested readers to the textbook of Zamir for a detailed history as well as a comprehensive treatment of lattice codes~\cite{zamir}.

For the sake of notational simplicity, we will state all of our results for real-valued channels. Analogous results can be obtained for complex-valued channels via real-valued decompositions. Recent efforts have shown that compute-and-forward can also be realized for more general algebraic structures~\cite{fsk13}. For instance, building lattices from the Eisenstein integers yields better approximations for complex numbers on average, and can increase the average performance of compute-and-forward~\cite{thbn14}. 

Here, we will assume that full channel state information (CSI) is available to the transmitter and receiver, in order to optimize the beamforming matrices and power allocations. However, CSI may not always be available, especially at the transmitter. The original integer-forcing paper~\cite{zneg14} numerically demonstrated the performance gains over conventional linear receivers in terms of outage rates. It also established that, if each antenna encodes an independent data stream, then integer-forcing attains the optimal diversity-multiplexing tradeoff~\cite{zt03}. Subsequently, it was shown that if the transmitter mixes the data streams using a space-time code with a non-vanishing determinant, then integer-forcing operates within a constant gap of the capacity~\cite{oe15}. Recent work has also studied the advantages of a random precoding matrix on the outage probability for integer-forcing~\cite{de16}.

Integer-forcing can also serve as a framework for distributed source coding, and can be viewed as the ``dual'' of integer-forcing channel coding in a certain sense. See~\cite{oe13b,hn16ISIT} for further details. Very recent work has also established uplink-downlink duality for compression-based strategies for cloud radio access networks~\cite{lpy16}. 

Finally, we note that there is a rich body of work on lattice-aided reduction~\cite{yw02,wfh04,tmk07,tmk07lll,glm09,je10,wsjm11} for MIMO channels. For instance, in the uplink version of this strategy, each transmitter employs a lattice-based constellation (such as QAM). The decoder steers the channel to a full-rank integer matrix using equalization, makes hard estimates of the resulting integer-linear combinations of lattice symbols, and then applies the inverse integer matrix to obtain estimates of the emitted symbols. Roughly speaking, integer-forcing can be viewed as lattice-aided reduction that operates on the codeword, rather than symbol, level. This in turn allows us to write explicit achievable rate expressions for integer-forcing, whereas rates for lattice-aided reduction must be evaluated numerically. 

\subsection{Paper Organization}

The remainder of this paper is organized  as follows. In Section~\ref{s:problemstatement}, we give problem statements for the uplink and downlink. Next, in Section~\ref{s:overview}, we give a high-level overview of our duality results. Section~\ref{s:nestedlattice} provides background results from nested lattice coding that will be needed for our achievability scheme. We present detailed achievability strategies for the uplink and downlink in Sections~\ref{s:uplink} and~\ref{s:downlink}, respectively. Section~\ref{s:duality} formally establishes an uplink-downlink duality relationship for integer-forcing, and handles the technical issue associated with different effective noise variance associations across the MIMO MAC and BC. In Section~\ref{s:optimization}, we propose an iterative algorithm that uses uplink-downlink duality for optimizing the integer-forcing beamforming, equalization, and integer matrices. We provide simulations in Section~\ref{s:numerical} and Section~\ref{s:conclude} concludes the paper.

\subsection{Notation}\label{ss: notation}

We will make use of the following notation. Column vectors will be denoted by boldface, lowercase font (e.g., $\mathbf{a}\in \mathbb{Z}^L$) and matrices with boldface, uppercase font (e.g., $\mathbf{A}\in \mathbb{Z}^{L\times L}$). Let $\mathbf{a}[i]$ denote the $i^{\text{th}}$ coordinate of the vector $\mathbf{a}$. We will use $\|\mathbf{a}\|$ to represent $\ell_2$-norm of $\mathbf{a}$, $\tr(\mathbf{A})$ to represent the trace of $\mathbf{A}$, and $\eig(\Am)$ to denote the set of eigenvalues (i.e., spectrum) of $\mathbf{A}$. We will also use $\mathrm{diag}(\mathbf{a})$ to denote the diagonal matrix formed by using the placing the elements of $\mathbf{a}$ along the diagonal. All logarithms are taken to base $2$ and we define $ \log^+(x) \triangleq \max(0, \log{x})$. We denote the identity matrix by $\mathbf{I}$,  the all-ones column vector of length $k$ by $\mathbf{1}_k$  and the all-zeros column vector of length $k$ by $\mathbf{0}_k$. 

We will work with both the real field $\mathbb{R}$ and prime-sized finite fields $\mathbb{Z}_p = \{0,1,\ldots,p-1\}$ where $p$ is prime.\footnote{We note that some mathematicians prefer to use the notation $\mathbb{Z} / p \mathbb{Z}$ or $\mathbb{Z} / (p)$ to avoid confusion with the $p$-adic integers, which are often denoted by $\mathbb{Z}_p$. However, since we do not invoke the $p$-adic integers and will often use superscripts to denote vector spaces, we prefer to use the notation $\mathbb{Z}_p$ for the finite field.}  We will denote addition and summation over the reals by $+$ and $\sum$, respectively. For a prime-sized finite field, we will use $\oplus$ and $\bigoplus$ to denote addition and summation, respectively.  Define $[a]\bmod{p}$ to be the modulo-$p$ reduction of $a$. For vectors and matrices, the modulo-$p$ operation is taken elementwise and denoted by $[\mathbf{a}]\bmod{p}$ and  $[\mathbf{A}]\bmod{p}$, respectively. Taking a linear combination over a prime-sized finite field can be linked to taking a linear combination over the reals as follows,
\begin{align}
q_1 w_1 \oplus q_2 w_2 = [q_1 w_1 + q_2 w_2] \bmod{p} \ .
\end{align} Note that, on the left-hand side, $q_1,q_2,w_1,w_2$ are elements of the finite field whereas, on the right-hand side, they are elements of the integers under the natural mapping. Finally, subscripts ``$\text{u}$'' and ``$\text{d}$'' will be used to denote variables associated with the uplink and downlink, respectively. 

\section{Problem Statement}\label{s:problemstatement}

\begin{figure*}[!h]
\psset{unit=.73mm}
\begin{center}
\begin{pspicture}(10,10)(228,64)
%\psframe(10,10)(228,64)
%\small

\rput(-10,0){
%user 1
\psframe(20,50)(33,60)
\rput(26.5,55){$\mathsf{Tx}~1$} \rput(44,58.75){$\mathbf{X}_{\ut,1}$}
\psline[linecolor=black]{->}(33,55)(55,55)
\psframe(55,50)(68,60) \rput(61.5,55){$\mathbf{H}_{\ut,1}$}
\psline{->}(68,55)(77,55)(84,42)

%user 2
\psframe(20,35)(33,45)
\rput(26.5,40){$\mathsf{Tx}~2$} \rput(44,43.75){$\mathbf{X}_{\ut,2}$}
\psline[linecolor=black]{->}(33,40)(55,40)
\psframe(55,35)(68,45) \rput(61.5,40){$\mathbf{H}_{\ut,2}$}
\psline{->}(68,40)(83,40)

%user L
\psframe(20,10)(33,20)
\rput(26.5,15){$\mathsf{Tx}~L$} \rput(44,18.75){$\mathbf{X}_{\ut,L}$}
\psline[linecolor=black]{->}(33,15)(55,15)
\psframe(55,10)(68,20) \rput(61.5,15){$\mathbf{H}_{\ut,L}$}
\psline{->}(68,15)(77,15)(84,38)

\rput(28,29){\large{$\vdots$}}
\rput(61.5,29){\large{$\vdots$}}

%Channel
\pscircle(85.5,40){2.5} \psline{-}(84.25,40)(86.75,40)
\psline{-}(85.5,38.75)(85.5,41.25) \psline{<-}(85.5,42.5)(85.5,50) \rput(85.5,53){$\mathbf{Z}_{\ut}$}

%Decoder
\psline{->}(88,40)(100,40) \rput(93.5,43.75){$\mathbf{Y}_{\ut}$}
\psframe(100,35)(113,45) \rput(106.5,40){$\mathsf{Rx}$}
}

\rput(110,0){

%transmitter
\psframe(20,35)(33,45)
\rput(26.5,40){$\mathsf{Tx}$} \rput(40,43.75){$\mathbf{X}_{\dt}$}
\psline{->}(47,40)(55,55)(63,55)
\psline{->}(33,40)(63,40)
\psline{->}(47,40)(55,15)(63,15)
\psframe(63,50)(76,60) \rput(69.5,55){$\mathbf{H}_{\dt,1}$}
\psframe(63,35)(76,45) \rput(69.5,40){$\mathbf{H}_{\dt,2}$}
\psframe(63,10)(76,20) \rput(69.5,15){$\mathbf{H}_{\dt,L}$}

\psline{->}(76,55)(83,55)
\psline{->}(76,40)(83,40)
\psline{->}(76,15)(83,15)

\rput(69.5,29){\large{$\vdots$}}
%\rput(61.5,29){\large{$\vdots$}}

%Channels
\pscircle(85.5,55){2.5} \psline{-}(84.25,55)(86.75,55)
\psline{-}(85.5,53.75)(85.5,56.25) \psline{<-}(85.5,57.5)(85.5,61) \rput(87,63.5){$\mathbf{Z}_{\dt,1}$}

\pscircle(85.5,40){2.5} \psline{-}(84.25,40)(86.75,40)
\psline{-}(85.5,38.75)(85.5,41.25) \psline{<-}(85.5,42.5)(85.5,46) \rput(87,48.5){$\mathbf{Z}_{\dt,2}$}

\pscircle(85.5,15){2.5} \psline{-}(84.25,15)(86.75,15)
\psline{-}(85.5,13.75)(85.5,16.25) \psline{<-}(85.5,17.5)(85.5,21) \rput(87,23.5){$\mathbf{Z}_{\dt,L}$}

%Decoder
\psline{->}(88,55)(104,55) \rput(97,58.5){$\mathbf{Y}_{\dt,1}$}
\psframe(104,50)(117,60) \rput(110.5,55){$\mathsf{Rx}~1$}

\psline{->}(88,40)(104,40) \rput(97,43.5){$\mathbf{Y}_{\dt,2}$}
\psframe(104,35)(117,45) \rput(110.5,40){$\mathsf{Rx}~2$}

\psline{->}(88,15)(104,15) \rput(97,18.5){$\mathbf{Y}_{\dt,L}$}
\psframe(104,10)(117,20) \rput(110.5,15){$\mathsf{Rx}~L$}
}

\psline[linewidth=3pt,linecolor=LineBlue,linestyle=dashed](116.5,8)(116.5,64)
\rput(97,25){\textbf{Uplink}}
\rput(97,20){\textbf{Channel}}

\rput(136.5,25){\textbf{Downlink}}
\rput(136.5,20){\textbf{Channel}}

\end{pspicture}
\end{center}
\caption{Block diagram of the uplink and downlink channel models. We say that the channels are duals of each other if their channel matrices satisfy $\mathbf{H}_{\dt,\ell} = \mathbf{H}_{\ut,\ell}\T$.}
\label{f:channelmodel}
\end{figure*}
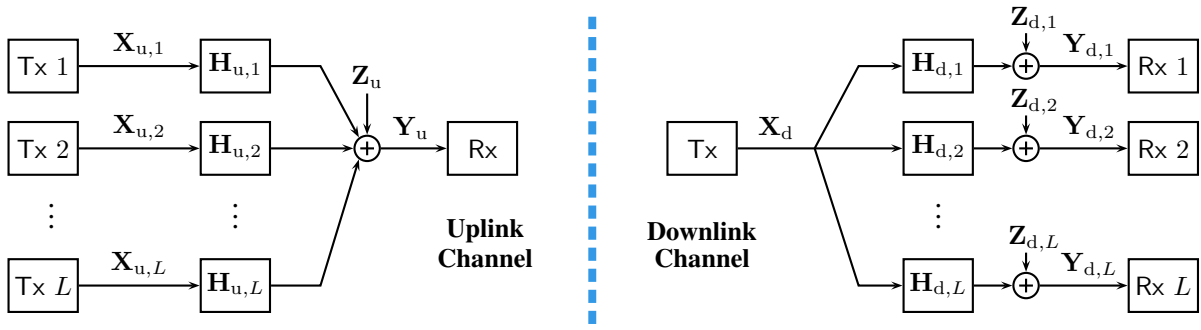

We now give problem statements for the uplink and downlink. See Figure~\ref{f:channelmodel} for a block diagram. We focus on real-valued channels, and note that our results are directly applicable to complex-valued channels by using a real-valued decomposition as in~\cite{zneg14}. Throughout the paper, we assume that full CSI is available at the transmitters and receivers.\footnote{For the case of CSI at the receivers only, our expressions can be suitably modified to obtain outage rate expressions, along the lines of~\cite{zneg14}.}

\noindent\textbf{Uplink Channel.} The uplink channel (i.e., MIMO MAC) consists of $L$ transmitters and a single $N$-antenna receiver. The $\ell^{\text{th}}$ transmitter is equipped with $M_{\ell}$ transmit antennas. It has a \textit{message} $w_{\ut,\ell}$ that is drawn independently and uniformly from $\{1,2,\ldots,2^{nR_{\ut,\ell}} \}$ and an \textit{encoder} $\mathcal{E}_{\ut,\ell}: \{1,2,\ldots,2^{nR_{\ut,\ell}} \} \rightarrow  \mathbb{R}^{M_{\ell} \times n}$ that maps this message into a \textit{channel input} $\mathbf{X}_{\ut,\ell} = \mathcal{E}_{\ut,\ell}(w_{\ut,\ell})$ of blocklength $n$. It will often be convenient to work with the concatenation of the channel inputs 
\begin{align}
\mathbf{X}_{\ut} \triangleq  \begin{bmatrix} \mathbf{X}_{\ut,1} \\ \vdots \\ \mathbf{X}_{\ut,L} \end{bmatrix}, 
\end{align} which is of dimension $M \times n$ where $M = \sum_{\ell}M_{\ell}$ denotes the total number of transmit antennas. The transmitters must satisfy a \textit{total power constraint} $\mathbb{E}\big[\tr(\mathbf{X}_{\ut} \mathbf{X}_{\ut}\T )\big] \le n \ptu$. 

The receiver observes a \textit{noisy linear combination} of the emitted signals, 
\begin{align} \label{e:uplinkchannel}
\mathbf{Y}_{\ut} = \sum_{\ell=1}^L \mathbf{H}_{\ut,\ell} \mathbf{X}_{\ut,\ell} + \mathbf{Z}_{\ut}
\end{align} where $\mathbf{H}_{\ut,\ell} \in\mathbb{R}^{N \times M_\ell}$ is the \textit{channel matrix}  from the $\ell^{\text{th}}$ transmitter to the receiver and the additive noise $\mathbf{Z}_{\ut} \in \mathbb{R}^{N \times n}$ is elementwise i.i.d. Gaussian with mean zero and variance one. We denote the concatenated channel matrices by 
\begin{align}
\mathbf{H}_{\ut} \triangleq \big[ \mathbf{H}_{\ut,1} ~ \cdots ~ \mathbf{H}_{\ut,L} \big] \ , 
\end{align} which lets us concisely write the channel output as 
\begin{align}\mathbf{Y}_{\ut} = \mathbf{H}_{\ut} \mathbf{X}_{\ut} + \mathbf{Z}_{\ut} \ .
\end{align} This channel output is sent through a \textit{decoder} $\mathcal{D}_{\ut}: \mathbb{R}^{N \times n} \rightarrow \{1,2,\ldots,2^{nR_1}\} \times \cdots \times \{1,2,\ldots,2^{nR_L}\}$ that produces estimates of the messages, $(\hat{w}_{\ut,1},\ldots,\hat{w}_{\ut,L}) = \mathcal{D}_{\ut}(\Ym_{\ut})$.

Overall, we say that the uplink rates $R_{\ut,1},\ldots,R_{\ut,L}$ are \textit{achievable} if, for any $\epsilon > 0$ and $n$ large enough, there exist encoders and decoder such that $\pr\big( \bigcup_{\ell = 1}^L \{ \hat{w}_{\ut,\ell} \neq w_{\ut,\ell} \} \big) < \epsilon$. The uplink capacity region is the closure of the set of all achievable rates. 

\noindent\textbf{Downlink Channel.} The downlink channel model mirrors the uplink channel model. There is a single $N$-antenna transmitter and $L$ receivers. Let $M_{\ell}$ represent the number of antennas at the $\ell^{\text{th}}$ receiver and let $M=\sum_{\ell}M_{\ell}$ be the total number of receive antennas. The transmitter has $L$ messages: the $\lth$ message $w_{\dt,\ell}$ is drawn independently and uniformly from $\{1,2,\ldots,2^{nR_{\dt,\ell}}\}$ and is intended for the $\lth$ receiver. The transmitter uses an \textit{encoder} $\mathcal{E}_{\dt}: \{1,2,\ldots,2^{nR_{\dt,1}}\} \times \{1,2,\ldots,2^{nR_{\dt,L}}\} \rightarrow \RR^{N \times n}$ to map these messages into a \textit{channel input} $\mathbf{X}_{\dt} = \mathcal{E}_{\dt}(w_{\dt,1},\ldots,w_{\dt,L})$ where $n$ represents the blocklength. This channel input must satisfy a \textit{total power constraint} $\mathbb{E}\big[\tr(\mathbf{X}_{\dt} \mathbf{X}_{\dt}\T )\big] \le n \ptd$. 

 For $m = 1,\ldots,L$, the \textit{channel output} observed by the $\mth$ receiver is
\begin{align} \label{e:downlinkchannel}
\mathbf{Y}_{\dt,m} = \mathbf{H}_{\dt,m} \mathbf{X}_{\dt} + \mathbf{Z}_{\dt,m}
\end{align} where $\mathbf{H}_{\dt,m} \in \mathbb{R}^{M_m \times N}$ is the channel matrix from the transmitter to the $m^{\text{th}}$ receiver and the noise $\mathbf{Z}_{\dt,m}  \in \mathbb{R}^{M_m \times n}$ is elementwise i.i.d.~Gaussian with mean zero and variance one. The receiver passes its channel output through a \textit{decoder} $\mathcal{D}_{\dt,m}: \RR^{M_m \times n} \rightarrow \{1,2,\ldots,2^{nR_{\dt,m}}\}$ in order to get an estimate $\hat{w}_{\dt,m} = \mathcal{D}_{\dt,m}(\Ym_{\dt,m})$ of its desired message. 

Overall, we say that the downlink rates $R_{\dt,1},\ldots,R_{\dt,L}$ are achievable if, for any $\epsilon > 0$ and $n$ large enough, there exist an encoder and decoders such that $\pr\big( \bigcup_{\ell = 1}^L \{ \hat{w}_{\dt,\ell} \neq w_{\dt,\ell} \} \big) < \epsilon$. The downlink capacity region is the closure of the set of all achievable rates. 

Finally, it will often be useful to work with the following concatenated matrices,
\begin{align}
\mathbf{Y}_{\dt} \triangleq \begin{bmatrix} \mathbf{Y}_{\dt,1} \\ \vdots \\ \mathbf{Y}_{\dt,L}\end{bmatrix} \qquad 
\mathbf{H}_{\dt} \triangleq \begin{bmatrix} \mathbf{H}_{\dt,1} \\ \vdots \\ \mathbf{H}_{\dt,L} \end{bmatrix} \qquad
\mathbf{Z}_{\dt} \triangleq \begin{bmatrix}\mathbf{Z}_{\dt,1} \\ \vdots \\ \mathbf{Z}_{\dt,L} \end{bmatrix} \ ,
\end{align} which enable us to compactly write the downlink channel output as \begin{align}\mathbf{Y}_{\dt} = \mathbf{H}_{\dt} \mathbf{X}_{\dt} + \mathbf{Z}_{\dt} \ . \end{align}

\begin{remark} Conventional MAC models impose a power constraint on each user individually. However, it is well-known that uplink-downlink duality can be established only if we are free to reallocate the power across transmitters~\cite{vjg03,vt03,yc04}. Note also that we use an expected power constraint rather than a hard power constraint of the form $\tr(\mathbf{X}_{\ut} \mathbf{X}_{\ut}\T ) \le n \ptu$. In order to impose a hard power constraint, we would first need to show that the nested lattice ensemble from~\cite{oe12eilat}, is also good for covering in the sense of~\cite{elz05}. This is currently an open question and beyond the scope of this paper. Alternatively, we could keep only a constant fraction of each codebook, throwing out the codewords with the highest powers. This would result in codebooks that meet hard power constraints and achieve the same rates, at the cost of disrupting the symmetry of the encoding scheme.
\end{remark}

\section{Overview of Main Results} \label{s:overview}

We now give a high-level overview of our main results. To put our results in context, we begin by stating the capacity regions for the uplink and downlink. We then give a quick summary of the rates achievable via conventional linear architectures and their uplink-downlink duality relationships. Finally, we overview our integer-forcing architectures for the uplink and downlink and state our uplink-downlink duality, capacity approximation, and algorithmic results. 

\subsection{Capacity Regions} \label{s:capacityregions}

\noindent\textbf{Uplink Channel.} The uplink (i.e., MIMO MAC) capacity region $\mathcal{C}_{\ut}$ is the set of rate tuples $(R_{\ut,1},\ldots,R_{\ut,L})$ satisfying 
\begin{align} \label{e:uplinkcapacity}
\sum_{\ell \in \mathcal{S}} R_{\ell} \leq \frac{1}{2} \log \det \bigg( \Id + \sum_{\ell \in \mathcal{S}} \Hm_{\ut,\ell} \Km_\ell \Hm_{\ut,\ell}\T \bigg)
\end{align} for all subsets $\mathcal{S} \subseteq \{1,2,\ldots,L\}$ and for some positive semi-definite matrices $\Km_1,\ldots,\Km_L$ satisfying the sum power constraint $\sum_{\ell = 1}^L \tr(\Km_{\ell}) \leq \ptu$. It can be attained with i.i.d. Gaussian encoding and simultaneous joint typicality decoding. Alternatively, it can be attained with i.i.d.~Gaussian encoding, successive interference cancellation decoding, and time sharing~\cite{vg97,wfgv98} or rate splitting~\cite{ru96}. See~\cite[\S 9.2.1]{elgamalkim} for more details. 

\noindent\textbf{Downlink Channel.} As shown by~\cite{wss06}, the downlink (i.e., MIMO BC) capacity region $\mathcal{C}_{\dt}$ is the convex hull of the set of rate tuples $(R_{\dt,1},\ldots,R_{\dt,L})$ satisfying 
\begin{align} \label{e:downlinkcapacity}
R_{\theta(\ell)} \leq \frac{1}{2} \log\left( \frac{ \det\Big(\Id + \displaystyle\sum_{m \geq \ell} \Hm_{\dt,\theta(m)} \Km_{\theta(m)} \Hm_{\dt,\theta(m)}\T \Big)}{ \det\Big(\Id + \displaystyle\sum_{m > \ell} \Hm_{\dt,\theta(m)} \Km_{\theta(m)} \Hm_{\dt,\theta(m)}\T \Big)}\right)~
\end{align} for some permutation $\theta$ of $\{1,2,\ldots,K\}$ and positive semi-definite matrices $\Km_1,\ldots,\Km_L$ satisfying the sum power constraint $\sum_{\ell = 1}^L \tr(\Km_\ell) \leq \ptd$. It can be attained using dirty-paper coding at the transmitter, joint typicality decoding at the receivers, and time sharing. See~\cite{wss06} or~\cite[\S 9.6.4]{elgamalkim} for more details.

\noindent\textbf{Uplink-Downlink Duality.} It can be argued that the uplink and downlink capacity regions described above are equal to one another, $\mathcal{C}_{\ut} = \mathcal{C}_{\dt}$. This was first shown for the sum-capacity~\cite{vjg03,vt03,yc04} and then for the full capacity region~\cite{wss06}.

\subsection{Conventional Linear Architectures} We begin with a summary of classical linear uplink and downlink architectures and their duality relationships. 

\noindent\textbf{Uplink Channel.} The $\lth$ transmitter has a codeword $\sv_{\ut,\ell} \in \RR^n$ with expected power $\frac{1}{n} \ex\| \sv_{\ut,\ell} \|^2 = \pul$. It uses a beamforming vector $\cv_{\ell} \in \RR^{M_\ell}$ to generate its channel input 
\begin{align}
\Xm_{\ut,\ell} = \cv_{\ut,\ell} \sv_{\ut,\ell}\T \ . 
\end{align} Collecting the beamforming vectors into the matrix
\begin{align} \label{e:uplinkbeamforming}
&\mathbf{C}_{\ut} \triangleq
\begin{bmatrix}
\cv_{\ut,1} & \mathbf{0}_{M_1} & \hdots & \mathbf{0}_{M_1}\\
\mathbf{0}_{M_2} & \cv_{\ut,2}  & \hdots & \mathbf{0}_{M_2}\\
\vdots & \vdots  & \ddots & \vdots\\
\mathbf{0}_{M_L} & \mathbf{0}_{M_L}& \hdots  & \cv_{\ut,L} \\
\end{bmatrix} 
\end{align} and the codewords into the matrix
\begin{align}
\Sm_{\ut} \triangleq \begin{bmatrix}
\sv_{\ut,1}\T \\ \vdots \\ \sv_{\ut,L}\T
\end{bmatrix} \ ,
\end{align}
we can write the beamforming operation as
\begin{align}
\Xm_{\ut} = \Cm_{\ut} \Sm_{\ut} \ . 
\end{align}

To recover the $\mth$ codeword, the receiver uses an equalization vector $\bv_{\ut,m} \in \RR^{N}$ to obtain the effective channel output 
\begin{align}
&\ytv_{\ut,m}\T\\  &= \bv_{\ut,m}\T \Ym_{\ut}  \\
&= \underbrace{\bv_{\ut,m}\T  \Hm_{\ut,m} \cv_{\ut,m} \sv_{\ut,m}\T}_{\text{signal}} + \underbrace{\sum_{\ell \neq m} \bv_{\ut,m}\T\Hm_{\ut,\ell} \cv_{\ut,\ell} \sv_{\ut,\ell}\T}_{\text{interference}} + \underbrace{\bv_{\ut,m}\T\Zm_{\ut}}_{\text{noise}} \ , 
\end{align} which is fed into a single-user decoder. By employing i.i.d.~Gaussian codewords, the transmitters can achieve the following rates
\begin{align} \label{e:uplinkzf}
R_{\ut,m} = \frac{1}{2} \log\Bigg( 1 + \frac{ P_{\ut,m} \big| \bv_{\ut,m}\T  \Hm_{\ut,m} \cv_{\ut,m} \big|^2}{ \sum_{\ell \neq m} \pul \big| \bv_{\ut,m}\T  \Hm_{\ut,\ell} \cv_{\ut,\ell} \big|^2}\Bigg)  
\end{align} for $m = 1,\ldots, L$.

\noindent\textbf{Downlink Channel.} The transmitter has a codeword $\sv_{\dt,\ell} \in \RR^n$ intended for the $\lth$ receiver with expected power $\frac{1}{n} \ex\| \sv_{\dt,\ell} \|^2 = \pdl$. It collects these codewords into a matrix 
\begin{align}
\Sm_{\dt} \triangleq \begin{bmatrix}
\sv_{\dt,1}\T \\ \vdots \\ \sv_{\dt,L}\T
\end{bmatrix}
\end{align} and applies a beamforming matrix $\Bm_{\dt} \in \RR^{N \times L}$ to create its channel input \begin{align}
\Xm_{\dt} = \Bm_{\dt} \Sm_{\dt} \ . 
\end{align}

The $\mth$ receiver uses an equalization vector $\cv_{\dt,m} \in \RR^{M_m}$ to form an effective channel output 
\begin{align}
&\ytv_{\dt,m}\T\\ &= \cv_{\dt,m}\T \Ym_{\dt} \\
&= \underbrace{\cv_{\dt,m}\T  \Hm_{\dt,m} \bv_{\dt,m} \sv_{\dt,m}\T}_{\text{signal}} + \underbrace{\sum_{\ell \neq m} \cv_{\dt,m}\T\Hm_{\dt,\ell} \bv_{\dt,\ell} \sv_{\dt,\ell}\T}_{\text{interference}} + \underbrace{\cv_{\dt,m}\T\Zm_{\dt,m}}_{\text{noise}} \ . 
\end{align} Using i.i.d.~Gaussian codewords, we can achieve the following rates for $m = 1,\ldots, L$:
\begin{align} \label{e:downlinkzf}
R_{\dt,m} = \frac{1}{2} \log\Bigg( 1 + \frac{ P_{\dt,m} \big| \cv_{\dt,m}\T  \Hm_{\dt,m} \bv_{\dt,m} \big|^2}{ \sum_{\ell \neq m} \pdl \big| \cv_{\dt,m}\T  \Hm_{\dt,\ell} \bv_{\dt,\ell} \big|^2}\Bigg) \ . 
\end{align}

\noindent\textbf{Uplink-Downlink Duality.} We can now state the uplink-downlink duality relationship for conventional linear architectures. Define the uplink and downlink equalization matrices 
\begin{align} \label{e:downlinkequalization}
\Bm_{\ut} \triangleq \begin{bmatrix}
\bv_{\ut,1}\T \\ \vdots \\ \bv_{\ut,L}\T
\end{bmatrix}\qquad \Cm_{\dt} \triangleq
\begin{bmatrix}
\cv_{\dt,1}\T & \mathbf{0}_{M_2}\T & \hdots & \mathbf{0}_{M_L}\T\\
\mathbf{0}_{M_1}\T & \cv_{\dt,2}\T  & \hdots & \mathbf{0}_{M_L}\T\\
\vdots & \vdots  & \ddots & \vdots\\
\mathbf{0}_{M_1}\T & \mathbf{0}_{M_2}\T& \hdots  & \cv_{\dt,L}\T \\
\end{bmatrix} \ , 
\end{align} respectively. Also, define the uplink and downlink power matrices,
\begin{align}
\upm = \diag(P_{\ut,1},\ldots,P_{\ut,L}) \qquad \dpm = \diag(P_{\dt,1},\ldots,P_{\dt,L}) \ , 
\end{align} respectively. The following theorem encapsulates the uplink-downlink result of~\cite{vt03} in our notation.

\begin{theorem}[{\cite{vt03}}] \label{t:dualityconventional}
For a given uplink channel matrix $\mathbf{H}_{\ut}$ and (diagonal) power matrix $\upm$ that meets the total power constraint $\tr(\mathbf{C}_{\ut}\T\mathbf{C}_{\ut}\upm) = \ptu$, let $R_{\ut,1},\ldots, R_{\ut,L}$ be a rate tuple that is achievable with equalization matrix $\mathbf{B}_{\ut}$ and precoding matrix $\mathbf{C}_{\ut}$. Then, for the downlink channel matrix $\mathbf{H}_{\dt} = \mathbf{H}_{\ut}\T$, there exists a unique (diagonal) power matrix $\dpm$ with total power usage $\tr(\mathbf{B}_{\dt}\T\mathbf{B}_{\dt}\dpm) = \ptu$, such that the rate tuple ${R}_{\dt,\ell}=R_{\ut,\ell}$ for $\ell=1,\ldots, L$  is achievable  using  equalization matrix $\mathbf{C}_{\dt} = \mathbf{C}_{\ut}\T$ and  precoding matrix $\mathbf{B}_{\dt} = \mathbf{B}_{\ut}\T$. The same relationship can be established starting from an achievable rate tuple for the downlink and going to the uplink. 
\end{theorem}

In other words, any rates that are achievable on an uplink channel can be achieved on a downlink channel with a transposed channel matrix by exchanging the roles of the equalization and beamforming matrices (and transposing them) as well as reallocating the powers.

\begin{remark} \label{r:ratesplitting}
In some cases, it may be desirable to employ rate splitting~\cite{ru96} at the transmitter(s). This can be viewed as creating virtual transmitters (in the uplink) or virtual receivers (in the downlink). In this setting, uplink-downlink duality continues to hold so long as the uplink and downlink users are split into virtual users in the same fashion. 
\end{remark}

\subsection{Integer-Forcing Linear Architectures}

The linear architectures discussed above fall short of achieving the MIMO MAC or BC capacity, due to noise amplification from the equalization step (which worsens as the condition number of the channel matrix increases). Integer-forcing linear architectures can substantially reduce this rate loss by allowing the single-user decoders to target integer-linear combinations, rather than individual codewords. These linear combinations can then be solved for the desired codewords. By carefully selecting the integer coefficients to match the interference presented by the channel, we can reduce the noise amplification caused by the equalization step. 

To streamline the overview, we will focus on the effective noise variances attained by the integer-forcing architecture as well as the resulting achievable rates. Due to the fact that the effective noise for a linear combination impacts all participating codewords, care is needed to ensure that this effective noise variance is only associated with the achievable rate of a single user. The technical details will be given in Sections~\ref{s:uplink} and~\ref{s:downlink} for the uplink and downlink, respectively.

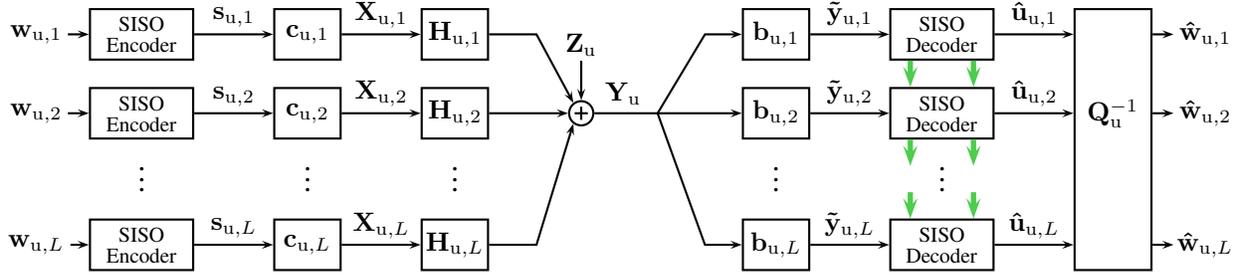
\begin{figure*}[h]
\psset{unit=.7mm}
\begin{center}
\begin{pspicture}(-2,10)(231,62)
%\psframe(-2,10)(231,62)
%\small

\rput(22,0){
%user 1
\rput(-18,55){$\mathbf{w}_{\ut,1}$}
\psline{->}(-12,55)(-8,55)
\psframe(-8,50)(12,60) \rput(2,57){\footnotesize{SISO}} \rput(2,53){\footnotesize{Encoder}}
\psline{->}(12,55)(27,55) \rput(19.5,58.25){$\mathbf{s}_{\ut,1}$}
\psframe(27,50)(40,60)
\rput(33.5,55){$\mathbf{c}_{\ut,1}$} \rput(47.5,58.75){$\mathbf{X}_{\ut,1}$}
\psline[linecolor=black]{->}(40,55)(55,55)
\psframe(55,50)(68,60) \rput(61.5,55){$\mathbf{H}_{\ut,1}$}
\psline{->}(68,55)(77,55)(84,42)

%user 2
\rput(-18,40){$\mathbf{w}_{\ut,2}$}
\psline{->}(-12,40)(-8,40)
\psframe(-8,35)(12,45) \rput(2,42){\footnotesize{SISO}} \rput(2,38){\footnotesize{Encoder}}
\psline{->}(12,40)(27,40) \rput(19.5,43.25){$\mathbf{s}_{\ut,2}$}
\psframe(27,35)(40,45)
\rput(33.5,40){$\mathbf{c}_{\ut,2}$} \rput(47.5,43.75){$\mathbf{X}_{\ut,2}$}
\psline[linecolor=black]{->}(40,40)(55,40)
\psframe(55,35)(68,45) \rput(61.5,40){$\mathbf{H}_{\ut,2}$}
\psline{->}(68,40)(83,40)

\rput(2,29){\large{$\vdots$}}
\rput(33.5,29){\large{$\vdots$}}
\rput(61.5,29){\large{$\vdots$}}

%user L
\rput(-18,15){$\mathbf{w}_{\ut,L}$}
\psline{->}(-11.5,15)(-8,15)
\psframe(-8,10)(12,20) \rput(2,17){\footnotesize{SISO}} \rput(2,13){\footnotesize{Encoder}}
\psline{->}(12,15)(27,15) \rput(19.5,18.25){$\mathbf{s}_{\ut,L}$}
\psframe(27,10)(40,20)
\rput(33.5,15){$\mathbf{c}_{\ut,L}$} \rput(47.5,18.75){$\mathbf{X}_{\ut,L}$}
\psline[linecolor=black]{->}(40,15)(55,15)
\psframe(55,10)(68,20) \rput(61.5,15){$\mathbf{H}_{\ut,L}$}
\psline{->}(68,15)(77,15)(84,38)

%Channel
\pscircle(85.5,40){2.5} \psline{-}(84.25,40)(86.75,40)
\psline{-}(85.5,38.75)(85.5,41.25) \psline{<-}(85.5,42.5)(85.5,50) \rput(85.5,53){$\mathbf{Z}_{\ut}$}
\psline{-}(88,40)(100,40) \rput(93.5,43.75){$\mathbf{Y}_{\ut}$}

%Decoder 1
\psline{->}(100,40)(109,55)(116,55)
\psframe(116,50)(129,60) \rput(122.5,55){$\mathbf{b}_{\ut,1}$}
\psline{->}(129,55)(144,55) \rput(136.5,58.75){$\mathbf{\tilde{y}}_{\ut,1}$}
\psframe(144,50)(164,60) \rput(154,57){\footnotesize{SISO}} \rput(154,53){\footnotesize{Decoder}}
\psline{->}(164,55)(179,55) \rput(171.5,58.75){$\mathbf{\hat{u}}_{\ut,1}$}
\psline[linewidth=2pt,linecolor=LineGreen]{->}(148,50)(148,45)
\psline[linewidth=2pt,linecolor=LineGreen]{->}(160,50)(160,45)

%Decoder 2
\psline{->}(100,40)(116,40)
\psframe(116,35)(129,45) \rput(122.5,40){$\mathbf{b}_{\ut,2}$}
\psline{->}(129,40)(144,40) \rput(136.5,43.75){$\mathbf{\tilde{y}}_{\ut,2}$}
\psframe(144,35)(164,45) \rput(154,42){\footnotesize{SISO}} \rput(154,38){\footnotesize{Decoder}}
\psline{->}(164,40)(179,40) \rput(171.5,43.75){$\mathbf{\hat{u}}_{\ut,2}$}
\psline[linewidth=2pt,linecolor=LineGreen]{->}(148,35)(148,30)
\psline[linewidth=2pt,linecolor=LineGreen]{->}(160,35)(160,30)

\rput(122.5,29){\large{$\vdots$}}
\rput(154,29){\large{$\vdots$}}

\psline[linewidth=2pt,linecolor=LineGreen]{->}(148,25)(148,20)
\psline[linewidth=2pt,linecolor=LineGreen]{->}(160,25)(160,20)

%Decoder L
\psline{->}(100,40)(109,15)(116,15)
\psframe(116,10)(129,20) \rput(122.5,15){$\mathbf{b}_{\ut,L}$}
\psline{->}(129,15)(144,15) \rput(136.5,18.75){$\mathbf{\tilde{y}}_{\ut,L}$}
\psframe(144,10)(164,20) \rput(154,17){\footnotesize{SISO}} \rput(154,13){\footnotesize{Decoder}}
\psline{->}(164,15)(179,15) \rput(171.5,18.75){$\mathbf{\hat{u}}_{\ut,L}$}

\psframe(179,10)(194,60) \rput(186.5,40){$\mathbf{Q}_{\ut}^{-1}$}
\psline{->}(194,55)(198,55) \rput(204,55){$\mathbf{\hat{w}}_{\ut,1}$}
\psline{->}(194,40)(198,40) \rput(204,40){$\mathbf{\hat{w}}_{\ut,2}$}
\psline{->}(194,15)(198,15) \rput(204,15){$\mathbf{\hat{w}}_{\ut,L}$}

}

\end{pspicture}
\end{center}
\caption{Block diagram of the integer-forcing uplink architecture. Each message vector $\mathbf{w}_{\ut,\ell}$ is encoded into a dithered lattice codeword $\mathbf{s}_{\ut,\ell}$ and mapped to a channel input $\mathbf{X}_{\ut,\ell} = \mathbf{c}_{\ut,\ell} \mathbf{s}_{\ut,\ell}\T$. For $m = 1,\ldots,L$, the receiver uses an equalized channel output $\ytv_{\ut,m} = \bv_{\ut,m}\T \Ym_{\ut}$ to make an estimate $\uhv_{\ut,m}$ of the linear combination $\uv_{\ut,m}$. At the $\mth$ decoder, algebraic successive cancellation is used to (digitally) cancel out $m-1$ codewords prior to applying a SISO decoder. These codewords are then restored to obtain an estimate of the $\mth$ linear combination. Finally, the linear combinations are inverted to recover estimates $\whv_{\ut,\ell}$ of the message vectors.}
\label{f:uplinkarch}
\end{figure*}

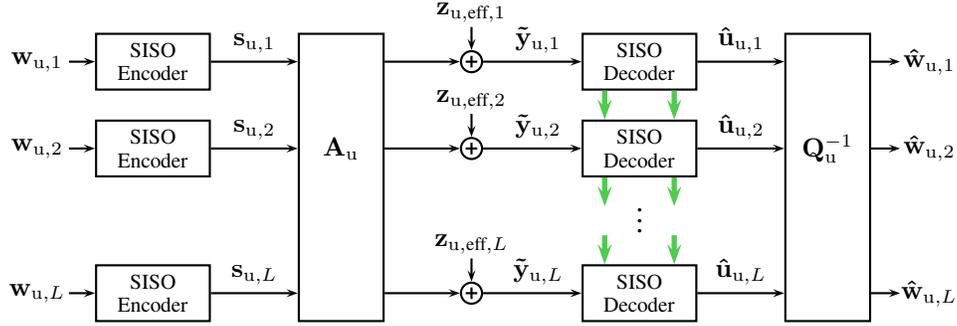
\begin{figure*}[h]
\psset{unit=.77mm}
\begin{center}
\begin{pspicture}(-3,10)(165,64)
%\psframe(-3,5)(165,64)
%\small

\rput(22,0){
%user 1
\rput(-18,55){$\mathbf{w}_{\ut,1}$}
\psline{->}(-12.5,55)(-8,55)
\psframe(-8,50)(12,60) \rput(2,57){\footnotesize{SISO}} \rput(2,53){\footnotesize{Encoder}}
\psline{->}(12,55)(27,55) \rput(19.5,58.25){$\mathbf{s}_{\ut,1}$}

\psframe(27,10)(42,60) \rput(34.5,40){$\mathbf{A}_{\ut}$}

%user 2
\rput(-18,40){$\mathbf{w}_{\ut,2}$}
\psline{->}(-12.5,40)(-8,40)
\psframe(-8,35)(12,45) \rput(2,42){\footnotesize{SISO}} \rput(2,38){\footnotesize{Encoder}}
\psline{->}(12,40)(27,40) \rput(19.5,43.25){$\mathbf{s}_{\ut,2}$}

%user L
\rput(-18,15){$\mathbf{w}_{\ut,L}$}
\psline{->}(-12.5,15)(-8,15)
\psframe(-8,10)(12,20) \rput(2,17){\footnotesize{SISO}} \rput(2,13){\footnotesize{Encoder}}
\psline{->}(12,15)(27,15) \rput(19.5,18.25){$\mathbf{s}_{\ut,L}$}

\rput(-68,0){
%Decoder 1
\psline{->}(110,55)(123,55)
\pscircle(125,55){2} \psline(124,55)(126,55) \psline(125,54)(125,56)
\psline{->}(125,61)(125,57) \rput(125,63.5){$\mathbf{z}_{\ut,\eff,1}$}
\psline{->}(127,55)(144,55) \rput(136.5,58.75){$\mathbf{\tilde{y}}_{\ut,1}$}
\psframe(144,50)(164,60) \rput(154,57){\footnotesize{SISO}} \rput(154,53){\footnotesize{Decoder}}
\psline{->}(164,55)(179,55) \rput(171.5,58.75){$\mathbf{\hat{u}}_{\ut,1}$}
\psline[linewidth=2pt,linecolor=LineGreen]{->}(148,50)(148,45)
\psline[linewidth=2pt,linecolor=LineGreen]{->}(160,50)(160,45)

%Decoder 2
\psline{->}(110,40)(123,40)
\pscircle(125,40){2} \psline(124,40)(126,40) \psline(125,39)(125,41)
\psline{->}(125,46)(125,42) \rput(125,48.5){$\mathbf{z}_{\ut,\eff,2}$}
\psline{->}(127,40)(144,40) \rput(136.5,43.75){$\mathbf{\tilde{y}}_{\ut,2}$}
\psframe(144,35)(164,45) \rput(154,42){\footnotesize{SISO}} \rput(154,38){\footnotesize{Decoder}}
\psline{->}(164,40)(179,40) \rput(171.5,43.75){$\mathbf{\hat{u}}_{\ut,2}$}
\psline[linewidth=2pt,linecolor=LineGreen]{->}(148,35)(148,30)
\psline[linewidth=2pt,linecolor=LineGreen]{->}(160,35)(160,30)

\rput(154,29){\large{$\vdots$}}

\psline[linewidth=2pt,linecolor=LineGreen]{->}(148,25)(148,20)
\psline[linewidth=2pt,linecolor=LineGreen]{->}(160,25)(160,20)

%Decoder L
\psline{->}(110,15)(123,15)
\pscircle(125,15){2} \psline(124,15)(126,15) \psline(125,14)(125,16)
\psline{->}(125,21)(125,17) \rput(125,23.5){$\mathbf{z}_{\ut,\eff,L}$}
\psline{->}(127,15)(144,15) \rput(136.5,18.75){$\mathbf{\tilde{y}}_{\ut,L}$}
\psframe(144,10)(164,20) \rput(154,17){\footnotesize{SISO}} \rput(154,13){\footnotesize{Decoder}}
\psline{->}(164,15)(179,15) \rput(171.5,18.75){$\mathbf{\hat{u}}_{\ut,L}$}

\psframe(179,10)(194,60) \rput(186.5,40){$\mathbf{Q}_{\ut}^{-1}$}
\psline{->}(194,55)(199,55) \rput(204,55){$\mathbf{\hat{w}}_{\ut,1}$}
\psline{->}(194,40)(199,40) \rput(204,40){$\mathbf{\hat{w}}_{\ut,2}$}
\psline{->}(194,15)(199,15) \rput(204,15){$\mathbf{\hat{w}}_{\ut,L}$}
}
}

\end{pspicture}
\end{center}
\caption{Block diagram of the effective channel induced by the integer-forcing uplink architecture. The $\mth$ decoder observes an integer-linear combination of the codewords plus effective noise, $\sum_\ell a_{\ut,m,\ell} \sv_{\ut,\ell} + \zv_{\ut,\eff,m}$ from which it makes an estimate of the linear combination $\uv_{\ut,\ell}$ with coefficients $q_{\ut,m,\ell} = [a_{\ut,m,\ell}] \bmod{p}$. Finally, it applies the inverse of the matrix $\Qm_{\ut} = \{q_{\ut,m,\ell}\}$ over $\ZZ_p$ to estimate the message. }
\label{f:uplinkeffective}
\end{figure*}
 
\noindent\textbf{Uplink Channel.} The operations at the transmitters mimic those of a conventional linear architecture, except that we use a nested lattice codebook to ensure that integer-linear combinations of codewords are themselves codewords. The goal is to recover $L$ integer-linear combinations of the form $\av_{\ut,1}\T\Sm_{\ut},\ldots,\av_{\ut,L}\T \Sm_{\ut}$ where the $\av_{\ut,m}\T$ are the rows of a full-rank integer matrix $\Am_{\ut} \in \ZZ^{L \times L}$, i.e., 
\begin{align}
\Am_{\ut} = \begin{bmatrix}
\av_{\ut,1}\T \\
\vdots \\
\av_{\ut,L}\T
\end{bmatrix} \ . 
\end{align} To recover the $\mth$ linear combination $\av_{\ut,m}\T \Sm_{\ut}$, the receiver applies an equalization vector $\bv_{\ut,m} \in \RR^{M_m}$ to form the effective channel output 
\begin{align}
\ytv_{\ut,m}\T &= \bv_{\ut,m}\T \Ym_{\ut}  \\
&= \av_{\ut,m}\T \Sm_{\ut} + \zv_{\ut,\eff,m}\T \\
\zv_{\ut,\eff,m}\T &\triangleq \bv_{\ut,m}\T \Zm_{\ut}  + \big( \bv_{\ut,m}\T \Hm_{\ut} \Cm_{\ut} - \av_{\ut,m}\T \big) \Sm_{\ut} \ . \end{align} We define the \textit{effective noise variance} as 
\begin{align}
\sigma_{\ut,m}^2 &\triangleq \frac{1}{n} \ex \| \zv_{\ut,\eff,m} \|^2 \\
&= \| \bv_{\ut,m} \|^2 + \Big\|  \big( \bv_{\ut,m}\T \Hm_{\ut} \Cm_{\ut} - \av_{\ut,m}\T \big) \upm^{1/2} \Big\|^2 \ . \label{e:zueff}
\end{align} Assuming the receiver can successfully recover all $L$ linear combinations, it  can now apply the inverse of the integer matrix to obtain the transmitted messages. A block diagram illustrating these operations and the resulting effective channels can be found in Figures~\ref{f:uplinkarch} and~\ref{f:uplinkeffective}, respectively. %As argued in~\cite{zneg14}, this inverse can be performed over the finite field from which the messages and nested lattice codes are drawn. 

In Section~\ref{s:uplink}, we will provide a detailed description of the uplink achievability scheme proposed in~\cite{ncnc16}. Overall, it establishes that the following rates are achievable 
\begin{align}
R_{\ut,m} = \frac{1}{2} \log^+\bigg(\frac{P_{\ut,m}}{\sigma_{\ut,\pi(m)}^2}\bigg) \qquad m = 1,\ldots,L \ . 
\end{align} for at least one permutation $\pi_{\ut}$.

\begin{remark} \label{r:identity} Although it is not immediately obvious, any rate tuple that is achievable via a conventional linear architecture is also achievable via an integer-forcing linear architecture by using the same beamforming matrix, setting the integer matrix to be the identity matrix, and scaling the equalization vectors by the appropriate MMSE coefficient~\cite[Lemma 3]{zneg14}. While~\cite{zneg14} only establishes this for the uplink, this follows naturally for the downlink as well. \end{remark}

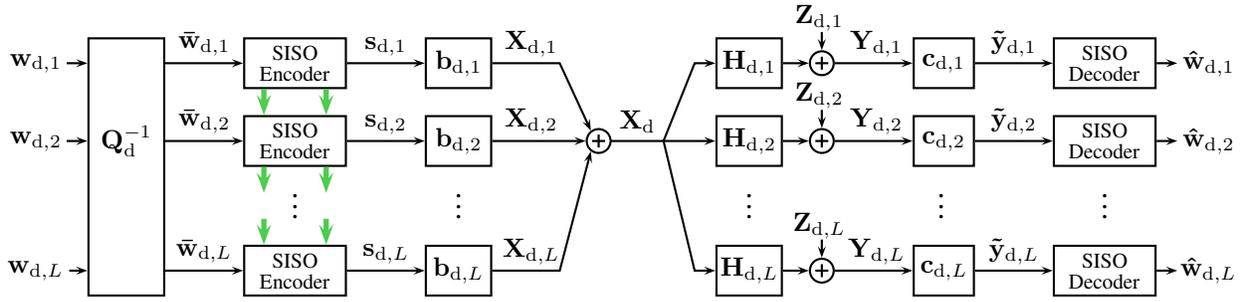
\begin{figure*}[h]
\psset{unit=.69mm}
\begin{center}
\begin{pspicture}(-31,10)(205,64)
%\psframe(-31,10)(205,64)
%\small

\rput(22,0){

\psframe(-38,10)(-23,60) \rput(-30.5,40){$\mathbf{Q}_{\dt}^{-1}$}

%user 1
\rput(-48,55){$\mathbf{w}_{\dt,1}$}
\psline{->}(-42,55)(-38,55)
\psline{->}(-23,55)(-8,55)\rput(-15.5,58.75){$\mathbf{\bar{w}}_{\dt,1}$}
\psframe(-8,50)(12,60) \rput(2,57){\footnotesize{SISO}} \rput(2,53){\footnotesize{Encoder}}
\psline{->}(12,55)(27,55) \rput(19.5,58.25){$\mathbf{s}_{\dt,1}$}
\psframe(27,50)(40,60)
\rput(33.5,55){$\mathbf{b}_{\dt,1}$} \rput(47.5,58.75){$\mathbf{X}_{\dt,1}$}
\psline[linecolor=black]{->}(40,55)(53,55)(59,42)
\psline[linewidth=2pt,linecolor=LineGreen]{->}(-4,50)(-4,45)
\psline[linewidth=2pt,linecolor=LineGreen]{->}(8,50)(8,45)

%user 2
\rput(-48,40){$\mathbf{w}_{\dt,2}$}
\psline{->}(-42,40)(-38,40)
\psline{->}(-23,40)(-8,40)\rput(-15.5,43.75){$\mathbf{\bar{w}}_{\dt,2}$}
\psframe(-8,35)(12,45) \rput(2,42){\footnotesize{SISO}} \rput(2,38){\footnotesize{Encoder}}
\psline{->}(12,40)(27,40) \rput(19.5,43.25){$\mathbf{s}_{\dt,2}$}
\psframe(27,35)(40,45)
\rput(33.5,40){$\mathbf{b}_{\dt,2}$} \rput(47.5,43.75){$\mathbf{X}_{\dt,2}$}
\psline[linecolor=black]{->}(40,40)(53,40)(58,40)

\rput(2,29){\large{$\vdots$}}
\rput(33.5,29){\large{$\vdots$}}
\rput(89.5,29){\large{$\vdots$}}

\psline[linewidth=2pt,linecolor=LineGreen]{->}(-4,35)(-4,30)
\psline[linewidth=2pt,linecolor=LineGreen]{->}(8,35)(8,30)

\psline[linewidth=2pt,linecolor=LineGreen]{->}(-4,25)(-4,20)
\psline[linewidth=2pt,linecolor=LineGreen]{->}(8,25)(8,20)

%user L
\rput(-48,15){$\mathbf{w}_{\dt,L}$}
\psline{->}(-41.5,15)(-38,15)
\psline{->}(-23,15)(-8,15)\rput(-15.5,18.75){$\mathbf{\bar{w}}_{\dt,L}$}
\psframe(-8,10)(12,20) \rput(2,17){\footnotesize{SISO}} \rput(2,13){\footnotesize{Encoder}}
\psline{->}(12,15)(27,15) \rput(19.5,18.25){$\mathbf{s}_{\dt,L}$}
\psframe(27,10)(40,20)
\rput(33.5,15){$\mathbf{b}_{\dt,L}$} \rput(47.5,18.75){$\mathbf{X}_{\dt,L}$}
\psline[linecolor=black]{->}(40,15)(53,15)(59,38)

%Channel
\pscircle(60.5,40){2.5} \psline{-}(59.25,40)(61.75,40)
\psline{-}(60.5,38.75)(60.5,41.25)
\psline{-}(63,40)(73,40)\rput(68,43.75){$\Xm_{\dt}$}

%Decoder 1
\psline{->}(73,40)(79,55)(83,55)
\psframe(83,50)(96,60) \rput(89.5,55){$\mathbf{H}_{\dt,1}$}
\psline{->}(96,55)(101,55)
\pscircle(103.5,55){2.5} \psline{-}(102.25,55)(104.75,55)
\psline{-}(103.5,53.75)(103.5,56.25)
\psline{<-}(103.5,57.5)(103.5,61) \rput(103,63.75){$\Zm_{\dt,1}$}
\psline{->}(106,55)(121,55) \rput(114,58.75){$\Ym_{\dt,1}$}
\psframe(121,50)(133,60) \rput(127,55){$\mathbf{c}_{\dt,1}$}
\psline{->}(133,55)(148,55) \rput(140.5,58.75){$\mathbf{\tilde{y}}_{\dt,1}$}
\psframe(148,50)(168,60) \rput(158,57){\footnotesize{SISO}} \rput(158,53){\footnotesize{Decoder}}

\rput(126.5,29){\large{$\vdots$}}
\rput(158,29){\large{$\vdots$}}

%Decoder 2
\psline{->}(73,40)(79,40)(83,40)
\psframe(83,35)(96,45) \rput(89.5,40){$\mathbf{H}_{\dt,2}$}
\psline{->}(96,40)(101,40)
\pscircle(103.5,40){2.5} \psline{-}(102.25,40)(104.75,40)
\psline{-}(103.5,38.75)(103.5,41.25)
\psline{<-}(103.5,42.5)(103.5,46) \rput(103,48.75){$\Zm_{\dt,2}$}
\psline{->}(106,40)(121,40) \rput(114,43.75){$\Ym_{\dt,2}$}
\psframe(121,35)(133,45) \rput(127,40){$\mathbf{c}_{\dt,2}$}
\psline{->}(133,40)(148,40) \rput(140.5,43.75){$\mathbf{\tilde{y}}_{\dt,2}$}
\psframe(148,35)(168,45) \rput(158,42){\footnotesize{SISO}} \rput(158,38){\footnotesize{Decoder}}

%Decoder L
\psline{->}(73,40)(79,15)(83,15)
\psframe(83,10)(96,20) \rput(89.5,15){$\mathbf{H}_{\dt,L}$}
\psline{->}(96,15)(101,15)
\pscircle(103.5,15){2.5} \psline{-}(102.25,15)(104.75,15)
\psline{-}(103.5,13.75)(103.5,16.25)
\psline{<-}(103.5,17.5)(103.5,21) \rput(103,23.75){$\Zm_{\dt,L}$}
\psline{->}(106,15)(121,15) \rput(114,18.75){$\Ym_{\dt,L}$}
\psframe(121,10)(133,20) \rput(127,15){$\mathbf{c}_{\dt,L}$}
\psline{->}(133,15)(148,15) \rput(140.5,18.75){$\mathbf{\tilde{y}}_{\dt,L}$}
\psframe(148,10)(168,20) \rput(158,17){\footnotesize{SISO}} \rput(158,13){\footnotesize{Decoder}}

\psline{->}(168,55)(172,55) \rput(178,55){$\mathbf{\hat{w}}_{\dt,1}$}
\psline{->}(168,40)(172,40) \rput(178,40){$\mathbf{\hat{w}}_{\dt,2}$}
\psline{->}(168,15)(172,15) \rput(178,15){$\mathbf{\hat{w}}_{\dt,L}$}

}

\end{pspicture}
\end{center}
\caption{Block diagram of the integer-forcing downlink architecture. The encoder applies the inverse of $\Qm_{\dt} = [ \Am_{\dt} ]\bmod{p}$ over $\ZZ_p$ to the message vectors $\mathbf{w}_{\dt,1},\ldots,\mathbf{w}_{\dt,L}$ and then maps the results to dithered lattice codewords $\mathbf{s}_{\dt,1},\ldots,\mathbf{s}_{\dt,L}$. {(The arrows between the encoders indicate that, as each codeword is formed, an additional linear precoding step is needed to eliminate potential interference from codewords with smaller powers.)} The channel input is formed by beamforming these codewords, $\Xm_{\dt} = \sum_\ell \bv_{\dt,\ell} \sv_{\dt,\ell}\T$. The $\mth$ decoder uses an equalized channel output $\ytv_{\dt,m} = \ctv_{\dt,m}\T \Ym_{\dt,m}$ to make an estimate of an integer-linear combination of the lattice codewords, which, due to the inverse operation corresponds to an estimate of its desired message.}
\label{f:downlinkarch}
\end{figure*}

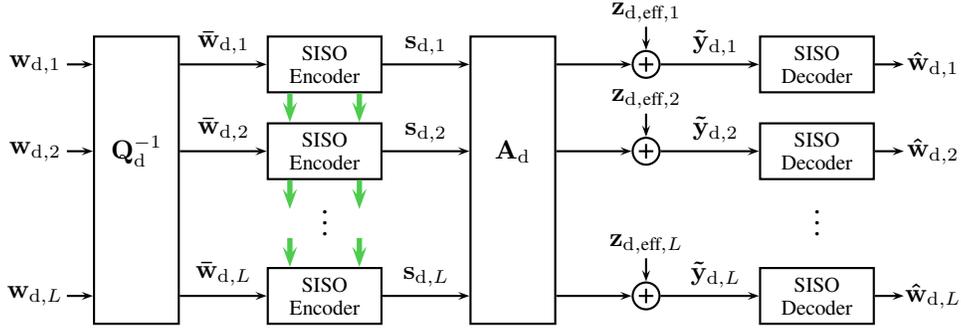
\begin{figure*}[h]
\psset{unit=.77mm}
\begin{center}
\begin{pspicture}(-31,10)(134,64)
%\psframe(-31,10)(134,64)
%\small

\rput(22,0){

\psframe(-38,10)(-23,60) \rput(-30.5,40){$\mathbf{Q}_{\dt}^{-1}$}

%user 1
\rput(-48,55){$\mathbf{w}_{\dt,1}$}
\psline{->}(-42.5,55)(-38,55)
\psline{->}(-23,55)(-8,55)\rput(-15.5,58.75){$\mathbf{\bar{w}}_{\dt,1}$}
\psframe(-8,50)(12,60) \rput(2,57){\footnotesize{SISO}} \rput(2,53){\footnotesize{Encoder}}
\psline{->}(12,55)(27,55) \rput(19.5,58.25){$\mathbf{s}_{\dt,1}$}
\psline[linewidth=2pt,linecolor=LineGreen]{->}(-4,50)(-4,45)
\psline[linewidth=2pt,linecolor=LineGreen]{->}(8,50)(8,45)

%user 2
\rput(-48,40){$\mathbf{w}_{\dt,2}$}
\psline{->}(-42.5,40)(-38,40)
\psline{->}(-23,40)(-8,40)\rput(-15.5,43.75){$\mathbf{\bar{w}}_{\dt,2}$}
\psframe(-8,35)(12,45) \rput(2,42){\footnotesize{SISO}} \rput(2,38){\footnotesize{Encoder}}
\psline{->}(12,40)(27,40) \rput(19.5,43.25){$\mathbf{s}_{\dt,2}$}

\rput(2,29){\large{$\vdots$}}

\psline[linewidth=2pt,linecolor=LineGreen]{->}(-4,35)(-4,30)
\psline[linewidth=2pt,linecolor=LineGreen]{->}(8,35)(8,30)

\psline[linewidth=2pt,linecolor=LineGreen]{->}(-4,25)(-4,20)
\psline[linewidth=2pt,linecolor=LineGreen]{->}(8,25)(8,20)

%user L
\rput(-48,15){$\mathbf{w}_{\dt,L}$}
\psline{->}(-42.5,15)(-38,15)
\psline{->}(-23,15)(-8,15)\rput(-15.5,18.75){$\mathbf{\bar{w}}_{\dt,L}$}
\psframe(-8,10)(12,20) \rput(2,17){\footnotesize{SISO}} \rput(2,13){\footnotesize{Encoder}}
\psline{->}(12,15)(27,15) \rput(19.5,18.25){$\mathbf{s}_{\dt,L}$}

\psframe(27,10)(42,60) \rput(34.5,40){$\mathbf{A}_{\dt}$}

\psline{->}(42,55)(55,55)
\psline{->}(42,40)(55,40)
\psline{->}(42,15)(55,15)

%Decoder 1
\pscircle(57.5,55){2.5} \psline{-}(56.25,55)(58.75,55)
\psline{-}(57.5,53.75)(57.5,56.25)
\psline{<-}(57.5,57.5)(57.5,61.5) \rput(57.5,64){$\zv_{\dt,\eff,1}$}
\psline{->}(60,55)(77,55) \rput(69.5,58.75){$\mathbf{\tilde{y}}_{\dt,1}$}
\psframe(77,50)(97,60) \rput(87,57){\footnotesize{SISO}} \rput(87,53){\footnotesize{Decoder}}

\rput(87,29){\large{$\vdots$}}

%Decoder 2
\pscircle(57.5,40){2.5} \psline{-}(56.25,40)(58.75,40)
\psline{-}(57.5,38.75)(57.5,41.25)
\psline{<-}(57.5,42.5)(57.5,46.5) \rput(57.5,49){$\zv_{\dt,\eff,2}$}
\psline{->}(60,40)(77,40) \rput(69.5,43.75){$\mathbf{\tilde{y}}_{\dt,2}$}
\psframe(77,35)(97,45) \rput(87,42){\footnotesize{SISO}} \rput(87,38){\footnotesize{Decoder}}

%Decoder L
\pscircle(57.5,15){2.5} \psline{-}(56.25,15)(58.75,15)
\psline{-}(57.5,13.75)(57.5,16.25)
\psline{<-}(57.5,17.5)(57.5,21.5) \rput(57.5,24){$\zv_{\dt,\eff,L}$}
\psline{->}(60,15)(77,15) \rput(69.5,18.75){$\mathbf{\tilde{y}}_{\dt,L}$}
\psframe(77,10)(97,20) \rput(87,17){\footnotesize{SISO}} \rput(87,13){\footnotesize{Decoder}}

\psline{->}(97,55)(102,55) \rput(107,55){$\mathbf{\hat{w}}_{\dt,1}$}
\psline{->}(97,40)(102,40) \rput(107,40){$\mathbf{\hat{w}}_{\dt,2}$}
\psline{->}(97,15)(102,15) \rput(107,15){$\mathbf{\hat{w}}_{\dt,L}$}

}

\end{pspicture}
\end{center}
\caption{Block diagram of the effective channel induced by the integer-forcing downlink architecture. The $\mth$ decoder observes an integer-linear combination of the codewords plus effective noise, $\sum_\ell a_{\dt,m,\ell} \sv_{\dt,\ell} + \zv_{\dt,\eff,m}$. Since the encoder applied the inverse of $\Qm_{\dt} = [ \Am_{\dt} ] \bmod{p}$ over $\ZZ_p$ to the message vectors prior to mapping them to lattice codewords, then the $\mth$ integer-linear combination corresponds to the $\mth$ message.   }
\label{f:downlinkeffective}
\end{figure*}

\noindent\textbf{Downlink Channel.} We use the same encoding operations at the transmitter as in a conventional linear architecture. We employ a nested lattice codebook to ensure that the codebook is closed under integer-linear combinations so that the users can decode linear combinations of the transmitted codewords. Additionally, as first proposed by Hong and Caire~\cite{hc12,hc13}, we apply a precoding step over the finite field in order to ``pre-invert'' the linear combinations before mapping the messages to codewords. This step guarantees that the integer-linear combinations recovered by the users correspond to their desired messages. These operations and the resulting effective channel are illustrated in Figures~\ref{f:downlinkarch} and~\ref{f:downlinkeffective}, respectively.

The $\mth$ receiver attempts to recover the linear combination $\av_{\dt,m}\T \Sm_{\dt}$ where $\av_{\dt,m}\T$ is the $\mth$ row of the full-rank, integer matrix $\Am_{\dt} \in \ZZ^{L \times L}$, i.e., 
\begin{align}
\Am_{\dt} = \begin{bmatrix}
\av_{\dt,1}\T \\
\vdots \\
\av_{\dt,L}\T
\end{bmatrix} \ . 
\end{align} To do so, it uses an equalization vector $\cv_{\dt,m} \in \RR^{M_m}$ to form the effective channel output
\begin{align}
\ytv_{\dt,m}\T &= \cv_{\dt,m}\T \Ym_{\dt,m}  \\
&= \av_{\dt,m}\T \Sm_{\dt} + \zv_{\dt,\eff,m}\T  \label{e:effectivedownlinkchannel}\\
\zv_{\dt,\eff,m}\T &\triangleq \big( \cv_{\dt,m}\T \Hm_{\dt,m} \Bm_{\dt} - \av_{\dt,m}\T \big) \Sm_{\dt} + \cv_{\dt,m}\T \Zm_{\dt,m} \label{e:zdeff} \ . \end{align} We define the \textit{effective noise variance} as 
\begin{align} \label{e:downlinknoisevar}
\sigma_{\dt,m}^2 &\triangleq \frac{1}{n} \ex \| \zv_{\dt,\eff,m} \|^2 \\
&=  \| \cv_{\dt,m} \|^2 + \Big\|  \big( \cv_{\dt,m}\T \Hm_{\dt,m} \Bm_{\dt} - \av_{\dt,m}\T \big) \dpm^{1/2} \Big\|^2\ . 
\end{align}

\noindent\textbf{Uplink-Downlink Duality.} The following theorem establishes uplink-downlink duality for integer-forcing in terms of the sum rate. 
\begin{theorem}\label{t:duality}
For a given uplink channel matrix $\mathbf{H}_{\ut}$, integer matrix $\mathbf{A}_{\ut}$, and power matrix $\upm$ that meets the total power constraint $\tr(\mathbf{C}_{\ut}\T\mathbf{C}_{\ut}\upm) = \ptu$, let $R_{\ut,1},\ldots, R_{\ut,L}$ be a rate tuple that is achievable via integer-forcing with equalization matrix $\mathbf{B}_{\ut}$ and precoding matrix $\mathbf{C}_{\ut}$. Then, for the downlink channel matrix $\mathbf{H}_{\dt} = \mathbf{H}_{\ut}\T$, integer matrix $\mathbf{A}_{\dt} = \mathbf{A}_{\ut}\T$, there exists a unique power matrix $\dpm$ with total power usage $\tr(\mathbf{B}_{\dt}\T\mathbf{B}_{\dt}\dpm) = \ptu$, such that the sum rate $\sum_{\ell} {R}_{\dt,\ell} \geq \sum_\ell R_{\ut,\ell}$  is achievable via integer-forcing using equalization matrix $\mathbf{C}_{\dt} = \mathbf{C}_{\ut}\T$ and  precoding matrix $\mathbf{B}_{\dt} = \mathbf{B}_{\ut}\T$. The same relationship can be established starting from an achievable rate tuple for the downlink and going to the uplink. 
\end{theorem} 
A full proof of this duality theorem will be given in Section~\ref{s:duality}.

\begin{remark} \label{r:permutations}
Note that Theorem~\ref{t:duality} only establishes duality for the sum rate whereas, for conventional linear architectures, Theorem~\ref{t:dualityconventional} establishes duality for the individual rates. This stems from the fact that, for a conventional linear architecture, the $\mth$ effective noise variance and $\mth$ effective power are always linked to the rates $R_{\ut,m}$ and $R_{\dt,m}$. However, for uplink integer-forcing, the effective noise variance for the $\mth$ linear combination may not correspond to $R_{\ut,m}$. Similarly, for downlink integer-forcing, the $\mth$ effective power may not correspond to $R_{\dt,m}$. While we can always find permutations that connect effective noise variances and power to rates, these permutations may differ between the uplink and downlink, which in turn limits the our approach in Section~\ref{s:duality} to sum-rate duality. 
\end{remark}

\begin{remark} As noted in Remark~\ref{r:ratesplitting}, uplink-downlink duality continues to hold for conventional linear architectures under rate splitting. The same is true for integer-forcing architectures, but in terms of the sum rate. \end{remark}

\noindent\textbf{Approximate Sum Capacity.} For the uplink channel, it is known that integer-forcing can operate with a constant gap of the sum capacity~\cite[Theorem 3]{oen14},~\cite[Theorem 4]{ncnc16}. Using uplink-downlink duality, we can establish a matching result for the downlink channel. 

\begin{theorem} \label{t:downlinkconstantgap} For any channel matrix $\Hm_{\dt} \in \RR^{M \times N}$ and total power constraint $P_{\text{total}}$, there is a choice of the power allocation $\Pm_{\dt}$, integer matrix $\Am_{\dt}$, beamforming matrix $\Bm_{\dt}$, and equalization vectors $\cv_{\dt,m}$ such that the integer-forcing beamforming architecture can operate within a constant gap of the downlink sum-capacity,
\begin{align}
\sum_{m = 1}^L R_{\dt,m} \geq \max_{\substack{\Km \succeq 0 \\ \tr(\Km) \leq P_{\text{total}}}} \frac{1}{2} \log\det\Big( \Id + \Hm_{\dt}\T \Km \Hm_{\dt} \Big) - \frac{L}{2} \log{L}
\end{align} where $\Km \succeq 0$ means that the matrix $\Km$ must be positive semidefinite.
\end{theorem} The proof is deferred to Appendix~\ref{a:downlinkconstantgap}.

\noindent\textbf{Iterative Optimization.} In Section~\ref{s:optimization}, we present an application of our uplink-downlink duality result for iteratively optimizing the beamforming and equalization matrices used in either an uplink or downlink integer-forcing architecture. This problem is non-convex in general as is the corresponding problem for conventional linear architectures. Our algorithm provably converges to a local optimum and performs well in simulations.

\section{Nested Lattice Codes} \label{s:nestedlattice}

Below, we review some basic lattice definitions as well as nested lattice code constructions that we will need for our achievability results. See~\cite{zamir} for a thorough introduction to lattice codes.

\subsection{Lattice Definitions}

A \textit{lattice} $\L$ is a discrete additive subgroup of $\RR^n$ such that, if $\lambdav_1,\lambdav_2 \in \L$, then $\lambdav_1 + \lambdav_2 \in \L$ and $-\lambdav_1, -\lambdav_2 \in \L$. The \textit{nearest neighbor quantizer} associated to $\L$ is defined as 
\begin{align}
Q_{\L}(\mathbf{x}) \triangleq \argmin_{\lambdav \in \L} \| \mathbf{x} - \lambdav\|
\end{align} (with ties broken in a systematic fashion). The \textit{fundamental Voronoi region} $\mathcal{V}$ of $\L$ is the set of all points in $\mathbb{R}^n$ that quantize to $\mathbf{0}$. We define the \textit{second moment} $\L$ as 
\begin{align}
\sigma^2(\L) \triangleq \frac{1}{n} \int_{\mathcal{V}} \| \xv \|^2 \frac{1}{\mathrm{Vol}(\mathcal{V})} d\xv  
\end{align} where $\mathrm{Vol}(\mathcal{V})$ denotes the volume of $\mathcal{V}$.

We also define the \textit{modulo operation} with respect to $\L$ as 
\begin{align}
[\mathbf{x}]\bmod\L \triangleq  \mathbf{x} - Q_{\L}(\mathbf{x}) \label{e:modlattice}
\end{align} and note that it satisfies a distributive law, $[a[\mathbf{x}]\bmod\L+b[\mathbf{y}]\bmod\L]\bmod\L=[a\mathbf{x}+b\mathbf{y}]\bmod\L$ for all $a,b\in \mathbb{Z}$ and $\mathbf{x},\mathbf{y}\in \RR^n$.

\begin{lemma}[Crypto Lemma]
Let $\xv$ be a random vector over $\RR^n$ and $\dv$ be an independent random vector drawn uniformly over the Voronoi region $\mathcal{V}$ of the lattice $\L$. The modulo sum $[ \xv + \dv] \bmod\L$ is independent of $\xv$ and uniform over $\mathcal{V}$. 
\end{lemma} See~\cite[Ch 4.1]{zamir} for a full proof.

The lattice $\L_\C$ is said to be \textit{nested} in the lattice $\L_F$ if $\L_\C \subset \L_\F$. In this case, $\L_\C$ is called the coarse lattice and $\L_\F$ the fine lattice. A \textit{nested lattice codebook} $\mathcal{L} = \L_\F \cap \mathcal{V}_{\C}$ consists of all fine lattice points that fall in the fundamental Voronoi region $\mathcal{V}_{\C}$ of the coarse lattice. Note that nested lattices satisfy the following quantization property:
\begin{align}\label{e:nestedquantization}
\big[ Q_{\L_{\F}}(\xv) \big] \bmod{\L_{\C}} = \big[ Q_{\L_{\F}}([\xv]\bmod{\L_{\C}}) \big] \bmod{\L_{\C}} \ .
\end{align}

\subsection{Nested Lattice Codes and Properties}

Our encoding strategies rely on the existence of good nested lattice codebooks. Below, we describe the nested lattice ensemble as well as properties that are central to our achievability proofs. Our notation closely follows that from~\cite[\S IV]{ncnc16}, which contains a more detailed exposition. 

Recall that $n$ denotes the blocklength of our coding scheme. Let $p$ represent a prime number and $\ZZ_p$ the finite field of size $p$. We will also need integer-valued parameters $0 \leq \kcl \leq \kfl,~ \ell = 1,\ldots,L$. Define $\kc \triangleq \min_\ell \kcl$, $\kf \triangleq \max_\ell \kfl$, and $k \triangleq \kf - \kc$. 

The construction begins with the generator matrix of a linear code $\Gm \in \ZZ_p^{\kf \times n}$. For $\ell = 1,\ldots,L$, define $\Gm_{\C,\ell}$ and $\Gm_{\F,\ell}$ to be the submatrices consisting of the first $\kcl$ and $\kfl$ rows of $\Gm$, respectively. Let 
\begin{align}
\Cc_{\C,\ell} &= \Big\{ \Gm_{\C,\ell}\T \wv : \wv \in \ZZ_p^{\kcl} \Big\} \\
\Cc_{\F,\ell} &= \Big\{ \Gm_{\F,\ell}\T \wv : \wv \in \ZZ_p^{\kfl} \Big\} 
\end{align} denote the resulting linear codebooks. For $\gamma > 0$ to be specified later, define the mapping $\phi(w) \triangleq \gamma p^{-1} w$ from $\ZZ_p$ to $\RR$. We also define the inverse mapping $\bar{\phi}(\kappa) \triangleq [ \gamma^{-1} p \kappa ] \bmod{p}$, which is only defined on the domain $\gamma p^{-1} \ZZ$. Both of these mappings are taken elementwise when applied to vectors and will be used to go back and forth between linear codebooks and lattices.

We now generate $L$ coarse lattices and $L$ fine lattices as follows:
\begin{align}
\L_{\C,\ell} &= \Big\{ \lambdav \in \gamma p^{-1} \ZZ^n : \bar{\phi}(\lambdav) \in \Cc_{\C,\ell} \Big\} \\
\L_{\F,\ell} &= \Big\{ \lambdav \in \gamma p^{-1} \ZZ^n : \bar{\phi}(\lambdav) \in \Cc_{\F,\ell}  \Big\} \ . 
\end{align} By construction, these lattices are nested according to the order for which the parameters $\kcl$ and $\kfl$ are increasing. Define $\L_\C$ and $\L_\F$ to be the coarsest and finest lattices in the ensemble, respectively. Let $\Vc_{\C,\ell}$ and $\Vc_{\F,\ell}$ denote the Voronoi regions of $\L_{\C,\ell}$ and $\L_{\F,\ell}$, respectively. Finally, we take the elements of the fine lattice $\L_{\F,\ell}$ that fall in the Voronoi region of the coarse lattice $\L_{\C,\ell}$ to be the nested lattice codebook
\begin{align}
\Lc_{\ell} &\triangleq \L_{\F,\ell} \cap \Vc_{\C,\ell} \\
&= [\L_{\F,\ell}] \bmod{\L_{\C,\ell}} \label{e:nestedlatticecodebook}
\end{align} for the $\lth$ user. 

The theorem below summarizes results from~\cite{oe12eilat} that demonstrate that this nested lattice construction exhibits good shaping and noise tolerance properties.

\begin{theorem}[{{\cite[Theorem 2]{oe12eilat}}}] \label{t:lattice}
For $\ell = 1,\ldots,L$, select parameters $P_\ell > 0$ and $0 < \sigma_{\eff,\ell}^2 < P_\ell$. Then, for any $\epsilon > 0$ and $n$ and $p$ large enough, there are parameters $\gamma$, $\kcl$, and $\kfl$ and a generator matrix $\Gm \in \ZZ_p^{\kf \times n}$ such that, for $\ell = 1,\ldots,L$
\begin{itemize}
\item[(a)] the submatrices $\Gm_{\C,\ell}$ and $\Gm_{\F,\ell}$ are full rank.
\item[(b)] the coarse lattices $\L_{\C,\ell}$ have second moments close to their power constraints
$$P_\ell - \epsilon < \sigma^2(\L_{\C,\ell}) < P_\ell \ . $$
\item[(c)] the lattices can tolerate the desired level of effective noise. Let $\zv_0,\zv_1,\ldots,\zv_L$ be independent noise vectors where $\zv_0 \sim \Nc(\zerov,\Id)$ and $\zv_\ell \sim \mathrm{Unif}(\Vc_{\C,\ell})$. For any $\beta_0,\beta_1,\ldots,\beta_L \in \RR$, let $\zv_{\eff} = \sum_{\ell = 0}^L \zv_\ell$. If $\beta_0^2 + \sum_{\ell=1}^L \beta_\ell^2 P_\ell \leq \sigma_{\eff,m}^2$, any fine lattice point $\lambdav \in \L_{\F,m}$ can recover from $\zv_{\eff}$ with high probability, 
$$\pr\Big( Q_{\L_{\F,m}}(\lambdav + \zv_{\eff}) \neq \lambdav \Big) < \epsilon \ . $$ Similarly, if $\beta_0^2 + \sum_{\ell=1}^L \beta_\ell^2 P_\ell \leq  P_m$, any coarse lattice point $\lambdav \in \L_{\C,m}$ can recover from $\zv_{\eff}$ with high probability, 
$$\pr\Big( Q_{\L_{\C,m}}(\lambdav + \zv_{\eff}) \neq \lambdav \Big) < \epsilon \ . $$
\item[(d)] the rates of the nested lattice codes satisfy
$$\frac{1}{n}\log|\Lc_{\ell}| = \frac{\kfl - \kcl}{n} \log_2{p} >  \frac{1}{2}\log\bigg(\frac{P_\ell}{\sigma_{\eff,\ell}^2}\bigg) - \epsilon \ . $$
\end{itemize}
\end{theorem}

Finally, it can be argued that we can label lattice codewords so that integer-linear combinations of codewords correspond to linear combinations of the messages over $\ZZ_p$. We recall the definition of a linear labeling from~\cite{fsk13}.

\begin{definition}\label{d:linearlabeling}
We say that a mapping $\varphi: \L_\F \rightarrow \ZZ_p^k$ is a \textit{linear labeling} if 
\begin{itemize}
\item[(a)]$\lambdav \in \L_{\F,\ell}$ if and only if the last $\kf - \kfl$ components of its label $\varphi(\lambdav)$ are zero. Similarly, $\lambdav \in \L_{\C,\ell}$ if and only if the last $\kf - \kcl$ components of its label $\varphi(\lambdav)$ are zero. 
\item[(b)] For all $a_\ell \in \ZZ$ and $\lambdav_\ell \in \L_{\F}$, we have that 
$$\varphi\bigg( \sum_{\ell = 1}^L a_\ell \lambdav_\ell \bigg) = \bigoplus_{\ell = 1}^L q_\ell \varphi(\lambdav_\ell) $$
\end{itemize} where $q_\ell = [a_\ell] \bmod{p}$.
\end{definition}

Consider the mapping that sets $\varphi(\lambdav)$ to be the last $k$ components of the unique vector $\vv \in \ZZ_p^{\kf}$ satisfying $\bar{\phi}(\lambdav) = \Gm\T \vv$. From~\cite[Theorem 10]{ncnc16}, $\varphi$ is a linear labeling. We also define the inverse map $$\bar{\varphi} \triangleq \phi\Bigg(\Gm\T \begin{bmatrix} \zerov_{\kc} \\ \wv \end{bmatrix} \Bigg) \ ,$$ which satisfies $\varphi\big(\bar{\varphi}(\wv)\big) = \wv$.

\subsection{Intuition via Signal Levels} \label{s:signallevels}

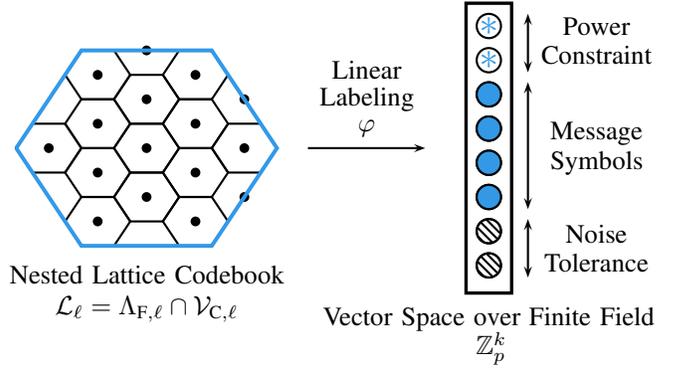
\begin{figure}[h]
\begin{center}
\psset{unit=0.65mm}
\begin{pspicture}(-30,-45)(105,32)
%\psframe(-30,-45)(105,32)

\rput(-20,0){
\psline(-3.333,5)(3.333,5)(6.667,0)(3.333,-5)(-3.333,-5)(-6.667,0)(-3.333,5)
\pscircle[fillstyle=solid,fillcolor=black](0,0){1}
}
\rput(-10,15){
\psline(-3.333,5)(3.333,5)(6.667,0)(3.333,-5)(-3.333,-5)(-6.667,0)(-3.333,5)
\pscircle[fillstyle=solid,fillcolor=black](0,0){1}
}
\rput(-10,5){
\psline(-3.333,5)(3.333,5)(6.667,0)(3.333,-5)(-3.333,-5)(-6.667,0)(-3.333,5)
\pscircle[fillstyle=solid,fillcolor=black](0,0){1}
}
\rput(-10,-5){
\psline(-3.333,5)(3.333,5)(6.667,0)(3.333,-5)(-3.333,-5)(-6.667,0)(-3.333,5)
\pscircle[fillstyle=solid,fillcolor=black](0,0){1}
}
\rput(-10,-15){
\psline(-3.333,5)(3.333,5)(6.667,0)(3.333,-5)(-3.333,-5)(-6.667,0)(-3.333,5)
\pscircle[fillstyle=solid,fillcolor=black](0,0){1}
}
\rput(20,0){
\psline(-3.333,5)(3.333,5)(6.667,0)(3.333,-5)(-3.333,-5)(-6.667,0)(-3.333,5)
\pscircle[fillstyle=solid,fillcolor=black](0,0){1}
}
\rput(0,10){
\psline(-3.333,5)(3.333,5)(6.667,0)(3.333,-5)(-3.333,-5)(-6.667,0)(-3.333,5)
\pscircle[fillstyle=solid,fillcolor=black](0,0){1}
}
\rput(0,0){
\psline(-3.333,5)(3.333,5)(6.667,0)(3.333,-5)(-3.333,-5)(-6.667,0)(-3.333,5)
\pscircle[fillstyle=solid,fillcolor=black](0,0){1}
}
\rput(0,-10){
\psline(-3.333,5)(3.333,5)(6.667,0)(3.333,-5)(-3.333,-5)(-6.667,0)(-3.333,5)
\pscircle[fillstyle=solid,fillcolor=black](0,0){1}
}
\rput(10,15){
\psline(-3.333,5)(3.333,5)(6.667,0)(3.333,-5)(-3.333,-5)(-6.667,0)(-3.333,5)
\pscircle[fillstyle=solid,fillcolor=black](0,0){1}
}
\rput(10,5){
\psline(-3.333,5)(3.333,5)(6.667,0)(3.333,-5)(-3.333,-5)(-6.667,0)(-3.333,5)
\pscircle[fillstyle=solid,fillcolor=black](0,0){1}
}
\rput(10,-5){
\psline(-3.333,5)(3.333,5)(6.667,0)(3.333,-5)(-3.333,-5)(-6.667,0)(-3.333,5)
\pscircle[fillstyle=solid,fillcolor=black](0,0){1}
}
\rput(10,-15){
\psline(-3.333,5)(3.333,5)(6.667,0)(3.333,-5)(-3.333,-5)(-6.667,0)(-3.333,5)
\pscircle[fillstyle=solid,fillcolor=black](0,0){1}
}
\pscircle[fillstyle=solid,fillcolor=black](0,20){1}
\pscircle[fillstyle=solid,fillcolor=black](20,10){1}
\pscircle[fillstyle=solid,fillcolor=black](20,-10){1}

\psline[linewidth=1.5pt,linecolor=LineBlue](-26.667,0)(-13.333,20)(13.333,20)(26.667,0)(13.333,-20)(-13.333,-20)(-26.667,0)

\rput(0,-26){Nested Lattice Codebook}
\rput(0,-33){$\Lc_{\ell} = \Lambda_{\text{F},\ell} \cap \mathcal{V}_{\text{C},\ell}$}

\psline{->}(33,0)(57,0) \rput(45,4){$\varphi$} \rput(45,16){Linear}\rput(45,10){Labeling}

\rput(15,0){
\rput(40,-40){
\psframe[linewidth=1.25pt](10,10)(20,70)

\rput(15,11){
\pscircle(0,5){2.7}
\pscircle[fillstyle=vlines,hatchsep=1.5pt,fillcolor=black](0,5){2.7}
\pscircle(0,12){2.7}
\pscircle[fillstyle=vlines,hatchsep=1.5pt,fillcolor=black](0,12){2.7}
\pscircle(0,19){2.7}
\pscircle[fillstyle=solid,fillcolor=LineBlue](0,19){2.7}
\pscircle(0,26){2.7}
\pscircle[fillstyle=solid,fillcolor=LineBlue](0,26){2.7}
\pscircle(0,33){2.7}
\pscircle[fillstyle=solid,fillcolor=LineBlue](0,33){2.7}
\pscircle(0,40){2.7}
\pscircle[fillstyle=solid,fillcolor=LineBlue](0,40){2.7}
\pscircle(0,47){2.7}
\rput(0,47){\Large{$\textcolor{LineBlue}{*}$}}
\pscircle(0,54){2.7}
\rput(0,54){\Large{$\textcolor{LineBlue}{*}$}}
}
}

\rput(55,-35){Vector Space over Finite Field}
\rput(55.5,-41){$\mathbb{Z}_p^k$}

\psline{<->}(63,27.5)(63,15.5) \rput(77,25){Power} \rput(77,19){Constraint}
\psline{<->}(63,13.5)(63,-12.5)  \rput(77,3){Message} \rput(77,-3){Symbols}
\psline{<->}(63,-14.5)(63,-26.5)  \rput(77,-17.5){Noise} \rput(77,-23.5){Tolerance}
}

\end{pspicture}
\end{center}
\caption{Illustration of the linear labeling of the $\ell^{\text{th}}$ nested lattice codebook. The first $\kcl - \kc$ elements of the linear label are ``don't care'' entries (denoted by the $*$ symbol) and correspond to the $\bmod{\ \L_{\C,\ell}}$ operation. The next $\kfl - \kcl$ elements are free to carry information symbols (denoted by solid circles). The last $\kf - \kfl$ elements are zero (denoted by {dashed lines}).} \label{f:latticesignallevels}
\end{figure}

We now develop some intuition by describing the linear labeling of our nested lattice construction in terms of ``signal levels'' over $\mathbb{Z}_p^k$. See Figure~\ref{f:latticesignallevels} for an illustration. Each codeword from the $\ell^{\text{th}}$ nested lattice codebook can be expressed as an element from the $\ell^{\text{th}}$ fine lattice, $\lambdav_{\F,\ell} \in \L_{\F,\ell}$ modulo the $\ell^{\text{th}}$ coarse lattice, $[\lambdav_{\F,\ell}] \bmod{\L_{\C,\ell}} \in \Lc_{\ell}$. We can thus write the linear label of any codeword in $\Lc_{\ell}$ as 
\begin{align}
\varphi\left([\lambdav_{\F,\ell}] \bmod{\L_{\C,\ell}} \right) &= \varphi\left(\lambdav_{\F,\ell} - Q_{\L_{\C,\ell}}\big(\lambdav_{\F,\ell} \big)\right) \\
&= \varphi(\lambdav_{\F,\ell}) \ominus \varphi\left(Q_{\L_{\C,\ell}}\big(\lambdav_{\F,\ell} \big)\right)
\end{align} where $\ominus$ denotes the subtraction operation over $\ZZ_p$. From Definition~\ref{d:linearlabeling}(a), we know that $\varphi\left(Q_{\L_{\C,\ell}}\big(\lambdav_{\F,\ell} \big)\right)$ only occupies the top $\kcl - \kc$ elements of the vector corresponding to the linear label. Similarly, we know that $\varphi(\lambdav_{\F,\ell})$ only occupies the top $\kf - \kfl$ elements of the vector. Therefore, the first $\kcl - \kc$ elements are determined by the shaping operation $\bmod{\Lc_{\ell}}$ and can be interpreted as enforcing the power constraint $P_\ell$. The next $\kfl - \kcl$ elements are occupied by information symbols and the final $\kf - \kfl$ elements are zero, which can be interpreted as enforcing the noise tolerance threshold $\sigma_{\eff,\ell}^2$.

Overall, we arrive at the following high-level intuitions:
\begin{itemize}
\item Only the encoder with the largest power (i.e., the coarsest lattice) can control the very top signal levels. For an encoder $\ell$ whose power is less than the maximum power, the top $\kcl - \kc$ signal levels are outside of its direct control. (In our strategy, these signal levels will be set during the decoding process.)
\item The lowest $\kf - \kfl$ are set to zero by the $\lth$ encoder so that all of its information symbols lie above the noise level (i.e, the corresponding fine lattice can tolerate the desired effective noise variance).
\end{itemize}

%%%%%%%%%%%%%%%%%%%%

\section{Uplink Integer-Forcing Architecture}\label{s:uplink}

Our uplink coding scheme is taken from~\cite[Section VI]{ncnc16}. Below, we summarize the encoding and decoding operations in order to highlight the similarities between the uplink and downlink integer-forcing schemes.

We begin by selecting a power allocation $\upm = \diag(P_{\ut,1},\ldots,P_{\ut,L})$ for the codewords and a beamforming matrix $\Cm_{\ut}$. Note that, in order to meet the total power constraint with equality, we require that $\tr(\mathbf{C}_{\ut}\T\mathbf{C}_{\ut}\upm) = \ptu$. We also select a full-rank integer matrix $\Am \in \ZZ^{L \times L}$ and an equalization matrix $\Bm_{\ut} = [\bv_{\ut,1} ~\cdots~\bv_{\ut,L}]\T \in \RR^{L \times N}$. These choices specify the effective noise variances $\sigma_{\ut,m}^2$ from~\eqref{e:zueff}.

The structure of the integer matrix $\Am_{\ut}$ determines which codewords can be cancelled out in each decoding step. In order to keep our notation manageable, we assume that $\Am_{\ut}$ is selected so that the $\mth$ user can be associated with the $\mth$ effective noise variance. The following definition describes when this is possible.

\begin{definition} \label{d:uplinkadmissible} We say that the \textit{identity permutation is admissible for the uplink} if \begin{itemize}
\item[(a)] the effective noise variances are in increasing order, $\sigma_{\ut,1}^2 \leq \cdots \leq \sigma_{\ut,L}^2$ and
\item[(b)] the leading principal submatrices of $\Am_{\ut}$ are full rank, $\mathrm{rank}(\Am_{\ut}^{[1:m]}) = m$ for $m=1,\ldots,L$. 
\end{itemize}
\end{definition} These conditions can always be met by permuting the rows and columns of the selected matrices. See the remark below for details.

\begin{remark} \label{r:uplinkpermutation} Assume that we have chosen uplink parameters $\Am_{\ut}$, $\Bm_{\ut}$, $\Cm_{\ut}$, $\Hm_{\ut}$, and $\Pm_{\ut}$, and now we wish to satisfy Definition~\ref{d:uplinkadmissible} by permuting row and column indices. First, select a permutation $\pi$ that places the effective noise variances in increasing order, $\sigma_{\ut,\pi(1)}^2 \leq \cdots \leq \sigma_{\ut,\pi(L)}^2$. Next, let $\theta$ be a permutation such that, after row permutation of $\Am_{\ut}$ by $\pi$ and column permutation by $\theta$, we obtain a matrix $\Atm = \{a_{\ut,\pi(m),\theta(\ell)}\}_{m,\ell}$ whose leading principal submatrices are full rank, $\mathrm{rank}(\Atm_{\ut}^{[1:m]}) = m$ for $m=1,\ldots,L$. Now, permute the rows of the equalization matrix $\Bm_{\ut}$ by $\pi$ to obtain $\mathbf{\tilde{B}}_{\ut,m}$. Finally, reindex the users by $\theta$ by setting $\mathbf{\tilde{C}}_{\ut}$ to be the beamforming matrix consisting of vectors $\ctv_{\ut,m} = \cv_{\ut,\theta(\ell)}$, $\mathbf{\tilde{H}}_{\ut}$ to be the channel matrix consisting of submatrices $\mathbf{\tilde{H}}_{\ut,\ell} = \mathbf{H}_{\ut,\theta(\ell)}$, and $\mathbf{\tilde{P}}_{\ut}$ to be the power matrix with diagonal entries $\tilde{P}_{\ut,\ell} = P_{\ut,\theta(\ell)}$. It can be verified that the effective noise variances $\tilde{\sigma}_{\ut,m}^2, m = 1, \ldots,L $ that result from these permuted parameters satisfies $\tilde{\sigma}_{\ut,m}^2 = \sigma_{\ut,\pi(m)}^2$. Overall, we find that the rates $R_{\ut,\pi(m)} = \frac{1}{2}\log^+\big(P_{\ut,\pi(m)}/\sigma_{\ut,\theta(m)}^2\big),~m=1,\ldots,L$ are achievable.
\end{remark}

For notational convenience, we assume going forward that the identity permutation is admissible.

In our decoding procedure, we will need to triangularize $\Am_{\ut}$ over $\ZZ_p$ in the following sense. We need a lower unitriangular matrix $\Lbm \in \ZZ_p^{L \times L}$ such that $\Abm = [ \Lbm \Am_{\ut}]\bmod{p}$ is upper triangular. First, note that the condition in Definition~\ref{d:uplinkadmissible}(b) is equivalent to the condition that there exists a lower unitriangular matrix $\Lm \in \RR^{L \times L}$ such that $\Lm \Am$ is upper triangular. Given the existence of such an $\Lm$, it follows from~\cite[Appendix A]{oen14} that, for $p$ large enough, we can always find an appropriate $\Lbm$. It also follows that $\Lbm$ has a lower unitriangular inverse $\Lbm^{(\text{inv})}$ over $\ZZ_p$. After our overview of the encoding and decoding steps, we provide an explicit example of such matrices in Example~\ref{ex:algebraicsic}.

Using the linear labeling $\varphi$, we can show that each nested lattice codebook $\Lc_\ell$ is isomorphic to the vector space $\ZZ_p^{\kfl - \kcl}$. Each user will take the $p$-ary expansion of its message index $w_{\ut,\ell}$ to obtain a message vector $\wv_{\ut,\ell} \in \ZZ_p^{\kfl - \kcl}$. The intermediate goal of the receiver is to recover $L$ linear combinations of the form
\begin{align} \label{e:uplinklinear}
\uv_{\ut,m} = \bigoplus_{\ell =1}^L q_{\ut,m,\ell} \wvt_{\ut,\ell} 
\end{align} where $q_{\ut,m,\ell} = [a_{\ut,m,\ell}] \bmod{p}$, $a_{\ut,m,\ell}$ is the $(m,\ell)^{\text{th}}$ entry of $\Am_{\ut}$, and $\wvt_{\ut,\ell} \in \llb \wv_{\ut,\ell} \rrb$ with
\begin{align} \label{e:msgcoset}
\scalemath{0.93}{
\llb \wv_{\ut,\ell} \rrb \triangleq \left\{ \wv \in \ZZ_p^k : \wv = \begin{bmatrix} \ev \\ \wv_{\ut,\ell} \\ \zerov_{\kf - \kfl} \end{bmatrix} \text{~for some~} \ev \in \ZZ^{\kcl - \kc} \right\} .}
\end{align} That is, the receiver attempts to recover $L$ linear combinations of cosets of the messages. The top elements $\kcl - \kc$ of $\wvt_{\ut,\ell}$ (denoted by $\ev$ in~\eqref{e:msgcoset}) can be thought of as ``don't care'' entries. That is, they can take any values, but are not directly set by the $\lth$ transmitter, since this would require it to exceed its power constraint. In our scheme, these entries are determined during the decoding process and can be recovered by the receiver, although this is not required to recover the original messages $\wv_{\ut,\ell}$.

We now state the encoding and decoding steps used in the uplink integer-forcing architecture. We select an ensemble of good nested lattices $\L_{\C,1},\ldots,\L_{\C,L},\L_{\F,1},\ldots,\L_{\F,L}$ with parameters $P_{\ut,1},\ldots,P_{\ut,L}$ and $\sigma_{\ut,1}^2,\ldots, \sigma_{\ut,L}^2$ using Theorem~\ref{t:lattice}.

\noindent\textbf{Encoding:} The $\lth$ transmitter starts by taking the $p$-ary expansion of its message index $w_{\ut,\ell}$ to obtain the message vector $\wv_{\ut,\ell} \in \ZZ_p^{\kfl - \kcl}$. It then uses the inverse linear labeling to map this to a lattice point
\begin{align}
\lambdav_{\ut,\ell} &= \left[ \bar{\varphi}\left( \begin{bmatrix} \zerov_{k_{\C,\ell} - \kcmin} \\ \mathbf{w}_{\ut,\ell} \\ \zerov_{\kfmax - k_{\F,\ell}} \end{bmatrix} \right)\right] \bmod{\L_{\C,\ell}} \label{e:uplinkencoding1} 
\end{align} and dithers it to produce the codeword 
\begin{align}
\sv_{\ut,\ell} = [ \lambdav_{\ut,\ell} + \dv_{\ut,\ell} ] \bmod{\L_{\C,\ell}}
\end{align} where the dither vector $\dv_{\ut,\ell}$ is drawn independently and uniformly over $\Vc_{\C,\ell}$. Thus, by the Crypto Lemma and Theorem~\ref{t:lattice}(b), $\sv_{\ut,\ell}$ is independent of $\lambdav_{\ut,\ell}$ and has expected power close to $P_{\ut,\ell}$. Finally, the $\ell^{\text{th}}$ transmitter uses its beamforming vector $\cv_{\ut,\ell}$ to produce its channel input
\begin{align}
\Xm_{\ut,\ell} = \cv_{\ut,\ell} \sv_{\ut,\ell}\T \ .
\end{align}

\noindent\textbf{Decoding:} The receiver attempts to recover linear combinations of the form~\eqref{e:uplinklinear} and then solve them to obtain estimates of the message vectors. As an intermediate step, the receiver will attempt to decode certain integer-linear combinations of the lattice codewords, i.e., 
\begin{align}
\muv_{\ut,m} = \bigg[ \sum_{\ell = 1}^L a_{\ut,m,\ell} \,\lambdatv_{\ut,\ell} \bigg] \bmod{\L_{\C}}
\end{align} where $\lambdatv_{\ut,\ell} \triangleq \lambdav_{\ut,\ell} - Q_{\L_{\C,\ell}}(\lambdav_{\ut,\ell} + \dv_{\ut,\ell})$. The linear labels of these integer-linear combinations correspond to the desired linear combinations, $\varphi(\muv_{\ut,m}) = \uv_{\ut,m}$. (This decoding step sets the top $\kcl - \kc$ ``don't care'' entries of $\wtv_{\ut,\ell}$ to the values specified by the linear label $\varphi\big(-Q_{\L_{\C,\ell}}(\lambdav_{\ut,\ell} + \dv_{\ut,\ell})\big)$.)

The main obstacle is that, in order to decode the $\mth$ integer-linear combination, the receiver must first cancel out the first $m-1$ codewords using the prior $m-1$ linear combinations. This is accomplished via the algebraic SIC technique from~\cite{oen14}, which takes advantage of the fact that the nested lattice codebook is isomorphic to a vector space over $\ZZ_p$. Specifically, by adding an integer-linear combination of $\muv_{\ut,1},\ldots,\muv_{\ut,m-1}$ it can ``digitally'' null out the lattice codewords $\lambdatv_{\ut,1},\ldots,\lambdatv_{\ut,m-1}$ without impacting effective noise. Define 
\begin{align}
\nuv_{\ut,m} &= \bigg[ \muv_{\ut,m} + \sum_{i=1}^{m-1} \bar{l}_{m,i} \muv_{\ut,i} \bigg] \bmod{\L_{\C}} \label{e:algebraicsic} \\ 
&=\bigg[ \sum_{\ell = 1}^L \bar{a}_{m,\ell} \,\lambdatv_{\ut,\ell} \bigg] \bmod{\L_{\C}}
\end{align} where $\bar{l}_{m,i}$ is the $(m,i)^{\text{th}}$ entry of $\Lbm$ and $\bar{a}_{m,\ell}$ is the $(m,\ell)^{\text{th}}$ entry of the upper triangular matrix $\Abm$ defined above. Note that $\nuv_{\ut,m} \in \L_{\F,m}$ and, given $\nuv_{1},\ldots,\nuv_{m}$, we can recover $\muv_{\ut,m}$:
\begin{align}
\muv_{\ut,m} = \bigg[ \sum_{i = 1}^m \bar{l}^{(\text{inv})}_{m,i} \nuv_i \bigg] \bmod{\L_{\C}} 
\end{align} where $\bar{l}^{(\text{inv})}_{m,i}$ is the $(m,i)^{\text{th}}$ entry of $\Lbm^{(\text{inv})}$.

For the $\mth$ decoding step, we assume that the receiver has already successfully recovered the previous $m-1$ integer-linear combinations, i.e., $\muhv_{\ut,1} = \muv_{\ut,1},\ldots,\muhv_{\ut,m-1} = \muv_{\ut,m-1}$. The receiver begins by equalizing its observation, \begin{align}
\ytv_{\ut,m}\T = \bv_{\ut,m}\T \Ym_{\ut}  \ .
\end{align} The receiver then removes the dithers\footnote{For lattice coding proofs, it is usually assumed that the dithers are available at the transmitters and receivers. Note that it can be shown that fixed dither vectors exist that attain the same performance~\cite[Appendix H]{ncnc16}. In other words, the randomness dithers should not be viewed as common randomness, but rather as part of the usual probabilistic method used in random coding proofs.}, nulls out the lattice codewords corresponding to the first $m-1$ users, and quantizes onto the $\mth$ fine lattice,
\begin{align}
\nuv_{\ut,m} &= \bigg[ Q_{\L_{\F,m}}\Big(\ytv_{\ut,m} + \sum_{i=1}^{m-1} \bar{l}_{m,i} \muhv_{\ut,i} \\
&~~~~~~~~~~~~~~~~-~ \sum_{\ell = 1}^L a_{\ut,m,\ell}\, \dv_{\ut,\ell} \bigg) \bigg] \bmod{\L_{\C}}\\ 
&=\bigg[ Q_{\L_{\F,m}}\bigg(\sum_{\ell = 1}^L a_{\ut,m,\ell} (\sv_{\ut,\ell} -  \dv_{\ut,\ell})  \\
& ~~~~~~~~~~~~~~~~+~ \sum_{i=1}^{m-1} \bar{l}_{m,i} \muhv_{\ut,i}  + \zv_{\ut,\eff,m} \bigg) \bigg] \bmod{\L_{\C}} \\
&=\bigg[ Q_{\L_{\F,m}}\bigg( \muv_{\ut,m} + \sum_{i=1}^{m-1} \bar{l}_{m,i} \muhv_{\ut,i}  + \zv_{\ut,\eff,m} \bigg) \bigg] \bmod{\L_{\C}} \\
&= \Big[Q_{\L_{\F,m}}(\nuv_{\ut,m} + \zv_{\ut,\eff,m})\Big] \bmod{\L_{\C}}
\end{align} where the last step follows from~\eqref{e:algebraicsic} and the distributive law. It then forms an estimate of its desired linear combination
\begin{align}
\muhv_{\ut,m} &= \bigg[ \sum_{i = 1}^m \bar{l}^{(\text{inv})}_{m,i} \nuhv_i \bigg] \bmod{\L_{\C}} \\
\uhv_{\ut,m} &= \varphi(\muhv_{m}) \ .
\end{align} 
%\begin{align}
%&\nuv_{\ut,m} \\&\scalemath{0.93}{= \bigg[ Q_{\L_{\F,m}}\bigg(\ytv_{\ut,m} + \sum_{i=1}^{m-1} \bar{l}_{m,i} \muhv_{\ut,i} - \sum_{\ell = 1}^L a_{\ut,m,\ell}\, \dv_{\ut,\ell} \bigg) \bigg] \bmod{\L_{\C}}} \\
%&\scalemath{0.8}{=\bigg[ Q_{\L_{\F,m}}\bigg(\sum_{\ell = 1}^L a_{\ut,m,\ell} (\sv_{\ut,\ell} -  \dv_{\ut,\ell})  + \sum_{i=1}^{m-1} \bar{l}_{m,i} \muhv_{\ut,i}  + \zv_{\ut,\eff,m} \bigg) \bigg] \bmod{\L_{\C}}} \\
%&\scalemath{1}{=\bigg[ Q_{\L_{\F,m}}\bigg( \muv_{\ut,m} + \sum_{i=1}^{m-1} \bar{l}_{m,i} \muhv_{\ut,i}  + \zv_{\ut,\eff,m} \bigg) \bigg] \bmod{\L_{\C}}} \\
%&\scalemath{1}{= \big[Q_{\L_{\F,m}}(\nuv_{\ut,m} + \zv_{\ut,\eff,m})\big] \bmod{\L_{\C}}}
%\end{align}  
Finally, if all $L$ linear combinations have been recovered correctly, we can solve the linear combinations to recover the original messages. This strategy leads to the following achievable rates.

\begin{theorem}[{{\cite[Lemma 13]{ncnc16}}}] \label{t:uplinkif}
Choose a power allocation $\upm = \diag(P_{\ut,1},\ldots,P_{\ut,L})$, beamforming matrix $\Cm_{\ut} \in \RR^{M \times L}$, channel matrix $\Hm_{\ut} \in \RR^{N \times M}$, full-rank integer matrix $\Am_{\ut} = [\av_{\ut,1} ~ \cdots ~ \av_{\ut,L}]\T \in \ZZ^{L \times L}$, and equalization vectors $\bv_{\ut,m} \in \RR^{N}$. Assume, without loss of generality, that the identity permutation is admissible for the uplink according to Definition~\ref{d:uplinkadmissible}. Then, the following rates are achievable
\begin{align}
R_{\ut,m} &= \frac{1}{2}\log^+\bigg(\frac{P_{\ut,m}}{\sigma_{\ut,m}^2} \bigg),~~~m = 1,\ldots,L, \\
\sigma_{\ut,m}^2 &= \| \bv_{\ut,m} \|^2 + \Big\|  \big( \bv_{\ut,m}\T \Hm_{\ut} \Cm_{\ut} - \av_{\ut,m}\T \big) \upm^{1/2} \Big\|^2  \ . 
\end{align} 
\end{theorem} For a full proof, see~\cite[\S VI]{ncnc16}.

\begin{example}\label{ex:algebraicsic}
To illustrate the algebraic SIC technique, consider the integer matrix 
$$\Am = \begin{bmatrix} 2 & 1 \\ 0 & 1 \end{bmatrix} $$ and assume that the underlying finite field is $\ZZ_3$ and $\sigma_{\ut,1}^2 < \sigma_{\ut,2}^2$. Define the following lower unitriangular matrices 
$$\Lbm_1 = \begin{bmatrix} 1 & 0 \\ 0 & 1 \end{bmatrix} \qquad \qquad \Lbm_2 = \begin{bmatrix} 1 & 0 \\ 2 & 1 \end{bmatrix} \ .$$ Each matrix can be used to cancel out one of the codewords, i.e., 
\begin{align*}
\Abm_1 &= [\Lbm_1 \Am] \bmod{3} = \begin{bmatrix} 2 & 1 \\ 0 & 1 \end{bmatrix}  \\ \Abm_2 &= [\Lbm_2 \Am] \bmod{3} = \begin{bmatrix} 2 & 1 \\ 1 & 0 \end{bmatrix}  \ .\end{align*} Since $\Abm_1$ is upper triangular, the first lattice codeword $\lambdatv_{\ut,1}$ only needs to tolerate effective noise variance $\sigma_{\ut,1}^2$, and the second lattice codeword $\lambdatv_{\ut,2}$ must tolerate the larger noise variance $\sigma_{\ut,2}^2$. In other words, the identity permutation is admissible according to Definition~\ref{d:uplinkadmissible}. Alternatively, using $\Lbm_2$, we obtain a matrix $\Abm_2$ that is upper triangular after permuting the two columns. In other words, the first lattice codeword must tolerate the larger noise variance $\sigma_{\ut,2}^2$ whereas the second lattice codeword only needs to tolerate the smaller one $\sigma_{\ut,1}^2$. In general, not all orderings are possible, as demonstrated by the following integer matrix, which admits only the identity permutation:
$$\Am = \begin{bmatrix} 1 & 0 \\ 1 & 1 \end{bmatrix} \ . $$

\end{example}

%%%%%%%%%%%%%%%%%%%%

\section{Downlink Integer-Forcing Architecture}\label{s:downlink}

The key idea underlying downlink integer-forcing is the fact that the transmitter can \textit{pre-invert the linear combinations} prior to encoding. This technique, which was first proposed by Hong and Caire~\cite{hc12,hc13}, allows each receiver to decode any integer-linear combination of the codewords in order to reduce the effective noise but still recover its desired message. These papers focused on the important special case where all users employ the same fine and coarse lattices, and thus have equal powers and must tolerate the worst effective noise across receivers. Specifically, as illustrated in Figure~\ref{f:downlinkarch}, in the equal power case, the transmitter should apply the inverse $\Qm_{\dt}^{-1}$ to the $p$-ary expansions of the messages, prior to generating the lattice codewords. As a result, each receiver's integer-linear combination of codewords corresponds to its desired message.

This basic strategy can be generalized to allow for unequal powers and a unique effective noise variance associated to each receiver. However, if each lattice codeword is generated using a different coarse lattice, it does not suffice to apply the inverse $\Qm_{\dt}^{-1}$ to the messages. As discussed in Section~\ref{s:signallevels}, the issue is that the top $\kcl - \kc$ elements (i.e., the ``don't care'' entries) cannot be set directly by the $\lth$ encoder. Rather, in our coding scheme, their values will be set as a function of the message and dither vectors, and will act as interference for encoders whose higher power levels allows them access to these entries. Thus, we will encode the messages in stages, starting with the signal levels accessible to all encoders and applying the inverse $\Qm_{\dt}^{-1}$. For the next set of signal levels, the encoder with the lower power will not participate, other than to add interference via its ``don't care'' entries. The remaining encoder will pre-cancel this interference as well as apply the inverse of the submatrix of $\Qm_{\dt}$ with row and column indices corresponding to the active encoders. This process will continue, removing an additional encoder at each stage, until all signal levels have been filled.

\subsection{Integer-Forcing Beamforming}

We begin by choosing a power allocation $\dpm = \diag(P_{\dt,1},\ldots,P_{\dt,L})$ for the codewords and a full-rank integer matrix $\Am_{\dt} \in \ZZ^{L \times L}$. We also select a beamforming matrix $\Bm_{\dt} \in \RR^{N \times L}$ and equalization vectors $\cv_{\dt,m} \in \RR^{M_m},~m = 1,\ldots,L$.  To meet the total power constraint with equality, we need that $\tr(\Bm_{\dt}\T \Bm_{\dt} \dpm) = \ptd$. Taken together, these choices specify the effective noise variances $\sigma_{\dt,m}^2$ from~\eqref{e:downlinknoisevar}.

As in the uplink case, the structure of the integer matrix $\Am_{\dt}$ will determine the order in which interference cancellation is possible via a digital variation on dirty-paper precoding that occurs at the message level. To simplify our notation, we will assume that $\Am_{\dt}$ is selected so that the $\mth$ user can be associated with the $\mth$ power $P_{\dt,m}$. We specify when this is possible below.

\begin{definition} \label{d:downlinkadmissible} We say that the \textit{identity permutation is admissible for the downlink} if \begin{itemize}
\item[(a)] the powers are in decreasing order, $P_{\dt,1} \geq \cdots \geq P_{\dt,L}$ and
\item[(b)] the leading principal submatrices of $\Am_{\dt}$ are full rank, $\mathrm{rank}(\Am_{\dt}^{[1:m]}) = m$ for $m=1,\ldots,L$. 
\end{itemize}
\end{definition} 
These conditions can be satisfied via reindexing the rows and columns of the chosen matrices, as demonstrated in the following remark.
\begin{remark} \label{r:downlinkpermute}
Assume that we have selected downlink parameters $\Am_{\dt}$, $\Bm_{\dt}$, $\Cm_{\dt}$, $\Hm_{\dt}$, and $\Pm_{\dt}$, and now wish to satisfy Definition~\ref{d:downlinkadmissible} by permuting row and column indices. We first choose a permutation $\pi$ that puts the powers in decreasing order $P_{\dt,\pi(1)} \geq \cdots \geq P_{\dt,\pi(L)}$. Next, we take a permutation $\theta$ such that, for $\Atm_{\dt} = \{a_{\dt,\theta(m),\pi(\ell)}\}_{m,\ell}$, the leading principal submatrices are all full rank, $\mathrm{rank}(\Atm_{\dt}^{[1:m]}) = m$ for $m = 1,\ldots,L$. Then, we permute the columns of the beamforming matrix $\Bm_{\dt}$ to get $\mathbf{\tilde{B}}_{\dt}$  as well as the powers $\tilde{P}_{\dt,\ell} = P_{\dt,\pi(\ell)}$ to get a permuted power matrix $\mathbf{\tilde{P}}_{\dt}$. Finally, we reindex the users by $\theta$, which in turns yields a permuted equalization matrix $\mathbf{\tilde{C}}_{\dt}$ consisting of equalization vectors $\ctv_{\dt,m} = \ctv_{\dt,\theta(m)}$ as well as permuted channel submatrices $\mathbf{\tilde{H}}_{\dt,m} = \Hm_{\dt,\theta(m)}$ to form overall channel matrix $\mathbf{\tilde{H}}_{\dt}$. It can be verified that the effective noise variances $\tilde{\sigma}_{\dt,m}^2, m = 1,\ldots,L$, that result from these permuted parameters satisfy $\tilde{\sigma}_{\dt,m}^2 = \sigma_{\dt,\theta(m)}^2$. Thus, the rates $R_{\dt,\theta(m)} = \frac{1}{2} \log^+\big(P_{\dt,\pi(m)} / \sigma^2_{\dt,\theta(m)} \big)$ are achievable.  
\end{remark}

To keep our notation manageable, we assume below that the identity permutation is admissible.

We now describe the encoding and decoding steps used in the integer-forcing beamforming architecture. Using the parameters $P_{\dt,1} \geq  \cdots \geq P_{\dt,L}$ and $\sigma_{\dt,1}^2,\ldots,\sigma_{\dt,L}^2$, we pick a good ensemble of nested lattices $\L_{\C,1},\ldots,\L_{\C,L},\L_{\F,1},\ldots,\L_{\F,L}$ via Theorem~\ref{t:lattice}. We will assume that the prime $p$ used in the lattice construction is large enough so that $\Qm_{\dt}^{[1:m]} = [ \Am_{\dt}^{[1:m]} ] \bmod{p}$ is full rank over $\ZZ_p$ for $m= 1,\ldots,L$. It is always possible to choose such a prime, as argued in~\cite[Lemmas 3, 4]{ncnc16}.

\noindent\textbf{Encoding:} Take the $p$-ary expansion of each message $w_{\dt,\ell}$ to obtain the message vector $\wv_{\ell} \in \ZZ_p^{\kfl - \kcl}$ for $\ell = 1,\ldots,L$. These vectors are then zero-padded to obtain
\begin{align} \label{e:msgzeropad}
\wbv_{\dt,\ell} = \begin{bmatrix}
\zerov_{k_{\C,\ell} - \kc} \\ 
\wv_{\dt,\ell} \\
\zerov_{\kf - k_{\F,\ell}} 
\end{bmatrix} \ .
\end{align} Recall from Section~\ref{s:signallevels} that the parameter $k_{\C,\ell} - \kc$ denotes how many of the top signal levels cannot be directly set by the $\lth$ encoder, due to its power constraint. Since the powers are assumed to be in decreasing order, it follows that these parameters are in increasing order, $k_{\C,1} - \kc \leq \cdots \leq k_{\C,L} - \kc$.

We now proceed to pre-invert the linear combinations in $L$ stages. Recall that the notation $\wv[i]$ refers to the $i^{\text{th}}$ entry of the vector $\wv$. 

\noindent\underline{Initialization Step, $k_{\C,L} -\kc + 1 \leq i \leq k$:} These signal levels are accessible by every encoder, meaning that we can simply apply the inverse,
\begin{align}
\begin{bmatrix}
\vv_{\dt,1}[i] \\ \vdots \\ \vv_{\dt,L}[i]
\end{bmatrix}
= \Qm_{\dt}^{-1} \begin{bmatrix}
\wbv_{\dt,1}[i] \\ \vdots \\ \wbv_{\dt,L}[i]
\end{bmatrix} \ .
\end{align} Note that this fully specifies all of the signal levels controlled by the $L^{\text{th}}$ encoder. Therefore, we set the remaining entries to zeros, {$\vv_{\dt,L}[1],\ldots,\vv_{\dt,L}[k_{\C,L} - \kc] = 0$}, apply the inverse linear labeling to obtain a fine lattice point
\begin{align}
\lambdav_{\dt,L} = \bar{\varphi}(\vv_{\dt,L}) \ ,
\end{align} and then generate our dithered codeword
\begin{align}
\sv_{\dt,L} = [ \lambdav_{\dt,L} + \dv_{\dt,L}] \bmod{\L_{\C,L}} 
\end{align}  where the dither vector $\dv_{\dt,L}$ is drawn independently and uniformly over $\Vc_{\C,L}$. This process fixes the ``don't care'' entries of the $L^{\text{th}}$ encoder (i.e., the top $k_{\C,L} - \kc$ signal levels) to  
\begin{align}
\ev_{\dt,L} = \varphi\big( Q_{\L_{\C,L}}(\lambdav_{\dt,L}  + \dv_{\dt,L}) \big) \ , \label{e:downlinkinterferenceL}
\end{align} which will act as interference towards the remaining $L-1$ encoders that send information in these levels.

For the rest of the signal levels, we proceed by induction for $m = 1,\ldots,L-1$, assuming that $\vv_{\dt,\ell},\lambdav_{\dt,\ell},\sv_{\dt,\ell},\ev_{\dt,\ell}$ have been set for $\ell = m+1,\ldots,L$.

\noindent\underline{Induction Step, $k_{\C,m} -\kc + 1 \leq i \leq k_{\C,m+1}$:} The first $m$ encoders can directly set these signal levels and the remaining $L-m$ encoders contribute interference via their ``don't care entries.'' Thus, the encoding task is to first cancel out the interference from these ``don't care'' entries and then apply the the inverse of the $m^{\text{th}}$ leading principal submatrix,
\begin{align} \label{e:msgsymbolinverse}
\begin{bmatrix}
\vv_{\dt,1}[i] \\ \vdots \\ \vv_{\dt,m}[i]
\end{bmatrix}
= \Big(\Qm_{\dt}^{[1:m]}\Big)^{-1} \begin{bmatrix}
\wbv_{\dt,1}[i] \oplus \bigoplus_{\ell =m+1}^L q_{\dt,1,\ell} \,\ev_{\dt,\ell}[i] \\ \vdots \\ \wbv_{\dt,m}[i] \oplus \bigoplus_{\ell =m+1}^L q_{\dt,m,\ell}\, \ev_{\dt,\ell}[i]
\end{bmatrix} \ .
\end{align} Note that this fully specifies all of the signal levels controlled by the $m^{\text{th}}$ encoder. Therefore, we set {$\vv_{\dt,m}[1],\ldots,\vv_{\dt,m}[k_{\C,m} - \kc] = 0$}, apply the inverse linear labeling to obtain a fine lattice point
\begin{align}
\lambdav_{\dt,m} = \bar{\varphi}(\vv_{\dt,m}) \ ,
\end{align} and then generate our dithered codeword
\begin{align}
\sv_{\dt,m} = [ \lambdav_{\dt,m} + \dv_{\dt,m}] \bmod{\L_{\C,m}} \ . \label{e:downlinkditheredcodeword}
\end{align}  where the dither vector $\dv_{\dt,m}$ is drawn independently and uniformly over $\Vc_{\C,m}$. This sets the ``don't care'' entries of the $m^{\text{th}}$ encoder (i.e., the top $k_{\C,m} - \kc$ signal levels) to 
\begin{align}
\ev_{\dt,m} = \varphi\big( Q_{\L_{\C,m}}(\lambdav_{\dt,m}  + \dv_{\dt,m}) \big) \ , 
\end{align}which will act as interference towards the remaining $m-1$ encoders that send information in these levels.

After all signal levels have been set, we stack the dithered codewords 
\begin{align}
\Sm_{\dt} = \begin{bmatrix}
\sv_{\dt,1}\T \\ 
\vdots \\ 
\sv_{\dt,L}\T 
\end{bmatrix} 
\end{align} and apply the beamforming matrix to create the channel input
\begin{align}
\Xm_{\dt} = \Bm_{\dt} \Sm_{\dt} \ .
\end{align}

\noindent\textbf{Decoding:} The goal of each receiver is to decode its message vector $\wv_{\dt,\ell}$. As a first step, it will make an estimate of the following integer-linear combination of the lattice codewords,
\begin{align}
\muv_{\dt,m} = \bigg[ \sum_{\ell = 1}^L a_{\dt,m,\ell}\, \lambdatv_{\dt,\ell} \bigg] \bmod{\L_{\C}} \label{e:integercombination}
\end{align} where $a_{\dt,m,\ell}$ is the $(m,\ell)^{\text{th}}$ entry of $\Am_{\dt}$ and $\lambdatv_{\dt,\ell} = \lambdav_{\dt,\ell} - Q_{\Lambda_{\C,\ell}}(\lambdav_{\dt,\ell} + \dv_{\dt,\ell})$. It forms its estimate by equalizing its observation
\begin{align}
\ytv_{\dt,m}\T = \cv_{\dt,m}\T \Ym_{\dt,m} \ ,
\end{align} removing the dither vectors, quantizing onto the $\mth$ fine lattice, and taking the modulus with respect to the coarsest lattice,
\begin{align}
\muhv_{\dt,m} = \bigg[ Q_{\L_{\F,m}}\bigg( \ytv_{\dt,m} - \sum_{\ell = 1}^L a_{\dt,m,\ell}\, \dv_{\dt,\ell} \bigg) \bigg] \bmod{\L_{\C}} \ .
\end{align} The linear label of this estimate can be viewed as an estimate of the desired message along with zero-padding,
\begin{align}
\varphi(\muhv_{\dt,m}) = \begin{bmatrix}
\mathbf{\tilde{e}}_{\dt,m} \\ \mathbf{\hat{w}}_{\dt,m} \\ \zerov_{\kf - k_{\F,m}} 
\end{bmatrix} \ . \label{e:inverselabel}
\end{align} for some $\mathbf{\tilde{e}}_{\dt,m} \in \ZZ_p^{k_{\C,m} - \kc}$. As we will argue below, if $\muhv_{\dt,m} = \muv_{\dt,m}$, then $\mathbf{\hat{w}}_{\dt,m} = \wv_{\dt,m}$.

\begin{theorem} \label{t:downlink}
Choose a power allocation $\dpm = \diag(P_{\dt,1},\ldots,P_{\dt,L})$, beamforming matrix $\Bm_{\dt} \in \RR^{N \times L}$, channel matrices $\Hm_{\dt,m} \in \RR^{M_m \times N}$, full-rank integer matrix $\Am_{\dt} \in \ZZ^{L \times L}$, and equalization vectors $\cv_{\dt,m} \in \RR^{M_m}$. Assume, without loss of generality, that the identity permutation is admissible for the downlink according to Definition~\ref{d:downlinkadmissible}. Then, the following rates are achievable 
\begin{align}
R_{\dt,m} &= \frac{1}{2}\log^+\bigg( \frac{P_{\dt,m}}{\sigma_{\dt,m}^2}\bigg) ,~~~ m = 1,\ldots,L, \\
\sigma_{\dt,m}^2 &=  \| \cv_{\dt,m} \|^2 + \Big\|  \big( \cv_{\dt,m}\T \Hm_{\dt,m} \Bm_{\dt} - \av_{\dt,m}\T \big) \dpm^{1/2} \Big\|^2 \ .
\end{align} 
\end{theorem}

\begin{IEEEproof} By the Crypto Lemma, each dithered codeword $\sv_{\dt,\ell}$ is uniformly distributed over $\Vc_{\C,\ell}$ and independent of the other dithered codewords. Thus, by Theorem~\ref{t:lattice}(b), we have that $\frac{1}{n}\ex \| \sv_{\dt,\ell} \|^2 \leq P_{\dt,\ell}$, which guarantees that the power constraint is met
\begin{align}
\frac{1}{n}\ex\big[\tr(\Xm_{\dt}\T \Xm_{\dt})\big] &= \frac{1}{n}\ex\big[\tr(\Sm_{\dt}\T \Bm_{\dt}\T \Bm_{\dt} \Sm_{\dt})\big] \\ 
&= \frac{1}{n}\ex\big[\tr( \Bm_{\dt}\T \Bm_{\dt} \Sm_{\dt}\T\Sm_{\dt})\big] \\
&= \frac{1}{n}\tr\big(\Bm_{\dt}\T \Bm_{\dt} \ex[\Sm_{\dt}\T\Sm_{\dt}]\big) \\
&\leq \frac{1}{n}\tr\big(\Bm_{\dt}\T \Bm_{\dt} \Pm_{\dt} \big) = \ptd  \ . 
\end{align}

At the receiver side, we need to argue that $\muhv_{\dt,m} = \muv_{\dt,m}$ with high probability and, if so, $\mathbf{\hat{w}}_{\dt,m} = \wv_{\dt,m}$. We begin by examining the linear label of $\muv_{\dt,m}$,
\begin{align}
\uv_{\dt,m} &= \varphi(\muv_{\dt,m})\\
&= \bigoplus_{\ell=1}^L q_{\dt,m,\ell} \Big( \varphi(\lambdav_{\dt,\ell}) \ominus \varphi\big(Q_{\Lambda_{\C,\ell}}(\lambdav_{\dt,\ell} + \dv_{\dt,\ell})\big) \Big) \\
&= \bigoplus_{\ell=1}^L q_{\dt,m,\ell} \Big( \varphi(\lambdav_{\dt,\ell}) \ominus \ev_{\dt,\ell} \Big) \ . 
\end{align} Now, we examine the $i^{\text{th}}$ symbol of this linear label for $k_{\C,m} - \kc +1 \leq i \leq k$,
\begin{align}
&\uv_{\dt,m}[i]\\ &= \bigoplus_{\ell = 1}^L q_{\dt,m,\ell}\, (\vv_{\dt,\ell}[i] \ominus \ev_{\dt,\ell}[i]) \\
&\overset{(a)}{=} \bigoplus_{\ell = 1}^m q_{\dt,m,\ell} \, \vv_{\dt,\ell}[i] \ominus \bigoplus_{\ell = m+1}^L q_{\dt,m,\ell}\, \ev_{\dt,\ell}[i] \\ 
&\overset{(b)}{=} \wbv_{\dt,m}[i] \oplus \bigoplus_{\ell = m+1}^L q_{\dt,m,\ell} \, \ev_{\dt,\ell}[i] \ominus \bigoplus_{\ell = m+1}^L q_{\dt,m,\ell}\, \ev_{\dt,\ell}[i] \\ 
&= \wbv_{\dt,m}[i]
\end{align} where $(a)$ uses the fact that $\vv_{\dt,\ell}[i] = 0$ for $\ell = m+1,\ldots,L$ by construction and $\ev_{\dt,\ell}[i] = 0$ for $\ell = 1,\ldots,m$ via Definition~\ref{d:linearlabeling}(a) since $\ev_{\dt,\ell}$ is the linear label of a lattice point from $\L_{\C,\ell}$ and $(b)$ follows from plugging in~\eqref{e:msgsymbolinverse}. From~\eqref{e:msgzeropad} it follows that, if $\muhv_{\dt,m} = \muv_{\dt,m}$, then $\mathbf{\hat{w}}_{\dt,m} = \wv_{\dt,m}$. Note that, since the last $\kf - k_{\F,m}$ entries of $\wbv_{\dt,m}$ are zero, we know from Definition~\ref{d:linearlabeling}(a) that $\muv_{\dt,m} \in \L_{\F,m}$. 

We need to argue that $\muhv_{\dt,m} = \muv_{\dt,m}$ with probability at least $1 - \epsilon$. Recall from~\eqref{e:effectivedownlinkchannel} and~\eqref{e:zdeff} that $\ytv_{\dt,m}\T = \av_{\dt,m}\T \Sm_{\dt} + \zv_{\dt,\eff,m}\T$. Thus, 
\begin{align}
\ytv_{\dt,m} &= \sum_{\ell=1}^L a_{\dt,m,\ell} \big( \lambdav_{\dt} + \dv_{\dt,m} -  Q_{\Lambda_{\C,\ell}}(\lambdav_{\dt,\ell} + \dv_{\dt,\ell})\big) + \zv_{\dt,\eff,m} \\
&= \sum_{\ell=1}^L a_{\dt,m,\ell} ( \lambdatv_{\dt,\ell} + \dv_{\dt,\ell} )+ \zv_{\dt,\eff,m} \ , 
\end{align} and, using~\eqref{e:nestedquantization},
\begin{align}
\muhv_{\dt,m} = \big[ Q_{\L_{\F,m}}( \muv_{\dt,m} + \zv_{\dt,\eff,m} ) \big] \bmod{\L_{\C}} \ .
\end{align} From Theorem~\ref{t:lattice}(c), we know that, since $\muv_{\dt,m} \in \L_{\F,m}$, the quantization step can tolerate noise with effective variance $\sigma_{\dt,m}^2$, which implies that $\pr( \muhv_m \neq \muv_m ) < \epsilon$. From Theorem~\ref{t:lattice}(d), we know that the rate satisfies $R_{\dt,m} > \frac{1}{2} \log^+(P_{\dt,m}/ \sigma_{\dt,m}^2) - \epsilon$. Finally, following the steps in~\cite[Appendix H]{ncnc16}, we can show that good fixed dither vectors exist. \end{IEEEproof}

\begin{example}\label{ex:algebraicdpc}
The choice of the integer matrix places constraints on the power levels that can be associated with each user. Consider the following integer matrices:
$$\Am_1 = \begin{bmatrix} 2 & 1 \\ 0 & 1 \end{bmatrix} \qquad \qquad \Am_2 = \begin{bmatrix} 0 & 1 \\ 2 & 1 \end{bmatrix}  \qquad \qquad \Am_3 = \begin{bmatrix} 2 & 1 \\ 1 & 1 \end{bmatrix} $$ and assume that $P_1 > P_2$. Note that $\mathrm{rank}{\big(\Am_1^{[1:m]}\big)} = m$ for $m = 1,2$, and thus the identity permutation is admissible, meaning that power $P_1$ can be associated with user $1$ and power $P_2$ with user $2$. Within our coding scheme, the key step is that we can invert $\Qm_1^{[1:1]} = [\Am_1^{[1:1}] \bmod{p}$ over $\ZZ_p$ since it is non-zero. In contrast, note that $\Am_2$ corresponds to exchanging the rows of $\Am_1$. If we also exchange the beamforming vectors, then, according to Remark~\ref{r:downlinkpermute}, this should correspond to associating power $P_1$ with user $2$ and power $P_2$ with user $1$. However, since $\Am_2^{[1:1]}=0$, we will not be able to invert $\Qm_1^{[1:1]} = [\Am_1^{[1:1}] \bmod{p}$, and this permutation is inadmissible. Finally, note that for $\Am_3$, both permutations will be admissible.
\end{example}

\section{Uplink-Downlink Duality} \label{s:duality}

As discussed in Section~\ref{s:capacityregions}, the uplink and downlink capacity regions are duals of one another\cite{vjg03,vt03,yc04,wss06}. Furthermore, for conventional linear architectures, we can achieve the same rate tuple on dual uplink and downlink channels by exchanging the roles of the beamforming and equalization matrices (and transposing them)~\cite{vt03}. (See Theorem~\ref{t:dualityconventional} for a precise statement in our notation.) For integer-forcing architectures, we can establish a similar form of uplink-downlink duality, but only for the sum rate.

Let
\begin{align}
\beta_{\ut,\ell} \triangleq \frac{P_{\ut,\ell}}{\sigma_{\ut,\ell}^2}
\end{align} denote the $\lth$ effective SINR for the uplink and let
\begin{align}
\beta_{\dt,\ell} \triangleq \frac{P_{\dt,\ell}}{\sigma_{\dt,\ell}^2}
\end{align} denote the $\lth$ effective SINR for the downlink. Our uplink-downlink duality results stem from showing that if the effective SINRs $\beta_{\ut,\ell}$ can be established on the uplink, then the effective SINRs $\beta_{\dt,\ell} = \beta_{\ut,\ell}$ can be established on the downlink, and vice versa. Unfortunately, this does not immediately translate to duality of the achievable rate tuples, since the rates $R_{\ut,\ell}=\frac{1}{2} \log^+(\beta_{\ut,\ell})$ and $R_{\dt,\ell} = \frac{1}{2}\log^+(\beta_{\dt,\ell})$ are only achievable within our integer-forcing framework if the identity permutation is admissible for both the uplink and downlink. 

In general, as discussed in Remark~\ref{r:permutations}, the identity permutation may not be admissible on both the uplink and downlink, even with the freedom to reindex the transmitters and receivers. However, we can always find permutations $\pi_{\ut}$ and $\pi_{\dt}$ such that the rates $R_{\ut,\ell} = 1/2\log^+(P_{\ut,\ell} / \sigma_{\ut,\pi_{\ut}(\ell)}^2)$ and $R_{\dt,\ell} = 1/2\log^+(P_{\dt,\pi_{\dt}(\ell)} / \sigma_{\dt,\ell}^2)$ are achievable via integer-forcing. Therefore, the duality of the effective SINRs allows us to establish sum-rate duality, as shown below.

\begin{lemma} \label{l:sumrateduality} Assume that the rates $R_{\ut,\ell} = \frac{1}{2} \log^+\big(P_{\ut,\ell} / \sigma_{\ut,\pi_{\ut}(\ell)}^2\big) > 0$, $\ell = 1,\ldots,L$, are achievable on the uplink for some permutation $\pi_{\ut}$. Also, assume that the rates $R_{\dt,\ell} = 1/2\log^+(P_{\dt,\pi_{\dt}(\ell)} / \sigma_{\dt,\ell}^2))$, $\ell = 1,\ldots,L$, are achievable on the uplink for some permutation $\pi_{\dt}$ and that the effective SINRs are equal, $\beta_{\ut,\ell} = \beta_{\dt,\ell}$, $\ell = 1,\ldots,L$.  Then, the downlink sum rate is at least as large as the uplink sum rate,
\begin{align}
\sum_{\ell = 1}^L R_{\dt,\ell} \geq \sum_{\ell = 1}^L R_{\ut,\ell} \ .
\end{align}
\end{lemma}
\begin{IEEEproof}
We have that
\begin{align}
\sum_{\ell = 1}^L R_{\ut,\ell} &= \sum_{\ell = 1}^L \frac{1}{2} \log^+\bigg( \frac{P_{\ut,\ell}}{\sigma_{\ut,\pi_{\ut}(\ell)}^2} \bigg) \\
&\overset{(a)}{=}  \sum_{\ell = 1}^L \frac{1}{2} \log\bigg( \frac{P_{\ut,\ell}}{\sigma_{\ut,\pi_{\ut}(\ell)}^2} \bigg) \\
&= \frac{1}{2}\log\bigg(\prod_{\ell = 1}^L \frac{P_{\ut,\ell}}{\sigma_{\ut,\pi_{\ut}(\ell)}^2} \bigg) \\
&= \frac{1}{2}\log\bigg(\prod_{\ell = 1}^L \frac{P_{\ut,\ell}}{\sigma_{\ut,\ell}^2} \bigg) \\
&\overset{(b)}{=} \frac{1}{2}\log\bigg(\prod_{\ell = 1}^L \frac{P_{\dt,\ell}}{\sigma_{\dt,\ell}^2} \bigg) \\
&= \frac{1}{2}\log\bigg(\prod_{\ell = 1}^L \frac{P_{\dt,\pi_{\dt}(\ell)}}{\sigma_{\dt,\ell}^2} \bigg) \\
&=  \sum_{\ell = 1}^L \frac{1}{2} \log\bigg( \frac{P_{\dt,\pi_{\dt}(\ell)}}{\sigma_{\dt,\ell}^2} \bigg)  \\
&\overset{(c)}{\leq}  \sum_{\ell = 1}^L \frac{1}{2} \log^+\bigg( \frac{P_{\dt,\pi_{\dt}(\ell)}}{\sigma_{\dt,\ell}^2} \bigg)  = \sum_{\ell = 1}^L R_{\dt,\ell}
\end{align} where (a) follows from the fact that all uplink rates are assumed to be positive, (b) from the assumption that $\beta_{\ut,\ell} = \beta_{\dt,\ell}$, and (c) from the fact that $\log(x) \leq  \log^+(x)$.
\end{IEEEproof}

We now recall the following basic results for non-negative matrices. A vector or a matrix is \textit{non-negative} (i.e., $\mathbf{F}\ge 0$) if all its entries are non-negative. A vector or a matrix is \textit{positive} (i.e., $\mathbf{F}> 0$) if all its entries are positive. A square matrix $\mathbf{F}$ is a \textit{Z-matrix} if all its off-diagonal elements are non-positive. An \textit{M-matrix} is a Z-matrix with eigenvalues whose real parts are positive. The following lemma is a special case of~\cite[Theorem 1]{plemmons77}.

\begin{lemma}[{\cite[Theorem 1]{plemmons77}}] \label{l:mmatrix}
Let  $\mathbf{F}$ be a square Z-matrix. The following statements are equivalent:
\begin{enumerate}[(a)]
\item $\mathbf{F}$ is a non-singular M-matrix.
\item $\mathbf{F}$ has a non-negative inverse. That is, $\mathbf{F}^{-1}$ exists and $\mathbf{F}^{-1}\ge 0$.
\item There exists $\mathbf{x}\ge 0$ satisfying $\mathbf{F}\mathbf{x}>0$.
\item Every real eigenvalue of $\mathbf{F}$  is positive.
\end{enumerate} 
\end{lemma} 

The lemma below establishes uplink-downlink duality for the effective SINRs.

\begin{lemma}\label{l:SINRduality}
Select a power matrix $\upm$ and beamforming matrix $\mathbf{C}_{\ut}$, and let  $\ptu = \tr(\mathbf{C}_{\ut}\T\mathbf{C}_{\ut}\upm)$ denote the total power consumption. Furthermore, select a channel matrix $\mathbf{H}_{\ut}$, full-rank integer matrix $\mathbf{A}_{\ut}$, and equalization matrix $\mathbf{B}_{\ut}$. Let $\beta_{\ut,\ell},~\ell = 1,\ldots,L$ denote the effective uplink SINRs and assume that $\beta_{\ut,\ell} > 0,~\ell = 1,\ldots,L$. Then, for beamforming matrix $\mathbf{B}_{\dt} = \mathbf{B}_{\ut}\T$, channel matrix $\mathbf{H}_{\dt} = \mathbf{H}_{\ut}\T$, integer matrix $\mathbf{A}_{\dt} = \mathbf{A}_{\ut}\T$, and equalization matrix $\mathbf{C}_{\dt} = \mathbf{C}_{\ut}\T$, there exists a unique power matrix $\dpm$ with total power usage $\tr(\mathbf{B}_{\dt}\T\mathbf{B}_{\dt}\dpm) = \ptu$, that yields effective downlink SINRs $\beta_{\dt,\ell} = \beta_{\ut,\ell},~\ell = 1,\ldots,L$. The same relationship can be established starting from effective SINRs for the downlink and going to the uplink. 
\end{lemma}

\begin{IEEEproof}
Our proof is inspired by the approach of~\cite{vt03}. We begin by defining vector notation for the powers and effective SINRs,
\begin{align}
\rhov_{\ut} \triangleq \begin{bmatrix} P_{\ut,1} \\ \vdots \\ P_{\ut,L} \end{bmatrix} ~~ \rhov_{\dt} \triangleq \begin{bmatrix} P_{\dt,1} \\ \vdots \\ P_{\dt,L} \end{bmatrix} ~~ \betav_{\ut} \triangleq \begin{bmatrix} \beta_{\ut,1} \\ \vdots \\ \beta_{\ut,L} \end{bmatrix} ~~ \betav_{\dt} \triangleq \begin{bmatrix} \beta_{\dt,1} \\ \vdots \\ \beta_{\dt,L}\end{bmatrix}  .
\end{align} 

Let $\mathbf{\bar{c}}_{\ut,\ell}$ denote the $\lth$ column of $\Cm_{\ut}$ (with zero-padding included) and ${a}_{\ut,m,\ell}$ denote the $(m,\ell)^{\text{th}}$ entry of $\Am_{\ut}$. Define the $L \times L$ non-negative matrix $\Mm_{\ut}$ whose $(m,\ell)^{\text{th}}$ entry is $[\Mm_{\ut}]_{m,\ell} = ( \bv_{\ut,m}\T \Hm_{\ut} \mathbf{\bar{c}}_{\ut,\ell} - a_{\ut,m,\ell} )^2$. Let $\Jm_{\ut} = \diag\big( [\| \bv_{\ut,1} \|^2~\cdots~\| \bv_{\ut,L} \|^2 ]\big)$. It can be verified that the relations 
\begin{align}
\sigma_{\ut,m}^2 = \| \bv_{\ut,m} \|^2 + \Big\| \big( \bv_{\ut,m}\T \Hm_{\ut} \Cm_{\ut} - \av_{\ut,m}\T \big) \Pm_{\ut}^{1/2} \Big\|^2 
\end{align} for $m = 1,\ldots,L$ can be equivalently expressed as 
\begin{align} \label{e:dualuplinkpower}
(\Id - \diag(\betav_{\ut}) \Mm_{\ut} ) \rhov_{\ut} = \Jm_{\ut} \betav_{\ut} \ .
\end{align}

We now repeat this process for the downlink. Let $\mathbf{\bar{b}}_{\dt,\ell}$ denote the $\lth$ column of $\Bm_{\dt}$ and $a_{\dt,m,\ell}$ denote the $(m,\ell)^{\text{th}}$ entry of $\Am_{\dt}$. Define the $L \times L$ non-negative matrix $\Mm_{\dt}$ whose $(m,\ell)^{\text{th}}$ entry is $[\Mm_{\dt}]_{m,\ell} = (\cv_{\dt,m}\T \Hm_{\dt,m} \mathbf{\bar{b}}_{\dt,\ell} - a_{\dt,m,\ell})^2$. Let $\Gm_{\dt} = \diag\big([ \| \cv_{\dt,1} \|^2 ~\cdots~ \| \cv_{\dt,L} \|^2] \big)$. It can be verified that the relations
\begin{align}
\sigma_{\dt,m}^2 = \| \cv_{\dt,m} \|^2 + \Big\| \big( \cv_{\dt,m}\T \Hm_{\dt,m} \Bm_{\dt} - \av_{\dt,m}\T  \big) \Pm_{\dt}^{1/2} \Big\|^2
\end{align} for $m = 1,\ldots, L$ can be equivalently expressed as 
\begin{align}
(\Id - \diag(\betav_{\dt}) \Mm_{\dt} ) \rhov_{\dt} = \Gm_{\dt} \betav_{\dt} \ .
\end{align}

By assumption, we have that $\mathbf{A}_{\dt}=\mathbf{A}_{\ut}\T$, $\mathbf{B}_{\dt}=\mathbf{B}_{\ut}\T$, $\mathbf{C}_{\dt}=\mathbf{C}_{\ut}\T$, and $\mathbf{H}_{\dt}=\mathbf{H}_{\ut}\T$. It follows that $\Mm_{\dt}\T = \Mm_{\ut}$ as well. Furthermore, since $\Mm_{\ut}$ and $\Mm_{\dt}$ are non-negative, we have that $(\Id - \diag(\betav_{\ut}) \Mm_{\ut})$ and $(\Id - \diag(\betav_{\dt}) \Mm_{\dt})$ are Z-matrices. Since, by assumption, $\betav_{\ut} > 0$, we have that $\Jm_{\ut} \betav_{\ut} > 0$ and thus $(\Id - \diag(\betav_{\ut}) \Mm_{\ut})$ satisfies condition (c) of Lemma~\ref{l:mmatrix}. This implies that every real eigenvalue of $(\Id - \diag(\betav_{\ut}) \Mm_{\ut})$ is positive. Setting $\betav_{\dt} = \betav_{\ut}$, we have that 
\begin{align*}
\eig(\diag(\betav_{\ut})\mathbf{M}_{\ut})&=\text{eig}(\diag(\betav_{\ut}) \mathbf{M}_{\dt}\T )\\
&=\eig(\mathbf{M}_{\dt}\diag(\betav_{\ut}))\\
&=\eig(\mathbf{M}_{\dt} \diag(\betav_{\dt}))\\
&=\eig(\diag(\betav_{\dt})\mathbf{M}_{\dt}) \ ,
\end{align*} which implies that all real eigenvalues of $(\Id - \diag(\betav_{\dt}) \Mm_{\dt})$ are also positive, satisfying condition (d) of Lemma~\ref{l:mmatrix}. This implies that the inverse $(\Id - \diag(\betav_{\dt}) \Mm_{\dt})^{-1}$ exists and is non-negative. Combining this with the fact that $\Gm_{\dt} \betav_{\dt} \geq 0$, we know that there exists a non-negative power vector
\begin{align} \label{e:dualdownlinkpower}
\rhov_{\dt} = (\Id - \diag(\betav_{\dt}) \Mm_{\dt})^{-1}\Gm_{\dt} \betav_{\dt} 
\end{align} that attains the desired effective downlink SINRs.

It remains to show that the total downlink power consumption is equal to the total uplink power consumption. Define $\Gm_{\ut} = \diag\big([ \| \cv_{\dt,1} \|^2 ~\cdots ~\| \cv_{\dt,L} \|^2]\big)$ and $\Jm_{\dt}$ as the $L \times L$ matrix with $(m,\ell)^{\text{th}}$ entry $[ \Jm_{\dt}]_{m,\ell} = b_{\dt,m,\ell}^2$ where $b_{\dt,m,\ell}$ is the $(m,\ell)^{\text{th}}$ entry of $\Bm_{\dt}$. The total uplink power consumption can be written as
\begin{align}
P_{\text{total}} &= \tr(\Cm_{\ut}\T \Cm_{\ut} \Pm_{\ut}) \\&= \mathbf{1}\T \Gm_{\ut} \rhov_{\ut} = \mathbf{1}\T \Gm_{\ut} (\Id - \diag(\betav_{\ut}) \Mm_{\ut} )^{-1}\Jm_{\ut} \betav_{\ut} 
\end{align} The total downlink power consumption can be written as \begin{align}
\tr(\Bm_{\dt}\T \Bm_{\dt} \Pm_{\dt}) &= \mathbf{1}\T \Jm_{\dt} \rhov_{\dt} = \mathbf{1}\T \Jm_{\dt} (\Id - \diag(\betav_{\dt}) \Mm_{\dt} )^{-1}\Gm_{\dt} \betav_{\dt} \ .
\end{align} We now demonstrate these quantities are equal:
\begin{align*}
&\mathbf{1}\T\mathbf{J}_{\dt}\big(\mathbf{I}_L-\diag(\dsinr)\mathbf{M}_{\dt}\big)^{-1}\mathbf{G}_{\dt}\dsinr\\
=&\mathbf{1}\T\mathbf{J}_{\dt}\Big(\mathbf{G}_{\dt}^{-1}\diag(\dsinr)^{-1}-\mathbf{G}_{\dt}^{-1}\mathbf{M}_{\dt}\Big)^{-1}\mathbf{1}\\
=&\mathbf{1}\T\Big(\diag(\dsinr)^{-1}\mathbf{G}_{\dt}^{-1}-\mathbf{M}_{\dt}\T\mathbf{G}_{\dt}^{-1}\Big)^{-1}\mathbf{J}_{\dt}\T\mathbf{1}\\
=&\mathbf{1}\T\Big(\diag(\dsinr)^{-1}\mathbf{G}_{\dt}^{-1}-\mathbf{M}_{\dt}\T\mathbf{G}_{\dt}^{-1}\Big)^{-1}\mathbf{J}_{\ut}\mathbf{1}\\
=&\mathbf{1}\T\Big(\mathbf{J}_{\ut}^{-1}\diag(\usinr)^{-1}\mathbf{G}_{\dt}^{-1}-\mathbf{J}_{\ut}^{-1}\mathbf{M}_{\ut}\mathbf{G}_{\dt}^{-1}\Big)^{-1}\mathbf{1}\\
\overset{(a)}{=}&\mathbf{1}\T\Big(\mathbf{J}_{\ut}^{-1}\diag(\usinr)^{-1}\mathbf{G}_{\ut}^{-1}-\mathbf{J}_{\ut}^{-1}\mathbf{M}_{\ut}\mathbf{G}_{\ut}^{-1}\Big)^{-1}\mathbf{1}\\
=&\mathbf{1}\T\mathbf{G}_{\ut}\big(\mathbf{I}_L-\diag(\usinr)\mathbf{M}_{\ut}\big)^{-1}\mathbf{J}_{\ut}\usinr=\ptu \ . 
\end{align*} where $(a)$ uses the fact that $\Gm_{\dt} = \Gm_{\ut}$ since $\Cm_{\dt}\T = \Cm_{\ut}$.
\end{IEEEproof}

\noindent{\textit{Proof of Theorem~\ref{t:duality}:}}
Without loss of generality, we assume that the identity permutation is admissible for the uplink and that all achievable rates are positive $R_{\ut,\ell} = \frac{1}{2} \log^+(\beta_{\ut,\ell}) > 0$, $\ell=1,\ldots,L$. From Lemma~\ref{l:SINRduality}, we know that we can establish effective downlink SINRs $\beta_{\dt,\ell} = \beta_{\ut,\ell}$ with the same total power consumption. Finally, it follows from Lemma~\ref{l:sumrateduality} that a sum rate satisfying $\sum_{\ell} R_{\dt,\ell} \geq \sum_\ell R_{\ut,\ell}$ is achievable on the downlink. The proof for starting from the downlink is identical. \hfill $\blacksquare$

\section{Iterative Optimization via Duality} \label{s:optimization}

In this section, we present an iterative optimization algorithm for the non-convex problem of optimizing the beamforming and equalization matrices in order to maximize the sum rate. Our algorithm exploits uplink-downlink duality to converge to a local optimum. We also explore algorithms for optimizing the integer matrix. We first present our algorithm for the uplink channel, and afterwards state the modifications needed to use it on a downlink channel.

\subsection{Uplink Optimization} \label{s:uplinkoptimization}

For a given uplink channel matrix $\Hm_{\ut}$ and total power constraint $P_{\text{total}}$, our task is to maximize the sum rate by selecting a full-rank integer matrix $\Am_{\ut}$, equalization matrix $\Bm_{\ut}$, beamforming matrix $\Cm_{\ut}$, and power allocation matrix $\Pm_{\ut}$. Assuming, without loss of generality, that the identity permutation is admissible, this corresponds to the following optimization problem:
\begin{align}
&\max_{\Am_{\ut},\Bm_{\ut},\Cm_{\ut},\Pm_{\ut}} \sum_{\ell = 1}^L \frac{1}{2} \log^+ \left(\frac{P_{\ut,\ell}}{ \sigma_{\ut,\ell}^2}\right) \label{e:uplinkoptimization} \\ 
&~~~~~~\text{subject to~~} \tr(\Cm_{\ut}\T \Cm_{\ut} \Pm_{\ut}) \leq P_{\text{total}}.  \nonumber
\end{align} 

  Note that, even for a fixed integer matrix, simultaneously optimizing the remaining parameters is a non-convex problem. We now develop a suboptimal iterative algorithm based on uplink-downlink duality. Assume, for now, that the integer matrix $\Am_{\ut}$, beamforming matrix $\Cm_{\ut}$, and power allocation matrix $\Pm_{\ut}$ are fixed. Consider the Cholesky decomposition 
\begin{align}
\Fm \Fm\T=\big(\Pm_{\ut}^{-1}+\mathbf{C}_{\ut}\T \mathbf{H}_{\ut}\T\mathbf{H}_{\ut}\mathbf{C}_{\ut}\big)^{-1} \label{e:successiveminimabasis}
\end{align} where $\Fm$ is a lower triangular matrix with strictly positive diagonal entries. It follows from~\cite[Lemma 2]{ncnc16} that the optimal equalization matrix is
 \begin{align}
\mathbf{B}_{\ut} = \mathbf{A}_{\ut}\upm\mathbf{C}_{\ut}\T\mathbf{H}_{\ut}\T\big(\mathbf{I}+\mathbf{H}_{\ut}\mathbf{C}_{\ut}\upm\mathbf{C}_{\ut}\T\mathbf{H}_{\ut}\T\big)^{-1} \ , \label{e:optB} \end{align} and, for this choice, the effective noises can be written as $\sigma_{\ut,m}^2 = \| \av_{\ut,m}\T \Fm \|^2$ where $\av_m\T$ is the $\mth$ row of $\Am_{\ut}$.

We are now prepared to create the dual downlink channel. Set $\Am_{\dt} = \Am_{\ut}\T$, $\Bm_{\dt} = \Bm_{\ut}\T$, $\Cm_{\dt} = \Cm_{\ut}\T$, and $\Hm_{\dt} = \Hm_{\ut}\T$. Use~\eqref{e:dualdownlinkpower} to solve for the downlink power vector $\rhov_{\dt}$ and set $\Pm_{\dt} = \diag(\rhov_{\dt})$. By Theorem~\ref{t:duality}, the downlink sum rate is at least as large as the uplink sum rate.

The $\mth$ downlink effective noise variance is
\begin{align}
\sigma_{\dt,m}^2 =\| \cv_{\dt,m} \|^2 + \Big\|  \big( \cv_{\dt,m}\T \Hm_{\dt,m} \Bm_{\dt} - \av_{\dt,m}\T   \big) \dpm^{1/2} \Big\|^2 \ , \label{e:downlinknoise2}
\end{align} which corresponds to a quadratic optimization problem in $\cv_{\dt,m}$ (holding all other parameters fixed). The optimal equalization vector is
\begin{align}
\cv_{\dt,m}\T =\mathbf{a}_{\dt,m}\T \dpm \mathbf{B}_{\dt}\T\mathbf{H}_{\dt,m}\T(\mathbf{I}+\mathbf{H}_{\dt,m}\T\mathbf{B}_{\dt}\dpm\mathbf{B}_{\dt}\T\mathbf{H}_{\dt,m}\T)^{-1} \ . \label{e:optC}
\end{align} If we use these equalization vectors in place of the original ones, we can only improve the effective noise variances, and hence the effective SINRs and sum rate. 

Finally, we are ready to return to the uplink channel. We update the beamforming matrix $\Cm_{\ut} = \Cm_{\dt}\T$ using the optimal equalization vectors found above and use \eqref{e:dualuplinkpower} to solve for the uplink power vector $\rhov_{\ut}$ and set $\Pm_{\ut} = \diag(\rhov_{\ut})$. We begin a new iteration on the uplink and continue until we converge to a local optimum. (Note that, since each step has a unique minimizer, this two-block coordinate descent algorithm will always converge. See, for instance,~\cite[Corollary 2]{gs00} for a convergence proof.)

The overall process is succinctly summarized in Algorithm~\ref{alg:uplink}.

\begin{algorithm}
\caption{Iterative Uplink Optimization via Duality}
\begin{algorithmic}\label{alg:uplink}
\STATE {Given $\mathbf{H}_{\ut}$ and $P_{\text{total}}$.}
\STATE {Set initial parameters $\mathbf{A}_{\ut}$, $\mathbf{B}_{\ut}$, $\mathbf{C}_{\ut}$, and $\Pm_{\ut}$.} 
\STATE  {Calculate initial uplink SINRs $\usinr$.}
\WHILE {$\usinr$ not converged}
\STATE {Set $\Bm_{\ut}$ using~\eqref{e:optB}.}
\STATE  {Create virtual dual downlink channel with $\mathbf{A}_{\dt} = \mathbf{A}_{\ut}\T$, $\mathbf{B}_{\dt} = \mathbf{B}_{\ut}\T$, and $\mathbf{C}_{\dt}=\mathbf{C}_{\ut}\T$.}
\STATE  Solve for $\rhov_{\dt}$ using~\eqref{e:dualdownlinkpower} and set $\Pm_{\dt} = \diag(\rhov_{\dt})$.
\STATE   Optimize $\mathbf{C}_{\dt}$ using~\eqref{e:optC}.
\STATE   Update $\mathbf{C}_{\ut}=\mathbf{C}_{\dt}\T$.
\STATE  Solve for $\rhov_{\ut}$ using~\eqref{e:dualuplinkpower} and set $\Pm_{\ut} = \diag(\rhov_{\ut})$.
\STATE Update $\usinr$.
\ENDWHILE
\STATE Output $\mathbf{A}_{\ut}$, $\mathbf{B}_{\ut}$, $\mathbf{C}_{\ut}$,  $\Pm_{\ut}$, and $\usinr$.
\end{algorithmic}
\end{algorithm}

The choice of a good integer matrix $\Am_{\ut}$ is critical for the performance of integer-forcing. Consider the lattice $\Gm\T \ZZ^L$. Assuming $\Bm_{\ut}$ is chosen according to~\eqref{e:optB}, respectively, the optimal integer matrix corresponds to finding the shortest set of $L$ linearly independent basis vectors (i.e., the successive minima) for the lattice $\Gm\T \ZZ^L$. This problem, which is known as the Shortest Independent Vector Problem in the theoretical computer science literature, is conjectured to be NP-hard~\cite{miccianciogoldwasser}. However, it is possible to find approximately optimal solutions in polynomial time via the LLL algorithm~\cite{lll82}. See~\cite{aevz02,bremner} for more details.

\begin{remark}
While it is possible to iteratively update the integer matrix $\Am_{\ut}$ as part of the algorithm, we have observed (numerically) that good performance is available by choosing a good basis at the beginning, and then refining the remaining matrices.
\end{remark}

\subsection{Downlink Optimization}

For a given downlink channel matrix $\Hm_{\dt}$ and total power constraint $P_{\text{total}}$, our task is to maximize the sum rate by selecting the power allocation matrix $\Pm_{\dt}$, beamforming matrix $\Bm_{\dt}$, full-rank integer matrix $\Am_{\dt}$, and equalization matrix $\Cm_{\dt}$. Assuming, without loss of generality, that the identity permutation is admissible, we have the following optimization problem:
\begin{align}
&\max_{\Am_{\dt},\Bm_{\dt},\Cm_{\dt},\Pm_{\dt}} \sum_{\ell = 1}^L \frac{1}{2} \log^+ \left(\frac{P_{\dt,\ell}}{\displaystyle \sigma_{\dt,\ell}^2}\right) \label{e:downlinkoptimization} \\ 
&~~~~~~\text{subject to~~} \tr(\mathbf{B}_{\dt}\T\mathbf{B}_{\dt}\Pm_{\dt}) \leq P_{\text{total}}.  \nonumber
\end{align} As in the uplink case, this is a non-convex optimization problem, even if $\Am_{\dt}$ is fixed. We will use iterative uplink-downlink optimization to converge to a local optimum. (As in the uplink, convergence follows from~\cite[Corollary 2]{gs00}.)

The overall process is summarized in Algorithm~\ref{alg:downlink}.

\begin{algorithm}
\caption{Iterative Downlink Optimization via Duality}
\begin{algorithmic}\label{alg:downlink}
\STATE {Given $\mathbf{H}_{\dt}$ and $P_{\text{total}}$.}
\STATE {Set initial parameters $\mathbf{A}_{\dt}$, $\mathbf{B}_{\dt}$, $\mathbf{C}_{\dt}$, and $\Pm_{\dt}$.} 
\STATE  {Calculate initial downlink SINRs $\dsinr$.}
\WHILE {$\dsinr$ not converged}
\STATE   Optimize $\mathbf{C}_{\dt}$ using~\eqref{e:optC}.
\STATE  Create virtual uplink channel with $\mathbf{A}_{\ut} = \mathbf{A}_{\dt}\T$, $\mathbf{B}_{\ut} = \mathbf{B}_{\dt}\T$, and $\mathbf{C}_{\ut}=\mathbf{C}_{\dt}\T$. 
\STATE  Solve for $\rhov_{\ut}$ using~\eqref{e:dualuplinkpower} and set $\Pm_{\ut} = \diag(\rhov_{\ut})$.
\STATE {Optimize $\Bm_{\ut}$ using~\eqref{e:optB}.}
\STATE  Update $\mathbf{B}_{\dt} = \mathbf{B}_{\ut}\T$.
\STATE  Solve for $\rhov_{\dt}$ using~\eqref{e:dualdownlinkpower} and set $\Pm_{\dt} = \diag(\rhov_{\dt})$.
\STATE Update $\dsinr$.
\ENDWHILE
\STATE Output $\mathbf{A}_{\dt}$, $\mathbf{B}_{\dt}$, $\mathbf{C}_{\dt}$, $\Pm_{\dt}$, and $\dsinr$.
\end{algorithmic}
\end{algorithm}

\section{Numerical Results} \label{s:numerical}

We now provide simulation results for our integer-forcing architecture and compare its performance to that of zero-forcing as well as capacity bounds.\footnote{MATLAB code to generate these figures is available on the second author's website.} Owing to uplink-downlink duality, we can simultaneously plot the sum rate for the uplink and downlink channel. For simplicity, we will state our notation in terms of the uplink channel. We draw the channel matrix $\Hm_{\dt}$ elementwise i.i.d.~$\mathcal{N}(0,1)$.

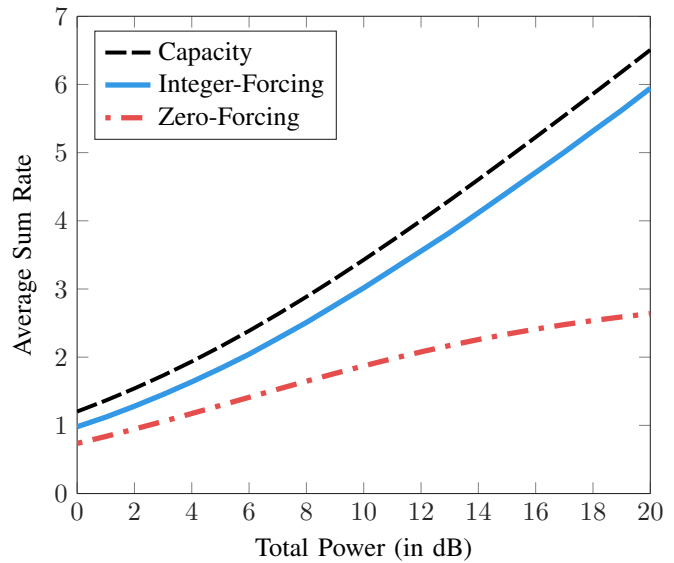
\begin{figure}[!h] 
\begin{center}
% This file was created by matlab2tikz v0.4.7 running on MATLAB 9.2.
% Copyright (c) 2008--2014, Nico Schlömer <nico.schloemer@gmail.com>
% All rights reserved.
% Minimal pgfplots version: 1.3
% 
% The latest updates can be retrieved from
%   http://www.mathworks.com/matlabcentral/fileexchange/22022-matlab2tikz
% where you can also make suggestions and rate matlab2tikz.
% 
\begin{tikzpicture}

\begin{axis}[%
compat=1.4,
width=3in,
height=2.5in,
scale only axis,
separate axis lines,
every outer x axis line/.append style={white!15!black},
every x tick label/.append style={font=\color{white!15!black}},
xmin=0,
xmax=20,
xlabel={Total Power (in dB)},
every outer y axis line/.append style={white!15!black},
every y tick label/.append style={font=\color{white!15!black}},
ymin=0,
ymax=7,
ylabel={Average Sum Rate},
legend style={at={(0.03,0.97)},anchor=north west,draw=white!15!black,fill=white,legend cell align=left}
]
\addplot [color=black,dash pattern=on 8pt off 2pt,line width=1.5pt]
  table[row sep=crcr]{%
0	1.20233658756391\\
1	1.36487041335984\\
2	1.54137922639918\\
3	1.73191103562103\\
4	1.93643345427746\\
5	2.15470796666425\\
6	2.38622913813353\\
7	2.63027098482403\\
8	2.88589261109631\\
9	3.15206527376635\\
10	3.42776229985767\\
11	3.71193718714586\\
12	4.00356674598127\\
13	4.30169733174085\\
14	4.60544556544029\\
15	4.91400236781543\\
16	5.22665104210662\\
17	5.54274397095714\\
18	5.86171806381907\\
19	6.18310419864124\\
20	6.50649027303475\\
};
\addlegendentry{Capacity};

\addplot [color=LineBlue,line width=2.0pt]
  table[row sep=crcr]{%
0	0.97944735789142\\
1	1.12091175871252\\
2	1.28086691174216\\
3	1.45320519509804\\
4	1.63836064704243\\
5	1.8349504407164\\
6	2.04375604232212\\
7	2.27478387126719\\
8	2.50896309290535\\
9	2.76275594791072\\
10	3.01649179886137\\
11	3.28550737581057\\
12	3.55626565832173\\
13	3.82745081002977\\
14	4.11790487912305\\
15	4.41248749352763\\
16	4.70928589549416\\
17	5.00665835438264\\
18	5.31714477048458\\
19	5.62125161083682\\
20	5.94398902269515\\
};
\addlegendentry{Integer-Forcing};

\addplot [color=LineRed,dash pattern=on 2pt off 4pt on 8pt off 4pt,line width=2.0pt]
  table[row sep=crcr]{%
0	0.735563756532414\\
1	0.837476532222781\\
2	0.945209891365331\\
3	1.05769784451923\\
4	1.17374767820824\\
5	1.2920803500128\\
6	1.41137176004837\\
7	1.53029447033784\\
8	1.64755944394779\\
9	1.76195671324376\\
10	1.87239299400261\\
11	1.97792364218635\\
12	2.07777635046464\\
13	2.17136467908901\\
14	2.25829072592338\\
15	2.33833761536514\\
16	2.41145365391703\\
17	2.47773069681191\\
18	2.53737941207938\\
19	2.59070380689387\\
20	2.63807679070543\\
};
\addlegendentry{Zero-Forcing};

\end{axis}
\end{tikzpicture}
\caption{Average sum rate under i.i.d.~Gaussian fading for integer-forcing and zero-forcing architectures with $L = 4$ single-antennas users and $N = 2$ basestation antennas.}
\label{fig:mimoN2L4}
\end{center}
\end{figure}

\begin{figure}[!h] 
\begin{center}
% This file was created by matlab2tikz v0.4.7 running on MATLAB 9.2.
% Copyright (c) 2008--2014, Nico Schlömer <nico.schloemer@gmail.com>
% All rights reserved.
% Minimal pgfplots version: 1.3
% 
% The latest updates can be retrieved from
%   http://www.mathworks.com/matlabcentral/fileexchange/22022-matlab2tikz
% where you can also make suggestions and rate matlab2tikz.
% 
\begin{tikzpicture}

\begin{axis}[%
compat=1.4,
width=3in,
height=2.5in,
scale only axis,
separate axis lines,
every outer x axis line/.append style={white!15!black},
every x tick label/.append style={font=\color{white!15!black}},
xmin=0,
xmax=20,
xlabel={Total Power (in dB)},
every outer y axis line/.append style={white!15!black},
every y tick label/.append style={font=\color{white!15!black}},
ymin=1,
ymax=11,
ylabel={Average Sum Rate},
legend style={at={(0.03,0.97)},anchor=north west,draw=white!15!black,fill=white,legend cell align=left}
]
\addplot [color=black,dash pattern=on 8pt off 2pt,line width=1.5pt]
  table[row sep=crcr]{%
0	1.85100235623796\\
1	2.10714078906329\\
2	2.38598712488494\\
3	2.68771128835417\\
4	3.01208388784988\\
5	3.35877251957809\\
6	3.72741717131934\\
7	4.11719828594087\\
8	4.52718435345407\\
9	4.95631892508085\\
10	5.40354979358825\\
11	5.86775644964149\\
12	6.3479335681774\\
13	6.84307541329273\\
14	7.35223814066537\\
15	7.87451412252375\\
16	8.4089286815306\\
17	8.95458933993762\\
18	9.51062897572727\\
19	10.0762220760256\\
20	10.6506278510757\\
};
\addlegendentry{Capacity};

\addplot [color=LineBlue,line width=2.0pt]
  table[row sep=crcr]{%
0	1.64385476749131\\
1	1.87973619087806\\
2	2.13677722459595\\
3	2.41828911550954\\
4	2.73160146354723\\
5	3.07816400548494\\
6	3.44693394589953\\
7	3.83574167469866\\
8	4.24481160802725\\
9	4.67650104334136\\
10	5.12505491138169\\
11	5.58655910101709\\
12	6.06773032027361\\
13	6.56337827976879\\
14	7.07568724334648\\
15	7.59920996506717\\
16	8.13721073025305\\
17	8.68277003342438\\
18	9.2423704829275\\
19	9.81031808245358\\
20	10.3884035479479\\
};
\addlegendentry{Integer-Forcing};

\addplot [color=LineRed,dash pattern=on 2pt off 4pt on 8pt off 4pt,line width=2.0pt]
  table[row sep=crcr]{%
0	1.43203666823924\\
1	1.63391249370827\\
2	1.85121466433955\\
3	2.08348648858261\\
4	2.33023033328431\\
5	2.59094282889576\\
6	2.86514348407411\\
7	3.15239735591986\\
8	3.45233274726095\\
9	3.76465454360593\\
10	4.08915314380621\\
11	4.42570839567099\\
12	4.77428780142037\\
13	5.1349385782171\\
14	5.50777382345725\\
15	5.89295380342123\\
16	6.29066400072077\\
17	6.70109184325198\\
18	7.12440395611088\\
19	7.56072542300784\\
20	8.01012208039235\\
};
\addlegendentry{Zero-Forcing};

\end{axis}
\end{tikzpicture}%
\caption{Average sum rate under i.i.d.~Gaussian fading for integer-forcing and zero-forcing architectures with $L = 4$ single-antennas users and $N = 4$ basestation antennas.}
\label{fig:mimoN4L4}
\end{center}
\end{figure}
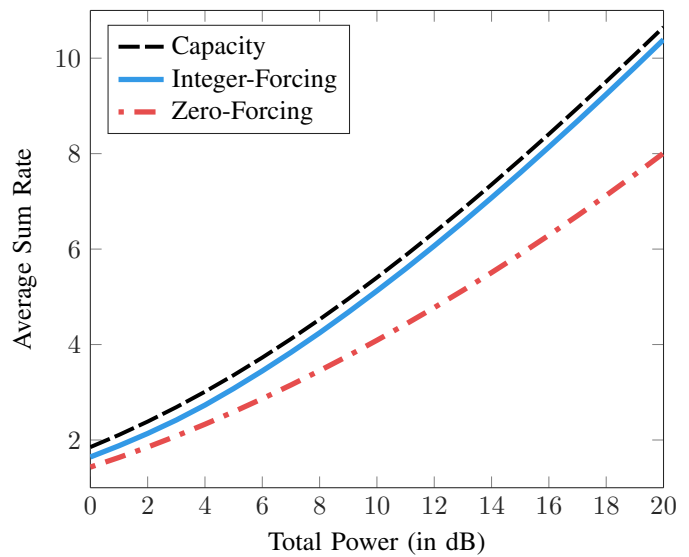

\begin{figure}[!h] 
\begin{center}
% This file was created by matlab2tikz v0.4.7 running on MATLAB 9.2.
% Copyright (c) 2008--2014, Nico Schlömer <nico.schloemer@gmail.com>
% All rights reserved.
% Minimal pgfplots version: 1.3
% 
% The latest updates can be retrieved from
%   http://www.mathworks.com/matlabcentral/fileexchange/22022-matlab2tikz
% where you can also make suggestions and rate matlab2tikz.
% 
\begin{tikzpicture}

\begin{axis}[%
compat=1.4,
width=3in,
height=2.5in,
scale only axis,
separate axis lines,
every outer x axis line/.append style={white!15!black},
every x tick label/.append style={font=\color{white!15!black}},
xmin=0,
xmax=20,
xlabel={Total Power (in dB)},
every outer y axis line/.append style={white!15!black},
every y tick label/.append style={font=\color{white!15!black}},
ymin=1,
ymax=8,
ylabel={Average Sum Rate},
legend style={at={(0.03,0.97)},anchor=north west,draw=white!15!black,fill=white,legend cell align=left}
]
\addplot [color=black,dash pattern=on 8pt off 2pt,line width=1.5pt]
  table[row sep=crcr]{%
0	1.42986756678583\\
1	1.61658064419392\\
2	1.81797374443427\\
3	2.03367468436008\\
4	2.26299885609765\\
5	2.5050493691875\\
6	2.75888266557111\\
7	3.0234737338303\\
8	3.29778870661315\\
9	3.58081375819287\\
10	3.87154003670104\\
11	4.16895043913702\\
12	4.47215533541093\\
13	4.78035166154917\\
14	5.09279962243386\\
15	5.40882149384209\\
16	5.72782678839225\\
17	6.04930657793191\\
18	6.3728266056251\\
19	6.6980197521795\\
20	7.02457829690936\\
};
\addlegendentry{Capacity};

\addplot [color=LineBlue,line width=2.0pt]
  table[row sep=crcr]{%
0	1.36993460076326\\
1	1.55955955864178\\
2	1.76387262829464\\
3	1.98417696758964\\
4	2.21705073627177\\
5	2.46203034345484\\
6	2.7183979274352\\
7	2.9853926682792\\
8	3.26055018751458\\
9	3.54435362021407\\
10	3.83597118986004\\
11	4.13366584901653\\
12	4.43688365581358\\
13	4.74512510946958\\
14	5.05754400692406\\
15	5.37345665618303\\
16	5.69237338287294\\
17	6.01371350944628\\
18	6.33702526109899\\
19	6.66211107624395\\
20	6.98858634698298\\
};
\addlegendentry{Integer-Forcing};

\addplot [color=LineRed,dash pattern=on 2pt off 4pt on 8pt off 4pt,line width=2.0pt]
  table[row sep=crcr]{%
0	1.30474217920656\\
1	1.48487784181825\\
2	1.67895460429346\\
3	1.88658606269944\\
4	2.10725483381177\\
5	2.34032772747425\\
6	2.58507403629687\\
7	2.84068658187806\\
8	3.10630460990027\\
9	3.38103725909105\\
10	3.66398622951333\\
11	3.95426642376755\\
12	4.25102364559735\\
13	4.55344881656503\\
14	4.86078853518734\\
15	5.17235210445581\\
16	5.4875153674959\\
17	5.80572181229683\\
18	6.12648144831105\\
19	6.44936794611813\\
20	6.77401449389521\\
};
\addlegendentry{Zero-Forcing};

\end{axis}
\end{tikzpicture}%
\caption{Average sum rate under i.i.d.~Gaussian fading for integer-forcing and zero-forcing architectures with $L = 2$ single-antennas users and $N = 4$ basestation antennas.}
\label{fig:mimoN4L2}
\end{center}
\end{figure}
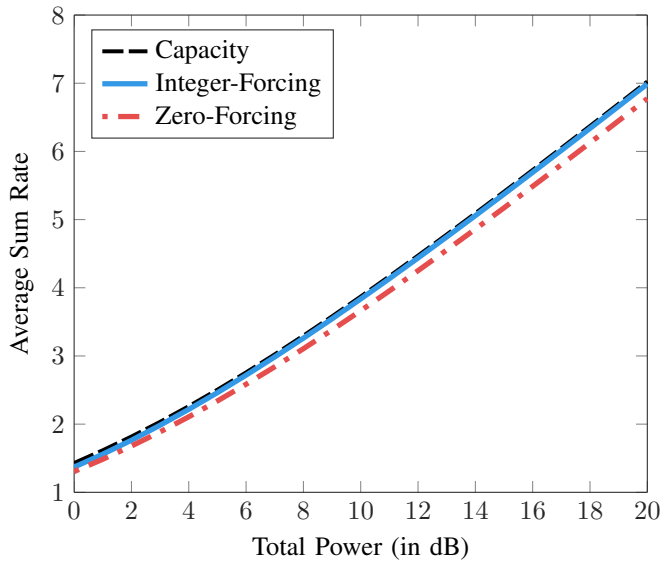

In our plots, the ``Capacity'' curves correspond to the MIMO MAC sum capacity (under a total power constraint) from~\eqref{e:uplinkcapacity} or to the MIMO BC sum capacity~\eqref{e:downlinkcapacity}. These expressions are evaluated following the dual decomposition approach from~\cite{yu06}.

The ``Integer-Forcing'' curves correspond to the sum rate for uplink integer-forcing from Theorem~\ref{t:uplinkif} or downlink integer-forcing from Theorem~\ref{t:downlink}. The integer matrix $\Am_{\ut}$ is chosen using the LLL algorithm to approximate the successive minima of the lattice $\mathbf{F}\T \mathbb{Z}^L$ where $\mathbf{F}$ is defined in~\eqref{e:successiveminimabasis} with the initial choice of $\Cm_{\ut} = \Id$.   Afterwards, we iteratively optimize $\Bm_{\ut}$, $\Cm_{\ut}$, and $\Pm_{\ut}$ using Algorithm~\ref{alg:uplink}.

The ``Zero-Forcing'' curves correspond to the sum rate of uplink zero-forcing from~\eqref{e:uplinkzf} or downlink zero-forcing from~\eqref{e:downlinkzf}. The matrices $\Bm_{\ut}$, $\Cm_{\ut}$, and $\Pm_{\ut}$ are iteratively optimized using Algorithm~\ref{alg:uplink} while holding $\Am_{\ut} = \Id$.

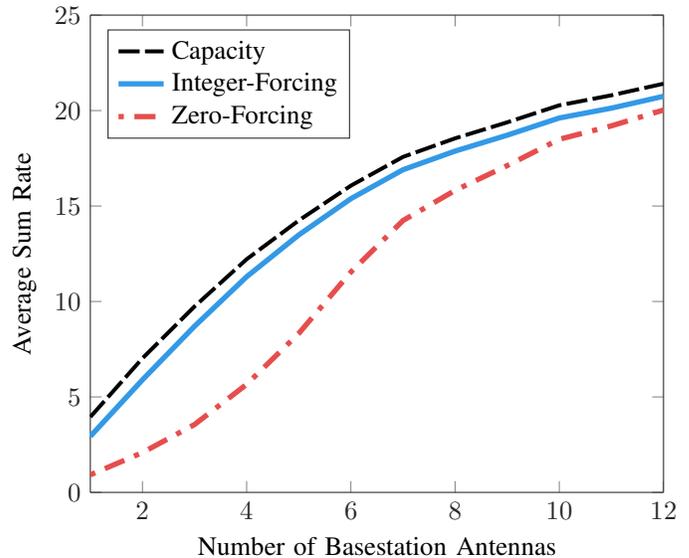
\begin{figure}[!h]
\begin{center}
% This file was created by matlab2tikz v0.4.7 running on MATLAB 9.2.
% Copyright (c) 2008--2014, Nico Schlömer <nico.schloemer@gmail.com>
% All rights reserved.
% Minimal pgfplots version: 1.3
% 
% The latest updates can be retrieved from
%   http://www.mathworks.com/matlabcentral/fileexchange/22022-matlab2tikz
% where you can also make suggestions and rate matlab2tikz.
% 
\begin{tikzpicture}

\begin{axis}[%
compat=1.4,
width=3in,
height=2.5in,
scale only axis,
separate axis lines,
every outer x axis line/.append style={white!15!black},
every x tick label/.append style={font=\color{white!15!black}},
xmin=1,
xmax=12,
xlabel={Number of Basestation Antennas},
every outer y axis line/.append style={white!15!black},
every y tick label/.append style={font=\color{white!15!black}},
ymin=0,
ymax=25,
ylabel={Average Sum Rate},
legend style={at={(0.03,0.97)},anchor=north west,draw=white!15!black,fill=white,legend cell align=left}
]
\addplot [color=black,dash pattern=on 8pt off 2pt,line width=1.5pt]
  table[row sep=crcr]{%
1	3.94978651755032\\
2	7.0233944798821\\
3	9.73914701537109\\
4	12.2146112917834\\
5	14.2436771475185\\
6	16.0658003396577\\
7	17.5720585859711\\
8	18.5449834031681\\
9	19.3834690013249\\
10	20.2757228051494\\
11	20.801038913811\\
12	21.4067705854914\\
};
\addlegendentry{Capacity};

\addplot [color=LineBlue,line width=2.0pt]
  table[row sep=crcr]{%
1	2.9374308447242\\
2	5.894877510487\\
3	8.70192349176429\\
4	11.2979223123379\\
5	13.4853179509548\\
6	15.3834637383809\\
7	16.8999116434424\\
8	17.8771963930698\\
9	18.7101360286246\\
10	19.6110686974389\\
11	20.1283622746958\\
12	20.7475049269959\\
};
\addlegendentry{Integer-Forcing};

\addplot [color=LineRed,dash pattern=on 2pt off 4pt on 8pt off 4pt,line width=2.0pt]  table[row sep=crcr]{%
1	0.919079665860557\\
2	2.0843281451363\\
3	3.5552932127663\\
4	5.66090165356379\\
5	8.32523441576738\\
6	11.5412995793207\\
7	14.2395865514832\\
8	15.817432473267\\
9	17.116056728182\\
10	18.4955766238375\\
11	19.1988200061506\\
12	20.0288731461188\\
};
\addlegendentry{Zero-Forcing};

\end{axis}
\end{tikzpicture}%
\caption{Average sum rate under i.i.d.~Gaussian fading for integer-forcing and zero-forcing architectures with $L = 6$ single-antennas users and $N$ basestation antennas at $P_{\text{total}} = 20$dB.}
\label{fig:varyantennasplot}
\end{center}
\end{figure}

In Figure~\ref{fig:mimoN2L4}, we have plotted the average sum rate with respect to $P_{\text{total}}$ for $L = 4$ single-antenna users and a basestation with $N=2$ antennas. In this scenario, there are not enough basestation antennas to invert the channel matrix, and thus the performance of zero-forcing saturates, whereas both the sum capacity and integer-forcing sum rate scale with $P_{\text{total}}$. In Figure~\ref{fig:mimoN4L4}, we increase the number of basestation antennas to $N=4$ while holding the number of users fixed at $L = 4$. The zero-forcing sum rate now scales with $P_{\text{total}}$, but there is still a significant gap to the sum capacity and integer-forcing performance. This gap can be nearly closed by reducing the number of users to $L = 2$ and keeping $N = 4$ basestation antennas.

Overall, we observe that the integer-forcing sum nearly matches the sum capacity. In contrast, zero-forcing operates near the sum capacity only when the number of basestation antennas $N$ is at least as large as the number of (single-antenna) users $L$. This is demonstrated in Figure~\ref{fig:varyantennasplot} by varying $N$ from $1$ to $12$ for $L = 12$ and $P_{\text{total}} = 20$dB.

\section{Conclusion} \label{s:conclude}

In this paper, we established an uplink-downlink duality relationship for integer-forcing. In the process, we extended prior work on downlink integer-forcing to allow for unequal powers and unequal rates. Using the duality relationship, we developed an iterative algorithm for the non-convex problem of optimizing the beamforming and equalization matrices. We also demonstrated that downlink integer-forcing can operate within a constant gap of the MIMO BC sum capacity.

An interesting direction for future work is utilizing uplink-downlink duality to optimize integer-forcing architectures for more complicated Gaussian networks. For instance, recent work~\cite{ehn15} has utilized uplink-downlink duality as a building block for optimizing the beamforming and equalization matrices used in integer-forcing interference alignment \cite{ncnc13ISIT}.

Another direction is to establish uplink-downlink duality between uplink integer-forcing enhanced by successive interference cancelation and downlink integer-forcing enhanced by dirty-paper coding. We investigated this relationship in an earlier conference paper~\cite{hns15}. Unfortunately, this result requires the identity permutation to be admissible on both the uplink and downlink without reindexing, which cannot be assumed without loss of generality. The key technical issue is that the SIC and DPC matrices must be lower and upper triangular, respectively, and this is not maintained under reindexing.

\begin{appendices}

%%%%%%%%%%%%%%%%%%%%%%%%%%%%%%%%%%%

\section{Proof of Theorem~\ref{t:downlinkconstantgap}}\label{a:downlinkconstantgap}

It is well-known~\cite{vjg03,vt03,yc04} that the sum capacity of the MIMO BC is equal to that of the dual MIMO MAC,
\begin{align}
\max_{\substack{\Km \succeq 0 \\ \tr(\Km) \leq P_{\text{total}}}} \frac{1}{2} \log\det\Big( \Id + \Hm_{\ut} \Km \Hm_{\ut}\T \Big) 
\end{align} where $\Hm_{\ut} = \Hm_{\dt}\T$. Select a covariance matrix $\Km_{\text{opt}}$ that attains the MIMO MAC sum capacity. Next, select a power allocation $\Pm_{\ut}$ and a beamforming matrix $\Cm_{\ut}$ satisfying $\Cm_{\ut} \Pm_{\ut} \Cm_{\ut}\T = \Km_{\text{opt}}$. From~\cite[Theorem 4]{ncnc16}, there exists an integer matrix $\Am_{\ut}$ such that integer-forcing via Theorem~\ref{t:uplinkif} attains the sum rate 
\begin{align}
\sum_{\ell = 1}^L R_{\ut,\ell} &= \frac{1}{2} \log\det\Big( \Id + \Hm_{\ut} \Cm_{\ut} \Pm_{\ut} \Cm_{\ut}\T \Hm_{\ut}\T \Big) - \frac{L}{2} \log{L} \\
&=  \frac{1}{2} \log\det\Big( \Id + \Hm_{\ut} \Km_{\text{opt}} \Hm_{\ut}\T \Big) - \frac{L}{2} \log{L} \ . 
\end{align} using the optimal equalization matrix $\Bm_{\ut}$ from~\eqref{e:optB}. From Theorem~\ref{t:duality}, we can attain the same sum rate on the downlink by using $\Am_{\dt} = \Am_{\ut}\T$, $\Bm_{\dt} = \Bm_{\ut}\T$, and $\Cm_{\dt} = \Cm_{\ut}\T$ as well as solving for the downlink power vector $\rhov_{\dt}$ using~\eqref{e:dualdownlinkpower} and setting $\Pm_{\dt} = \diag(\rhov_{\dt})$.

\end{appendices}

\section*{Acknowledgment}
The authors would like to thank Or Ordentlich for valuable discussions, Islam El Bakoury for help with the simulations, and the associate editor and reviewers for their thoughtful comments.

\bibliographystyle{ieeetr}

\begin{IEEEbiographynophoto}{Wenbo He} received the B.Sc. in electrical engineering from Polytechnic Institute of New York University and the M.Sc. and Ph.D. degrees in electrical engineering from Boston University, in 2011, 2014, and 2016, respectively. In 2016, Wenbo He joined MathWorks as a senior software engineer working in the sampling and scheduling field for Simulink core group. 

During his time in Boston University, his research interests focus on topics in information
theory and wireless communications, especially on multi-terminal coding for uplink and downlink channel.
\end{IEEEbiographynophoto}

\begin{IEEEbiographynophoto}{Bobak Nazer} (S '02 -- M '09) received the B.S.E.E. degree from Rice University, Houston, TX, in 2003, the M.S. degree from the University of California, Berkeley, CA, in 2005, and the Ph.D degree from the University of California, Berkeley, CA, in 2009, all in electrical engineering.

He is currently an Assistant Professor in the Department of Electrical and Computer Engineering at Boston University, Boston, MA. From 2009 to 2010, he was a postdoctoral associate in the Department of Electrical and Computer Engineering at the University of Wisconsin, Madison, WI. His research interests include  information theory, communications, signal processing, and neuroscience.

Dr. Nazer received the Eli Jury Award from the EECS Department at UC Berkeley in 2009, the Dean's Catalyst Award from the College of Engineering at BU in both 2011 and 2017, the NSF CAREER Award in 2013, the IEEE Communications Society and Information Theory Society Joint Paper Award in 2013, and the  the BU ECE Faculty Service Award in 2017. He was one of the co-organizers for the Spring 2016 Thematic Program at the Institut Henri Poincar\'{e} on the Nexus of Information and Computation Theories. 
\end{IEEEbiographynophoto}

\begin{IEEEbiographynophoto} {Shlomo Shamai (Shitz)} received the B.Sc., M.Sc., and Ph.D. degrees in
electrical engineering from the Technion---Israel Institute of Technology,
in 1975, 1981 and 1986 respectively.

During 1975-1985 he was with the Communications Research Labs,
in the capacity of a Senior Research Engineer. Since 1986 he is with
the Department of Electrical Engineering, Technion---Israel Institute of
Technology, where he is now a Technion Distinguished Professor,
and holds the William Fondiller Chair of Telecommunications.
His research interests encompasses a wide spectrum of topics in information
theory and statistical communications.

Dr. Shamai (Shitz) is an IEEE Fellow, an URSI Fellow, a member of the
Israeli Academy of Sciences and Humanities and a foreign member of the
US National Academy of Engineering. He is the recipient of the 2011
Claude E. Shannon Award, the 2014 Rothschild Prize in
Mathematics/Computer Sciences and Engineering and the
2017 IEEE Richard W. Hamming Medal.

He has been awarded the 1999 van der Pol Gold Medal of the Union Radio
Scientifique Internationale (URSI), and is a co-recipient of the 2000 IEEE
Donald G. Fink Prize Paper Award, the 2003, and
the 2004 joint IT/COM societies paper award, the 2007 IEEE Information
Theory Society Paper Award, the 2009 and 2015 European Commission FP7,
Network of Excellence in Wireless COMmunications (NEWCOM++, NEWCOM\#)
Best Paper Awards, the 2010 Thomson Reuters Award for International Excellence
in Scientific Research, the 2014 EURASIP Best Paper Award (for
the EURASIP Journal on Wireless Communications and Networking),
and the 2015 IEEE Communications Society Best Tutorial Paper Award.
He is also the recipient of 1985 Alon Grant for distinguished young scientists
and the 2000 Technion Henry Taub Prize for Excellence in Research.
He has served as Associate Editor for the Shannon Theory of the IEEE
Transactions on Information Theory, and has also served twice on the
Board of Governors of the Information Theory Society.
He has also served on the Executive Editorial Board of the IEEE Transactions
on Information Theory and on the IEEE Information Theory Society Nominations
and Appointments Committee.
\end{IEEEbiographynophoto}

\end{document}